\documentclass[twocolumn]{aastex63}
\usepackage{float}
\usepackage{tablefootnote}

\received{\today}
\submitjournal{ApJSS}

\shorttitle{On the Relation between $\kappa$ and $\beta$ }
\shortauthors{Eyelade et al}

\begin{document}
\title{On the Relation between Kappa Distribution Functions and the Plasma Beta Parameter \\ in the Earth Magnetosphere: THEMIS observations. }

\correspondingauthor{Adetayo V. Eyelade}
\email{adetayo.eyelade@usach.cl}
%

\author[0000-0002-2301-307X]{Adetayo V. Eyelade}
\affiliation{Departmento de F{\'i}sica, 
Universidad de Santiago de Chile (USACH),
Santiago, Chile}

\author[0000-0002-1053-3375]{Marina Stepanova}
\affiliation{Departmento de F{\'i}sica, 
Universidad de Santiago de Chile (USACH),
Santiago, Chile}

\author[0000-0003-2481-2348]{Crist{\'o}bal M. Espinoza}
\affiliation{Departmento de F{\'i}sica, 
Universidad de Santiago de Chile (USACH),
Santiago, Chile}

\author[0000-0002-9161-0888]{Pablo S. Moya}
\affiliation{Departmento de F{\'i}sica, Facultad de Ciencias, Universidad de Chile, 
Santiago, Chile}

\begin{abstract}
The Earth's magnetosphere represents a natural plasma laboratory that allows us to study the behavior of particle distribution functions in the absence of Coulomb collisions, typically described by the Kappa distributions. We have investigated the properties of these functions for ions and electrons in different magnetospheric regions, thereby making it possible to reveal the $\kappa$-parameters for a wide range of plasma beta ($\beta$) values (from $10^{-3}$ to $10^{2}$). This was done using simultaneous ion and electron measurements from the five Time History of Events and Macroscale Interactions during Substorms (THEMIS) spacecraft spanning the years 2008 to 2018. It was found that for a fixed plasma $\beta$, the $\kappa$-index and core energy ($E_c$) of the distribution can be modeled by the power-law $\kappa=AE_c^\gamma$ for both species, and the relation between $\beta$, $\kappa$, and $E_c$ is much more complex than earlier reported: both $A$ and $\gamma$ exhibit systematic dependencies with $\beta$. Our results indicate that $\beta \sim 0.1-0.3$ is a range where the plasma is more dynamic since it is influenced by both the magnetic field and temperature fluctuations, which suggests that the transition between magnetically dominated plasmas to kinetically dominated plasmas occurs at these values of $\beta$. For $\beta > 1 $, both $A$ and $\gamma$ take nearly constant values, a feature that is especially notable for the electrons and might be related to their demagnetization. The relation between $\beta$, $\kappa$, and $E_c$ that we present is an important result that can be used by theoretical models in the future.
\end{abstract}



\section{Introduction} \label{sec:intro}
Understanding the dynamics of charged energetic particle interactions in space and astrophysical plasmas has been one of the main challenges in the space physics community over several decades. 
Owing to the lack of adequate collisions, these plasmas are usually observed in quasi-equilibrium stationary states different from thermodynamic equilibrium. 
The kinetics of the relaxation process needed to reach a stationary state and the properties of such state in which the plasma and electromagnetic turbulence coexist, are still not well understood for a long list of space and astrophysical objects \citep{marsch2006kinetic, bruno2013solar,yoon2017kinetic}. 
These unsolved problems border on a lack of sufficient understanding of the interaction of charged particles in plasma environments, such as the solar wind and Earth's magnetosphere, which are essentially collisionless plasma systems in non-equilibrium stationary states. 

Over half a century since the introduction of the Kappa distribution function by \citet{Montgomery_et_al_1965}, several studies have shown that the properties of collisionless plasmas can be well modeled by distributions with enhanced suprathermal power-law tails, rather than Maxwellian distributions.  
The Kappa distributions play a crucial role in the description of plasma objects such as the solar wind \citep{collier1996neon, pierrard1999electron, mann2002electron, livadiotis2011ainfluence, Yoon2014, pierrard2014coronal},  the Earth's magnetosheath \citep{Vasyliunas_1968, ogasawara2013characterizing, ogasawara2015interplanetary}, and several regions in the magnetosphere like the magnetotail \citep{grabbe2000generation}, the ring current \citep{Pisarenko_2002}, and the plasma sheet \citep{Christon_et_al_1988, kletzing2003auroral, Marina15, espinoza18GRL, kirpichev2020dependencies}. 

The general form of the Kappa distribution is denoted by $f$:
%
\begin{eqnarray}
\label{eq:1}
f(E;n_{\alpha},\kappa_{\alpha},E_{c_{\alpha}}) =& n_{\alpha} \left( \frac{m_{\alpha}}{2 \pi E_{c_{\alpha}}} \right)^{\frac{3}{2}}\frac{\Gamma(\kappa_{\alpha})}{\Gamma(\kappa_{\alpha}-\frac{3}{2})\sqrt{\kappa_{\alpha}}} \nonumber \\
& \times \left[ 1+ \frac{E}{\kappa_{\alpha} E_{c_{\alpha}}} \right]^{- \kappa_{\alpha}-1} \quad, 
\end{eqnarray}
%
where $f$ is the phase space density, $E$ is the kinetic energy, the sub-index $\alpha$ corresponds to the particle index, which can be electron (e) or ion (i); $n_{\alpha}$ is the particle density, $m_{\alpha}$ is the particle mass, $\Gamma$ is the Euler gamma function, and $\kappa_{\alpha}$ and $E_{c_{\alpha}}$ are the $\kappa$-parameter and characteristic or core energy, respectively. Kappa distributions given by Equation~(\ref{eq:1}) exhibit a thermal core with characteristic energy $E_{c_{\alpha}}$ and suprathermal tails, such that the total characteristic particle kinetic energy $E_{\rm{total}}$ is given by
\begin{equation}
\label{eq:2}
E_{\rm{total}} = E_{c{\alpha}}
 \,\frac{\kappa_{\alpha}}{\kappa_{\alpha} - 3/2}\,,
 \label{eq:energy}
\end{equation}
which enables a straightforward comparison between Kappa and Maxwellian distributions, and to outline the effects of suprathermals as shown by~\citet{lazar2015destabilizing, lazar2016interpretation}. 
The spectral index $\kappa_{\alpha}$ is a measure of the slope of the energy spectrum of the suprathermal particles that form the tail of the velocity distribution function. Hence $\kappa_{\alpha}$ primarily provides a measure of the departure of the stationary states from thermal equilibrium \citep{burlaga2005tsallis}. For $\kappa_{\alpha} \rightarrow \infty$, equation (\ref{eq:1}) becomes identical to the Maxwell distribution and approaches the quasi-thermal core of the observed distribution.

\begin{eqnarray}
\label{eq:3}
f(E;n_{\alpha},E_{c_{\alpha}})= n_{\alpha}\left( \frac{m_{\alpha}}{2 \pi E_{c_{\alpha}}}\right)^{3/2}\exp\left(-\frac{E}{E_{c_{\alpha}}} \right) 
\end{eqnarray}

In the solar wind, several mechanisms have been reported in the literature that lead to the generation of Kappa distributions \citep{livadiotis2018generation}. 
For instance, \citet{lazar2017firehose} observed the presence of suprathermal electron fluxes, which can be well modeled by Kappa distributions.
They argued that the use of Kappa and bi-Kappa distributions allows a realistic interpretation of non-thermal electrons and their effects on the electron firehose instability, where growth rates are observed to increase while the instability thresholds and electron kappa ($\kappa_e$) are seen to decrease.

The study by \citet{maksimovic1997kinetic} describes the first exospheric model of the solar wind based on Kappa Velocity Distribution Functions for protons and electrons escaping from the corona. 
Their model provides a possible hint regarding key features of the solar wind flow.
It indicates that the fastest solar wind flows that originate from the corona holes are high-speed streams with an enhanced high-velocity tail simulated by a Kappa function with a small electron kappa $\kappa_{e}$ value. Whereas, for hot equatorial regions where the slow solar wind originates, the electron velocity distribution functions are closer to the Maxwellian equilibrium, corresponding to $\kappa_{e} = \infty$.

More recently, a significant relationship has been found between the $\kappa$-index and other plasma parameters. For instance, it was established that $\kappa$-index correlates with the solar wind density and temperature \citep{livadiotis2018generation}. 
Besides, the $\kappa$-index was also found to be connected with the polytropic index and magnetic field \citep{livadiotis2017kappa}. 
In addition, \citet{livadiotis2018generation} remarked that the $\kappa$-index decreases when the magnetic field's long-range interactions induce correlations among particles, as the system is turned away from thermal equilibrium. 
They found a strong correlation between the plasma $\beta$ parameter, (defined as $\left( p /B^{2}/2\mu_{0} \right)$, where $p$ is the plasma pressure, $B$ is the magnetic field, and $\mu_{0}$ is the magnetic permeability) and kappa. This correlation takes place when $\beta$ increases, as thermal pressure becomes dominant.
Similarly, as the $\kappa$-index increases, the long-range interactions due to the magnetic field become weaker. This observation further revealed that kappa regulation in the low beta regime is due to the magnetic field, which induces the correlation between particles. 

Furthermore, there are many studies in some specific regions of the Earth magnetosphere that utilized Kappa distributions. 
For instance,  \citet{Christon_et_al_1989, Christon_et_al_1991} obtained ion and electron Kappa distributions in the plasma sheet using the particle instruments onboard the International Sun-Earth Explorer 1 (ISEE 1). 
They found that the $\kappa$-index ranges between 4 and 8 for both ions and electrons, with a most probable value between 5 and 6, which shows that the spectral shape is distinctly non-Maxwellian.
Later, \citet{haaland10} found that the $\kappa$-index ranges between 3 and 6, using data of the Cluster satellites. \citet{Marina15} utilized Kappa distributions to fit ion and electron flux spectra for five events in which the THEMIS satellites were aligned along the plasma sheet. They obtained snapshots of kappa properties that show a tendency for the $\kappa$-index to increase in the tailward direction.  \citet{espinoza18GRL} also used the Kappa distribution to model ions and electrons flux spectra along the plasma sheet. 
Their results reveal that $\kappa_{i}>\kappa_{e}$, 
which suggests that non-thermal properties of the electrons are stronger than ions. 
Besides, their results show a persistent dawn-dusk asymmetry in the relative numbers of energetic ions, which increases during substorms. This is consistent with the previous study of \citet{wing2005dawn}. 

Recently, \citet{kirpichev2020dependencies} measured the $\kappa$-parameters for ions in different magnetospheric regions and during quiet magnetospheric conditions. 
They found that kappa depends on the core energy ($E_{c}$) for a wide energy range and a broad range of the plasma beta ($\beta$) parameter. Their results support earlier findings, which showed that $\kappa$-index increases with $E_{c}$ in the magnetosphere of the Earth \citep{Christon_et_al_1989} and the solar wind \citep{Collier_1999}. 
However, despite the aforementioned studies, there is no systematic experimental analysis that focuses on the coupling between Kappa distribution parameters (density, core energy, and $\kappa$-index) and the plasma beta $\beta$ parameter in the Earth's magnetosphere. This study, for the first time, considers simultaneous electron and ion measurements, we will explore more precisely the relationship between $\kappa$-index, core energy $E_{c}$, and plasma beta $\beta$, which will provide a better understanding of plasma thermalization, the relation between ion and electron properties, and the importance of the level of magnetization of each species within the Earth's magnetosphere, regardless of the details of a given magnetospheric region. 

The paper is organized as follows: In section \ref{sec:dataanalysis} we describe the data and methodology for obtaining the ion and electron Kappa distribution parameters and plasma beta; 
In section \ref{presentation} we present the results of the analyses and explore the relationship between kappa, beta, and core energy $E_{c}$. 
In section \ref{discussion} we discuss the observed results, and in section \ref{conclusion} we summarize and conclude our findings.

\section{Instrumentation and Data Analysis} 
\label{sec:dataanalysis}

\begin{figure}
\gridline{\fig{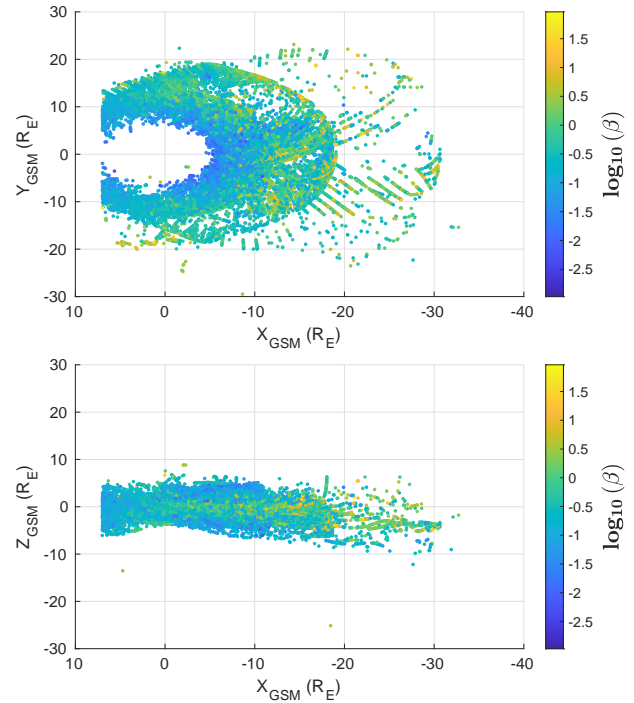}{0.45\textwidth}{}
          }
\caption{The spatial coverage of the measured spectra for which both electron and ion Kappa fits were successful. 
The color code is used to represent the value of the electron plasma $\beta_{e}$. 
Upper panel: $X$ and $Y$gsm plane. Lower panel: $X$ and $Z$gsm plane.}
\label{fig:BetaSpatialCoverage}
\end{figure}

The present study combines data sets of the multi-satellite mission Time History of Events and Macroscale Interactions during Substorms (THEMIS), using all its satellites (TH-A, TH-B, TH-C, TH-D, and TH-E), and spanning the years 2008 to 2018. 
The data was downloaded via the THEMIS ftp website\footnote{http://themis.ssl.berkeley.edu/index.shtml}. 
All measurements were constrained to the following Geocentric Solar Magnetospheric (GSM) coordinate system: $-35 \leq X \leq 7$ R$_{E}$, $-30 \leq Y \leq 30$ R$_{E}$, $-10 \leq Z \leq 10$ R$_{E}$,  and at distances larger than $5\,$R$_{E}$ from the center of the Earth. 
This region is depicted in Figure \ref{fig:BetaSpatialCoverage}, where panels (a) and (b) show the spatial coverage in the $X-Y_{GSM}$ and $X-Z_{GSM}$ planes, respectively. Each position is color-coded with the measured electron plasma beta $(\beta_e)$.
All measurements used in this study were averaged over 12 minute long intervals, which is long enough to make stable measurements of the particle fluxes, and at the same time, short enough to ensure that the distributions do not change significantly over this time \citep{Marina15,espinoza18GRL}.

The magnetic field data used in this study were obtained from the Flux Gate Magnetometer (FGM) onboard the THEMIS satellites \citep{auster2008themis}. 
The plasma particle data were obtained from the Electrostatic Analyzers \citep[ESA][]{McFadden_2008}, with an energy range from a few eV up to $30$\,keV for electrons and $25$\,keV for ions, and the Solid State Telescopes \citep[SST][]{Angelopoulos_2008} with an energy range from $25$\,keV to $6$\,MeV. 
Further, we used level 2 full mode particle energy fluxes, which are averaged over the particle instruments satellite rotation thereby significantly increasing the statistical measurements for each energy channel.
Magnetospheric plasmas are composed mainly of protons and electrons but we should also expect a low fraction of heavy ions. Unfortunately, THEMIS particle instruments do not distinguish protons from other ion species, hence, we refer to all of them as ions.

In our study, only the central energies of the particle instruments combined range were considered for fitting.
This was done to avoid contamination from the spacecraft potential and photoelectrons at low energies ($<40$\,eV) and cosmic rays and low statistics at high energies. 
The analyses were limited to the ranges $1.75$ to $210$\,keV for ions, and $0.36$ and $203.5$\,keV for electrons.  
However, our statistics at both ends of the distribution were low for very low fluxes. Hence we included sometimes fewer channels, as a function of particle number density, by setting some control conditions for high energy channel SST and low energy channel ESA.

The observed ion and electron energy spectrum were fitted by transforming the three-dimensional kappa distribution from equation (1) to particle energy fluxes, denoted as $F$, as shown below:
\begin{equation}
F_{\alpha}(E) = \frac{n_{\alpha}}{\sqrt{2 \pi^3 m_{\alpha}}}\frac{E^{2}}{E^{3/2}_{c_{\alpha}}}\frac{\Gamma(\kappa_{\alpha})}{\Gamma(\kappa_{\alpha}-\frac{1}{2}) \sqrt{\kappa_{\alpha}}}\left[ 1+ \frac{E}{\kappa E_{c_{\alpha}}} \right]^{-\kappa_{\alpha}-1} \label{eq:3}
\end{equation}

\begin{figure*}
\gridline{\fig{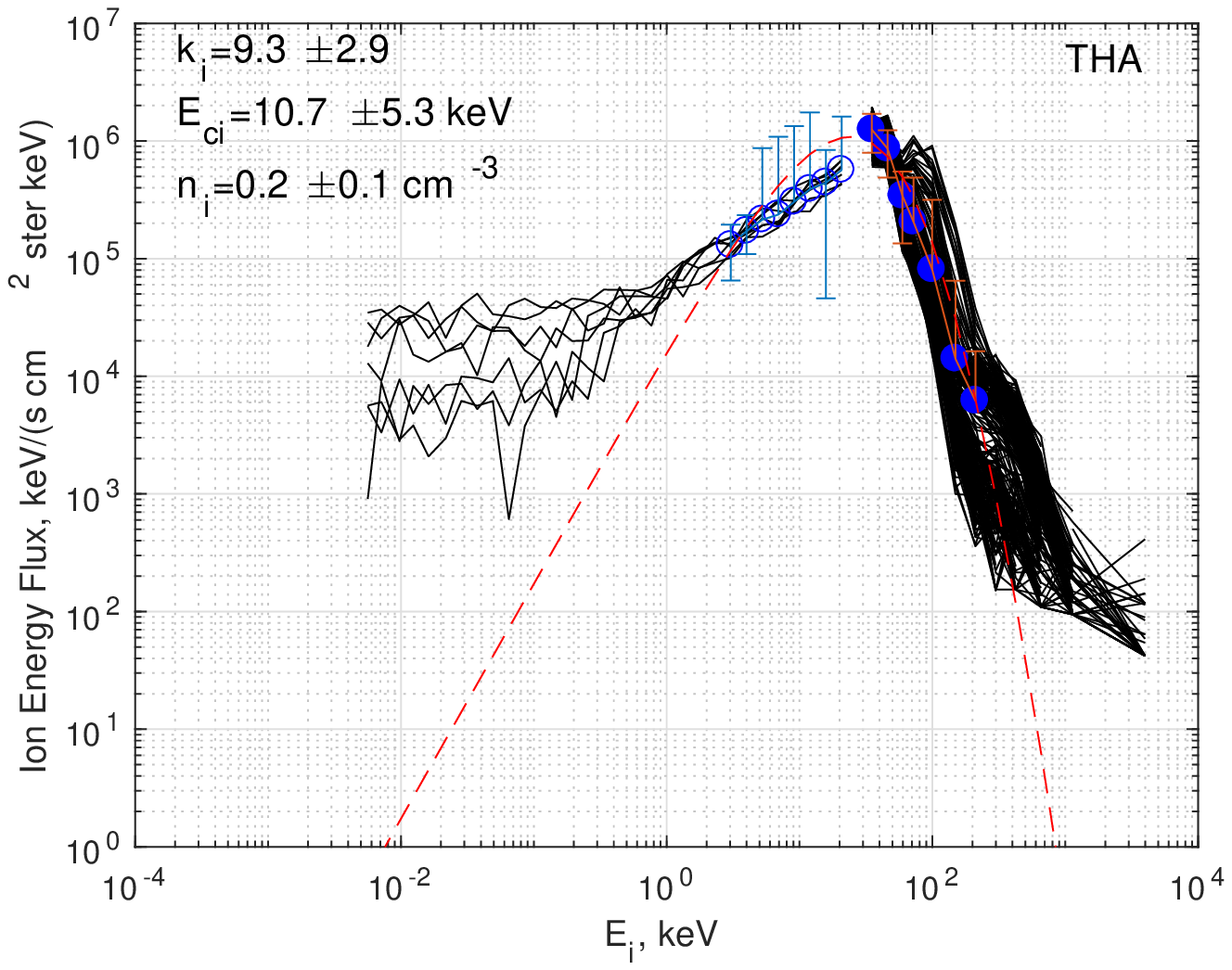}{0.32\textwidth}{(a)}
          \fig{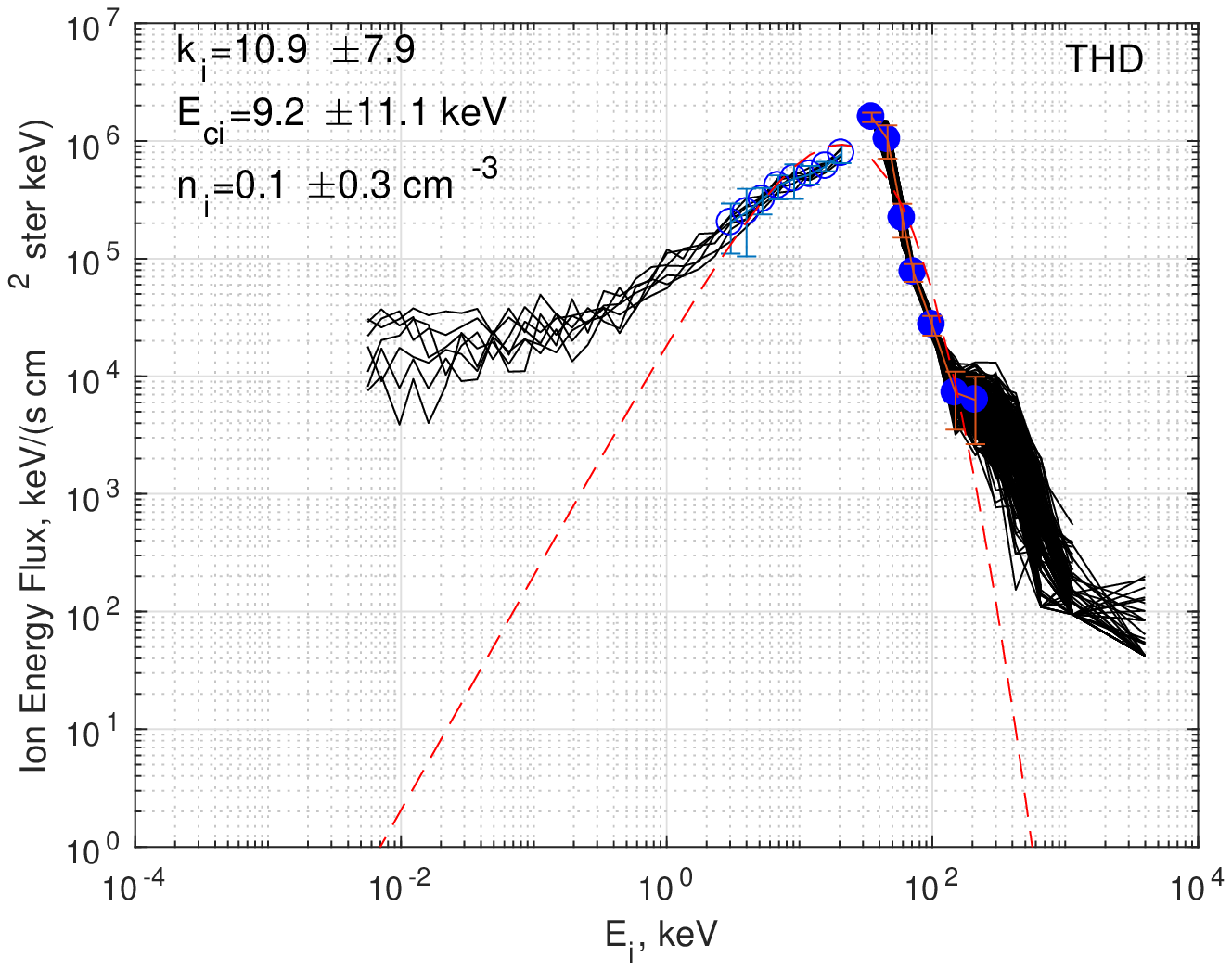}{0.32\textwidth}{(b)}
          \fig{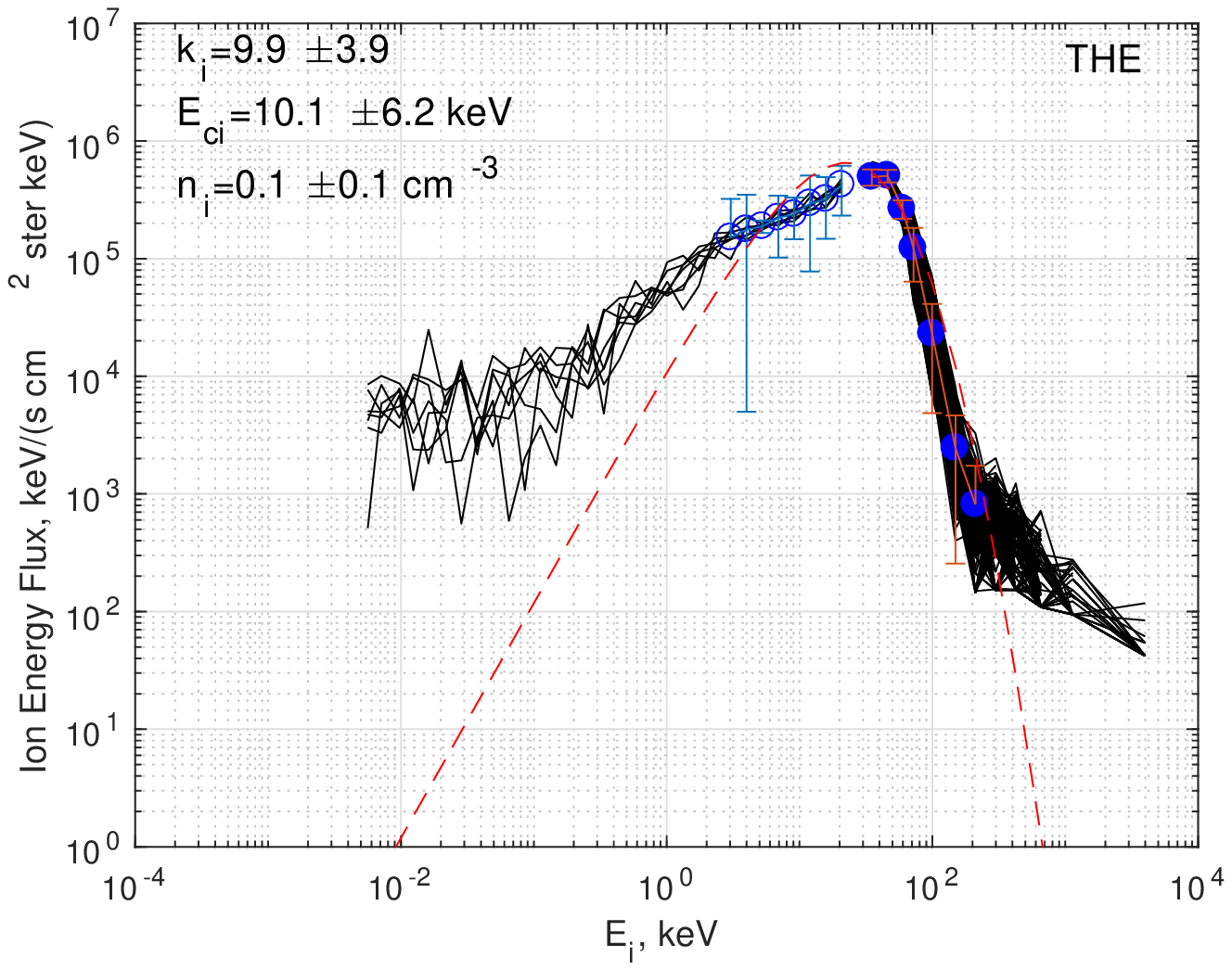}{0.32\textwidth}{(c)}
          }
\gridline{\fig{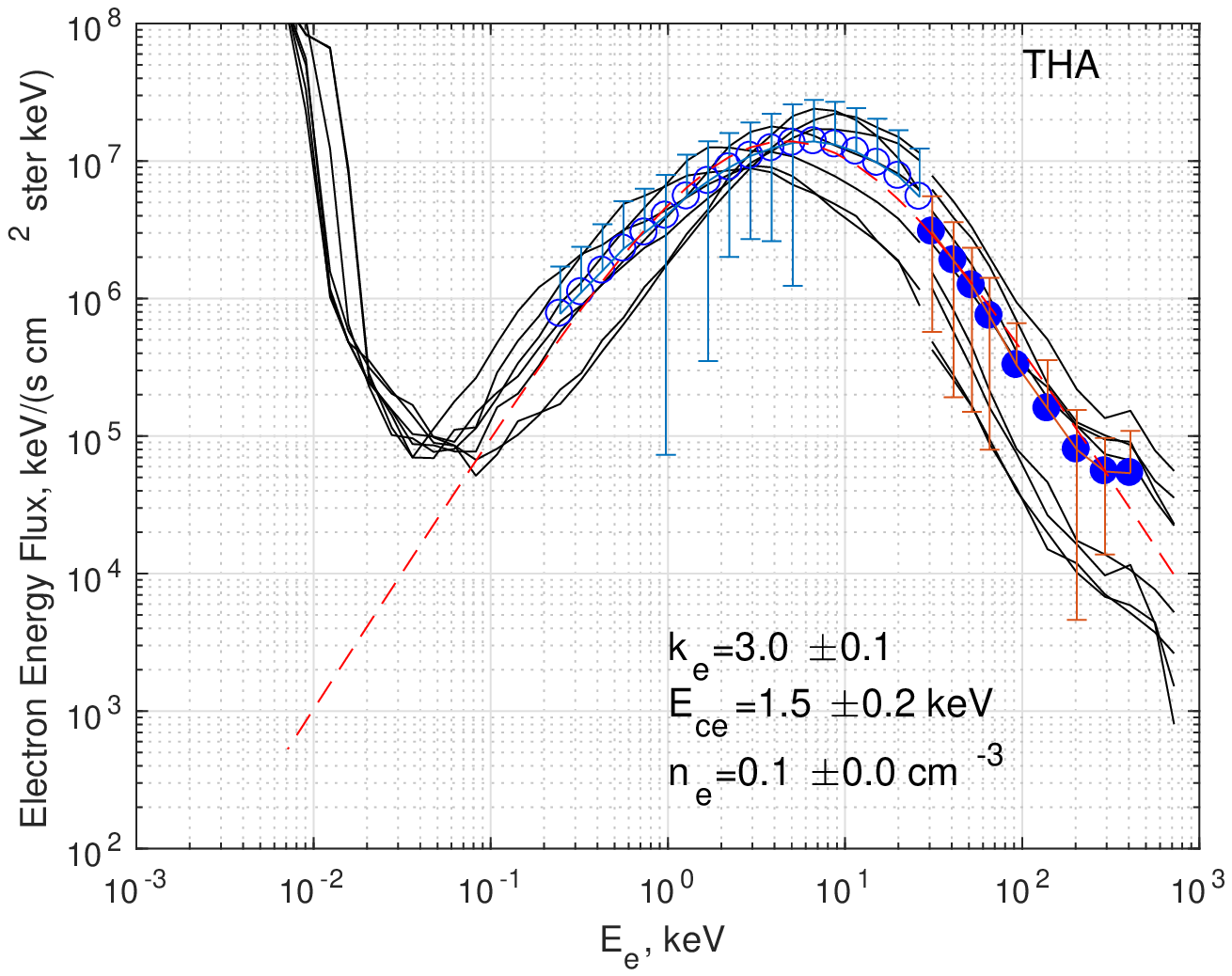}{0.32\textwidth}{(d)}
          \fig{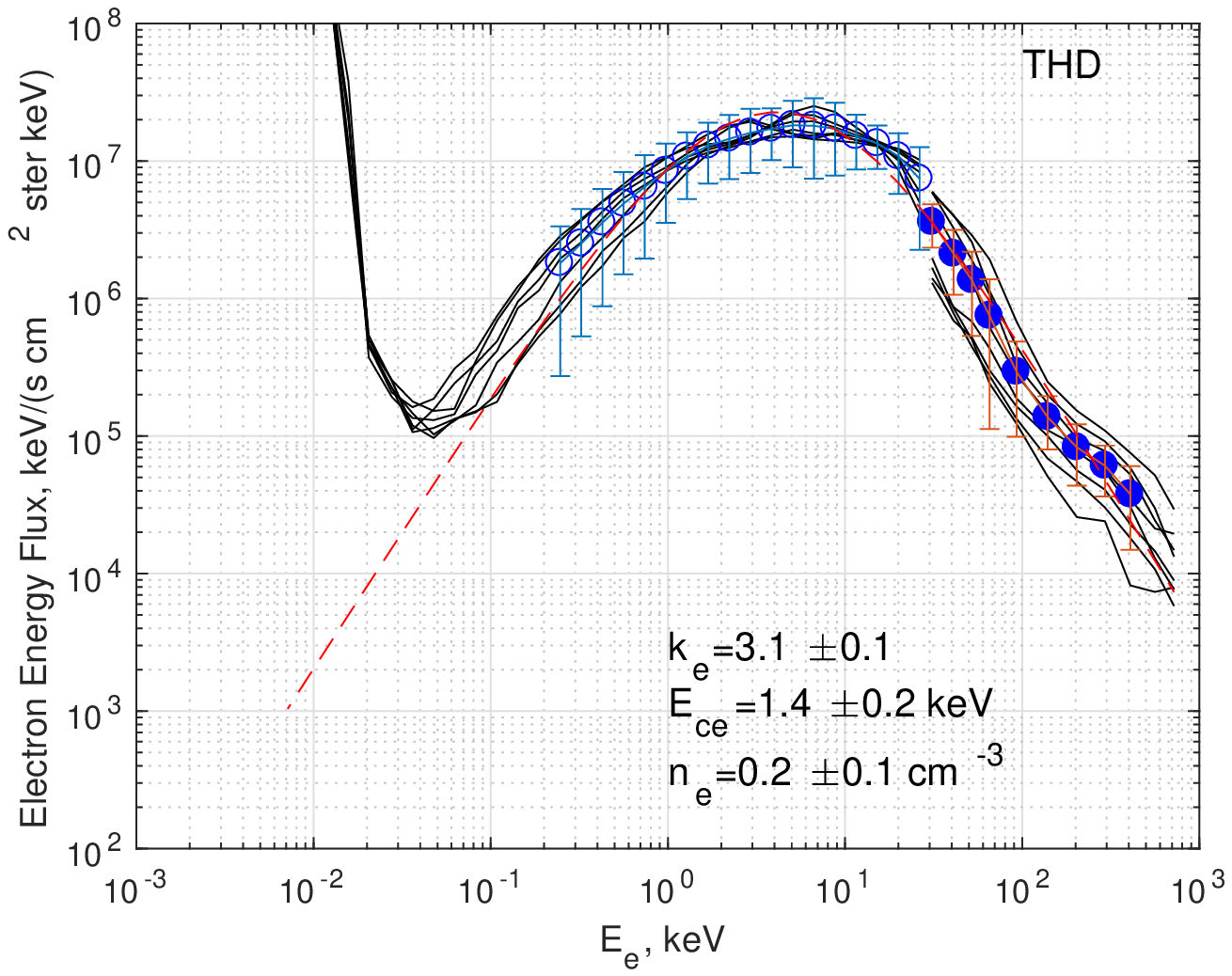}{0.32\textwidth}{(e)}
          \fig{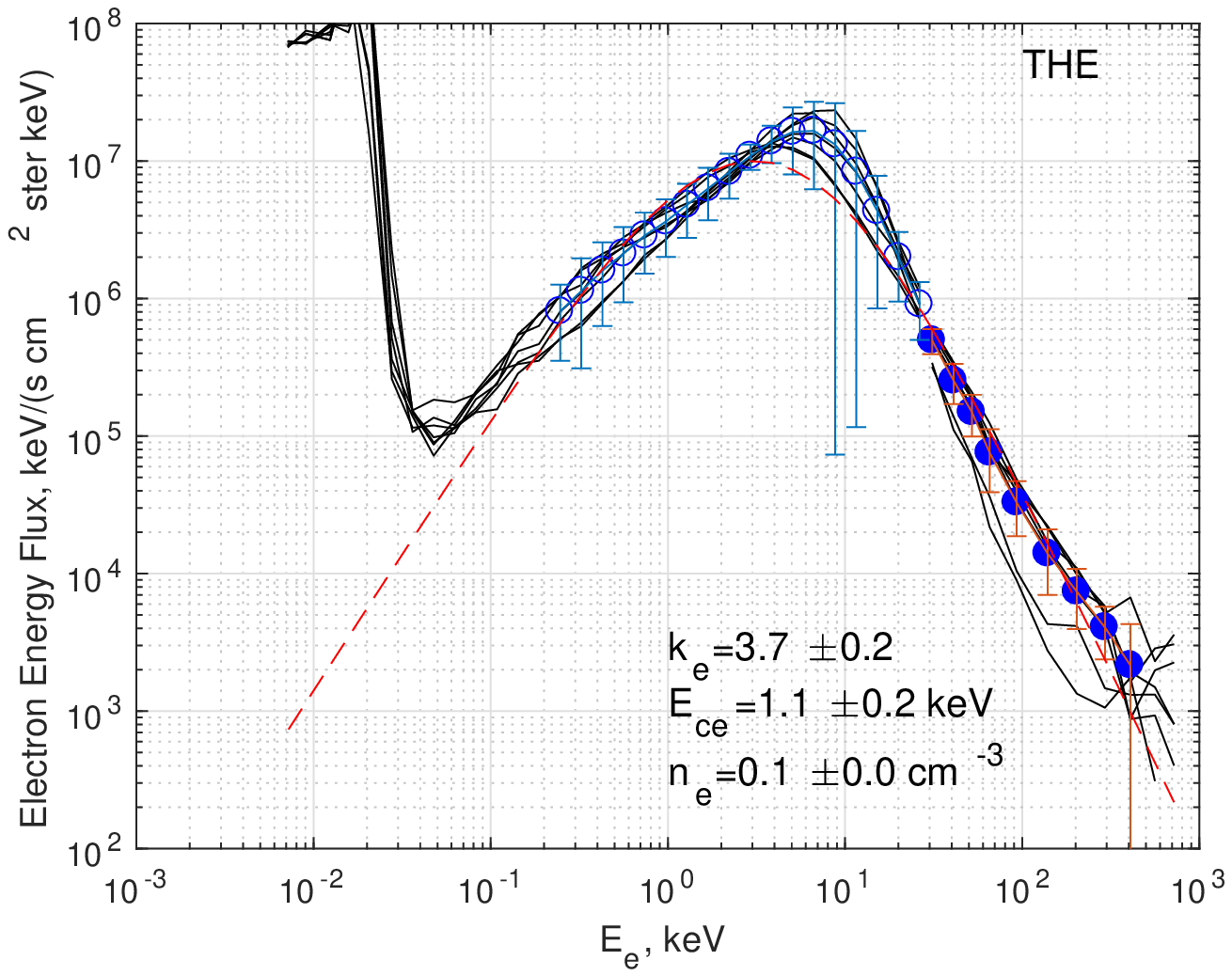}{0.32\textwidth}{(f)}
          }
\caption{Examples of fits of Kappa distributions to the flux spectra of ions (a-c) and electrons (d-f), which are simultaneously measured on 29th March 2008 between 10.4 UT and 10.5 UT by THA (first column), THD (second column), and THE (third column). In each panel, the black solid lines are sub-datasets taken over a total of 12 minutes. Open circles represent averages of the subsets measurements of low energy particles (ESA) and filled circles represent averages of the subsets of measurements of high energy particles (SST).
The red dash-line is the kappa function curve fitted to the open and filled circles.} 
\label{fig:iandespectra}
\end{figure*}

Figure \ref{fig:iandespectra} illustrates examples of ion (upper panels) and electron (bottom panels) energy flux spectra measured by combining both particle instruments (ESA and SST) onboard the THEMIS satellites (solid lines). The circles on the plots are the average of the spectra obtained for the 12 minutes time windows. The open circles represent measurements from the ESA, while the filled circles represent the SST. 
The error bars for the averaged flux data represent the spread between the maximum and minimum observed values.
They were found to vary significantly between the ESA and SST data, so they were normalized in the same way as \citet{espinoza18GRL}.
The inverse squared of the error bars is used to define weights for the fits, which were performed using a non-linear least-squares method combined with the Levenberg-Mardquart algorithm.

We visually inspected hundreds of spectra and decided to work only with the fits that give a reduced chi-squared $\chi^{2}<100$. In addition, only measurements for which both electron and ion fits were successful in the same time interval were used in the subsequent analysis. In order to ensure that we study only plasmas in the magnetosphere, we restricted the particle density, the magnetic field in the X component, and ion bulk velocity to the values shown in Table ~\ref{table:parametersrestrictions}. In all, we found 47,058-time intervals that satisfied these restrictions and could be well fitted by a Kappa distribution function for both ions and electrons. The final data set, analyzed in the next section, includes the three kappa parameters from the fits and the plasma beta for ions and electrons.

\begin{deluxetable}{lcl}
\tablecaption{Criteria applied to select magnetospheric plasmas. \label{table:parametersrestrictions}}
\tablehead{
\colhead{Parameter} & \colhead{Symbol} & \colhead{Condition}
}
\startdata
\hline 
Plasma ion density (\textrm{cm}$^{-3}$) & $n_{i}$ & $ \geqslant 0.1$ \\
Magnetic field (nT) & $B_{x}$ & $ \leqslant \; 100$  \\
Ion bulk velocity (\textrm{km\,s}$^{-1}$)& $v_{T}$ & $\leqslant 250$ \\
\enddata
\end{deluxetable}

\begin{deluxetable*}{lccccccc}
  \tablecaption{Statistics of ion $\beta$ and kappa parameters.
  \label{table:ionstarvalues}
  }
  \tablehead{
    \colhead{Parameter} & \colhead{Minimum}  & \colhead{Maximum} & \colhead{Mean} & \colhead{Median} & \colhead{$Q_{1}$} & \colhead{$Q_{3}$} & \colhead{$\sigma$}
  }
  \startdata
\multicolumn{8}{c}{For all $\beta_{i}$: 47,058 spectra} \\
$\kappa_{i}$ & 1.51 & 42.9 & 6.67 & 6.44 & 4.96 & 8.08 & 2.28 \\
$E_{c_{i}}$ & 0.25 & 10.17 & 3.45 & 2.97 & 1.91 & 4.73 & 2.01 \\
$E^T_{i}$ & 0.25 & 15.25 & 4.78 & 4.16 & 2.93 & 6.43 & 2.63 \\
$n_{i}$ & 0.10 & 3.70 & 0.65 & 0.37 & 0.23 & 0.94 & 0.60 \\
$S_{k_{i}}$ & -1.60 & 4.09 & 0.80 & 0.73 & 0.01 & 1.50 & 0.96\\
$\beta_{i}$ & 0.001 & 92.2 & 0.86 & 0.29 & 0.13 & 0.71 & 2.69\\
\hline
\multicolumn{8}{c}{For low $\beta_{i}$ regime, $\beta_{i} \leqslant  1$: 38,708 spectra} \\
$\kappa_{i}$ & 1.51 & 42.9 & 6.64 & 6.38 & 4.87 & 8.08 & 2.33 \\
$E_{c_{i}}$ & 0.25 & 10.17 & 3.84 & 3.31 & 2.13 & 5.50 & 2.14 \\
$E^T_{i}$ & 0.25 & 15.25 & 5.54 & 5.16 & 3.53 & 7.20 & 2.74 \\
$n_{i}$ & 0.10 & 2.80 & 0.61 & 0.32 & 0.21 & 0.94 & 0.56 \\
$S_{k_{i}}$ & -1.59 & 4.09 & 0.82 & 0.80 & 0 & 1.65 & 1.06 \\
$\beta_{i}$ & 0.001 & 1.00 & 0.29 & 0.22 & 0.11 & 0.42 & 0.24 \\
\hline
\multicolumn{8}{c}{For high $\beta_{i}$ regime, $\beta_{i} \geqslant 1$: 8,350 spectra} \\
$\kappa_{i}$ & 1.75 & 25.2 & 6.82 & 6.67 & 5.40 & 8.06 & 2.00 \\
$E_{c_{i}}$ & 0.25 & 7.50 & 2.71 & 2.51 & 1.61 & 3.46 & 1.48 \\
$E^T_{i}$ & 0.25 & 8.37 & 3.26 & 3.17 & 2.14 & 4.06 & 1.49 \\
$n_{i}$ & 0.10 & 3.70 & 0.75 & 0.46 & 0.31 & 1.04 & 0.67 \\
$S_{k_{i}}$ & -0.42 & 3.42 & 0.78 & 0.71 & 0.26 & 1.16 & 0.72 \\
$\beta_{i}$ & 1.00 & 92.2 & 3.47 & 1.82 & 1.30 & 3.23 & 5.69 \\
\enddata
  \tablecomments{ Energies are given in keV, and densities in cm$^{-3}$. $E^T_i$ is the ion total energy, $S_{k_{i}}$ is the skewness of the ion $E_c$ distribution. $Q_1$, $Q_3$, and $\sigma$ represent the lower quartile, upper quartile, and standard deviation of each quantity, respectively.
  }
\end{deluxetable*}
\begin{deluxetable*}{lccccccc}
  \tablecaption{Statistics for electron $\beta$ and kappa parameters.
  \label{table:electronstarvalues}
  }
  \tablehead{
    \colhead{Parameter} & \colhead{Minimum}  & \colhead{Maximum} & \colhead{Mean} & \colhead{Median} & \colhead{$Q_{1}$} & \colhead{$Q_{3}$} & \colhead{$\sigma$}
  }
  \startdata
\multicolumn{8}{c}{For all $\beta_{e}$: 47,058 spectra} \\
$\kappa_{e}$ & 1.51 & 34.3 & 4.60 & 4.41 & 3.75 & 5.20 & 1.37\\
$E_{c_{e}}$ & 0.25 & 6.25 & 1.33 & 0.83 & 0.42 & 1.74 & 1.25\\
$E^T_{e}$ & 0.25 & 7.12 & 1.85 & 1.41 & 0.77 & 2.51 & 1.44 \\
$n_{e}$ & 0.10 & 3.30 & 0.57 & 0.35 & 0.25 & 0.61 & 0.55\\
$S_{k_{e}}$ & -0.98 & 5.07 & 1.24 & 1.07 & 0.38 & 1.97 & 1.09 \\
$\beta_{e}$ & 0.002 & 87.5 & 0.31 & 0.11 & 0.04 & 0.27 & 1.29\\
\hline
\multicolumn{8}{c}{For low $\beta_{e}$ regime, $\beta_{e} \leqslant  1$:  44,774 spectra} \\
$\kappa_{e}$ & 1.51 & 26.5 & 4.59 & 4.40 & 3.74 & 5.19 & 1.36 \\
$E_{c_{e}}$ & 0.25 & 6.25 & 1.33 & 0.86 & 0.44 & 1.72 & 1.24 \\
$E^T_{e}$ & 0.25 & 7.12 & 1.98 & 1.48 & 0.87 & 2.67 & 1.48 \\
$n_{e}$ & 0.10 & 3.30 & 0.56 & 0.34 & 0.25 & 0.64 & 0.54 \\
$S_{k_{e}}$ & -0.98 & 5.07 & 1.41 & 1.35 & 0.53 & 2.10 & 1.13 \\
$\beta_{e}$ & 0.002 & 1.00 & 0.17 & 0.09 & 0.04 & 0.23 & 0.19\\
\hline
\multicolumn{8}{c}{For high $\beta_{e}$ regime, $\beta_{e} \geqslant 1$: 2284 spectra} \\
$\kappa_{e}$ & 2.21 & 34.3 & 4.82 & 4.54 & 3.91 & 5.41 & 1.59 \\
$E_{c_{e}}$ & 0.25 & 5.62 & 1.34 & 0.81 & 0.38 & 1.96 & 1.28\\
$E^T_{e}$ & 0.25 & 5.00 & 1.38 & 0.95 & 0.50 & 1.65 & 1.20 \\
$n_{e}$ & 0.1 & 3.30 & 0.59 & 0.38 & 0.27 & 0.60 & 0.58 \\
$S_{k_{e}}$ & -0.47 & 2.64 & 0.71 & 0.59 & 0 & 1.18 & 0.71 \\
$\beta_{e}$ & 1.00 & 87.5 & 3.08 & 1.68 & 1.23 & 2.78 & 5.06 \\
\enddata
  \tablecomments{Energies are given in keV, and densities in cm$^{-3}$.  $E^T_e$ is the electron total energy, $S_{k_{e}}$ is the skewness of the electron $E_c$ distribution. $Q_1$, $Q_3$, and $\sigma$ represent the lower quartile, upper quartile, and standard deviation of each quantity, respectively.   
  }
\end{deluxetable*}

\begin{figure}
\gridline{\fig{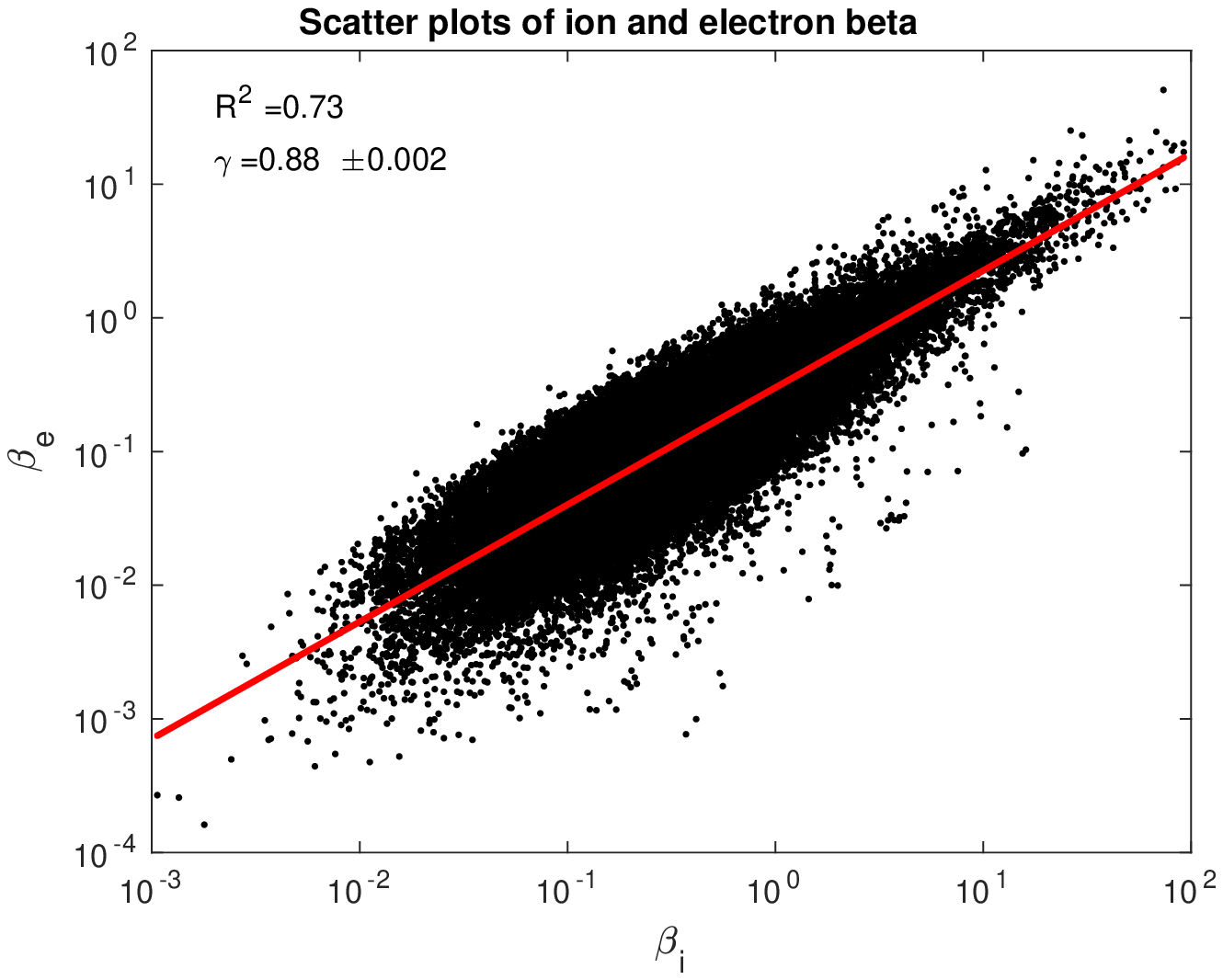}{0.43\textwidth}{(a)}
          }
\gridline{\fig{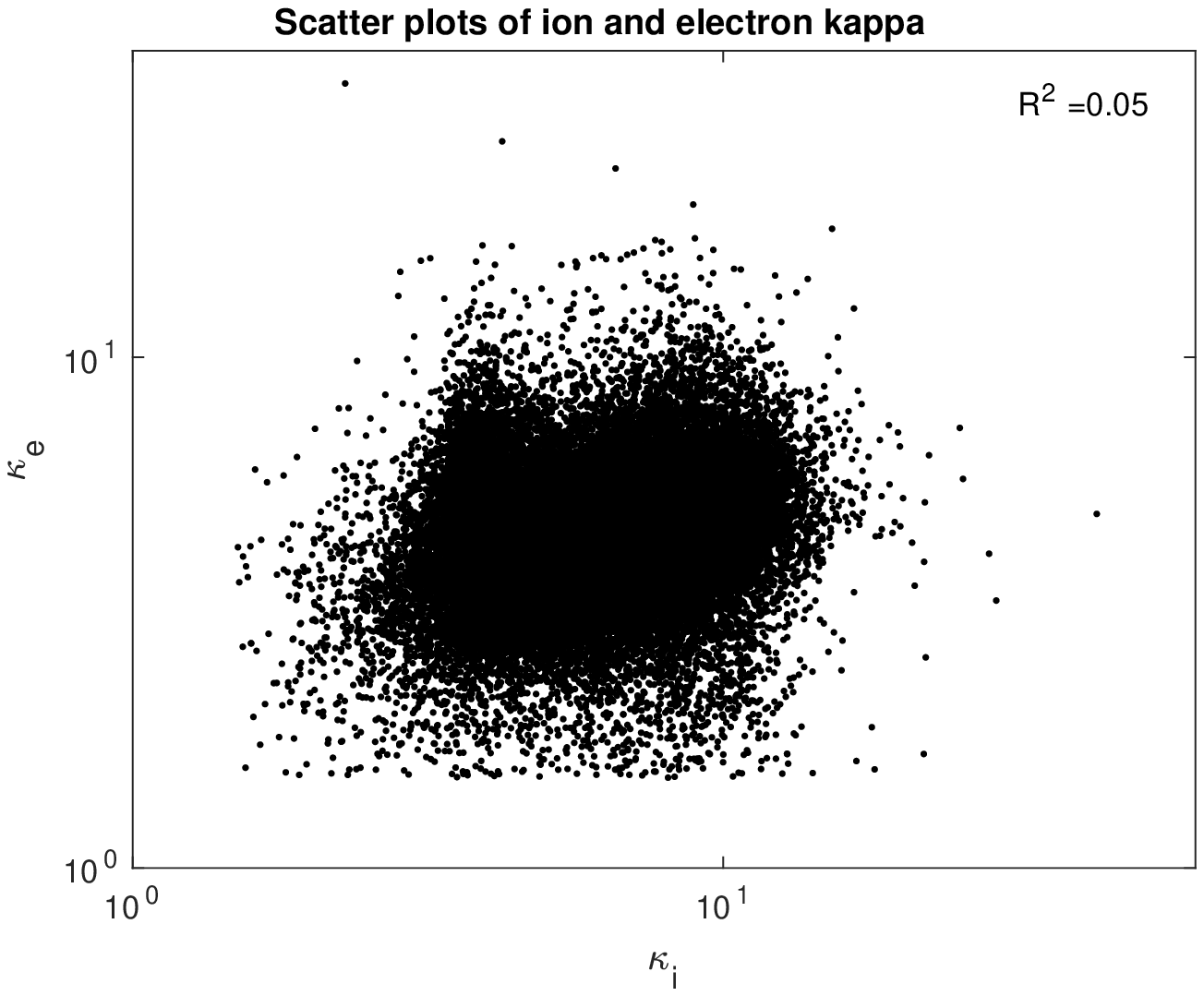}{0.43\textwidth}{(b)}
          }
\gridline{\fig{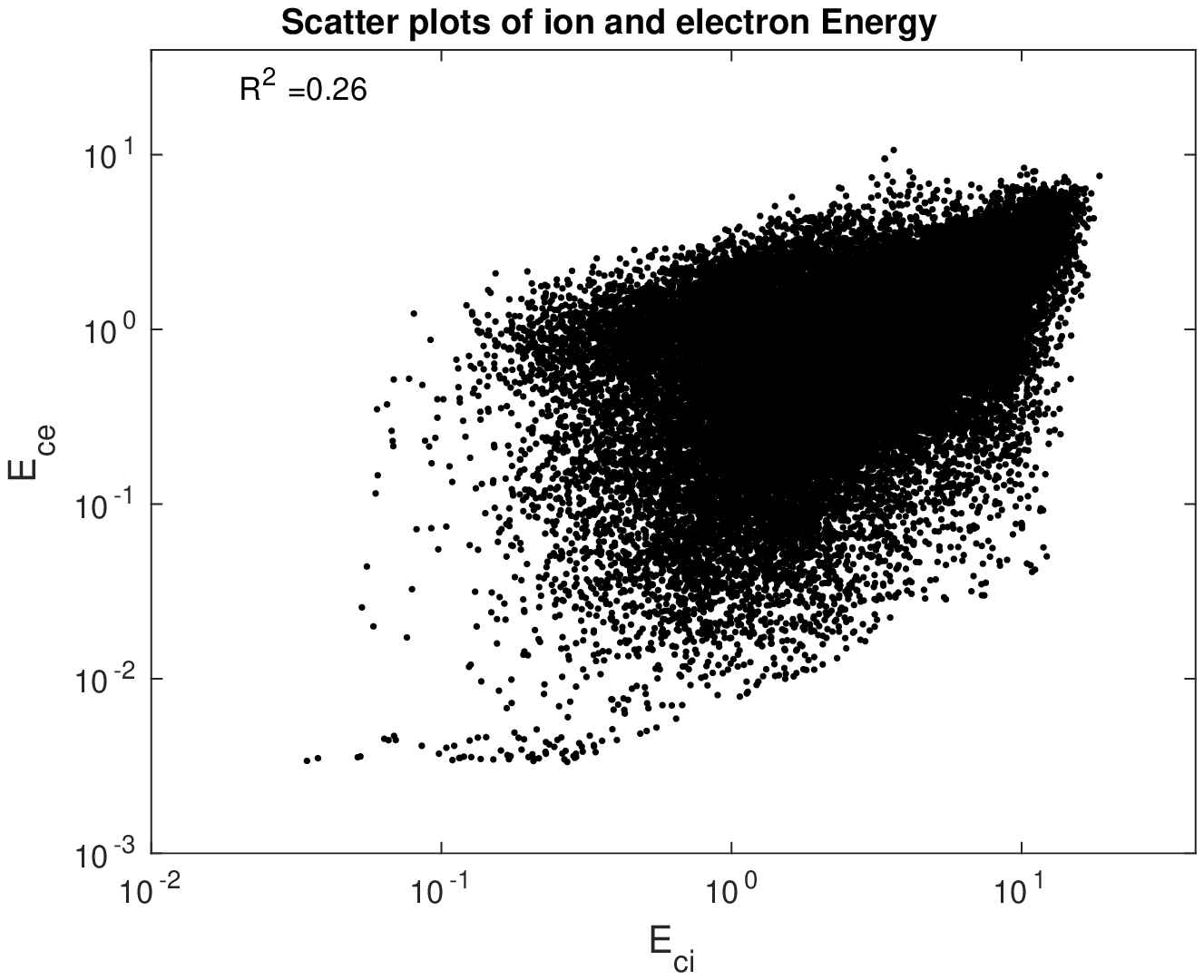}{0.43\textwidth}{(c)}
          }
\caption{Interrelationship of ion and electron parameters $\beta$ (a), $\kappa$ (b), and $E_{c}$ (c).}
\label{fig:kvEscatter}
\end{figure}

Tables \ref{table:ionstarvalues} and  \ref{table:electronstarvalues} give some statistical properties of the distributions of all the beta and kappa parameters obtained from the fits for ions and electrons, respectively.  
The data set was also divided into two groups, depending on beta, and the same quantities are given for each of these groups ($\beta<1$ or $\beta>1$).
Figure \ref{fig:kvEscatter}(a), (b) and (c), show the relation between $\beta_{e}$ versus $\beta_{i}$,  $\kappa_{e}$ versus $\kappa_{i}$, and $E_{c_{e}}$ versus $E_{c_{i}}$, respectively. 
As expected, $\beta_{i}$ and $\beta_{e}$ are correlated since the plasma on large scales is quasi-neutral, and the majority of the data we have are from the plasma sheet and from the region that surrounds the Earth, which is filled with plasma sheet-like plasma. 
In this case we expect ion temperatures to be higher than electron temperatures, with typical ion-to-electron temperature ratios $ T_{i}/ T_{e}$ between $4$ and $6$ according to \citet{baumjohann1993near, borovsky1997earth, espinoza18GRL}.

Further, the relation between $\kappa_{e}$ and $\kappa_{i}$ that we observe is uncorrelated, which is consistent with low-statistic results by \citet{Christon_et_al_1989}. 
The correlation we find between $E_{ce}$ and $E_{ci}$ is rather low, and less than what was obtained by  \citet{Christon_et_al_1989}. 
This could be related to the fact that we did not limit our study to the plasma sheet only. Nevertheless, these results suggest that a more detailed analysis is needed to establish whether there is a relation between $\beta$, $\kappa$, and $E_{c}$, which is done in the next section.

\section{Relationship between Beta ($\beta$), kappa ($\kappa$) and Core Energy ($E_{c}$)}
\label{presentation}
\begin{figure*}
\gridline{\fig{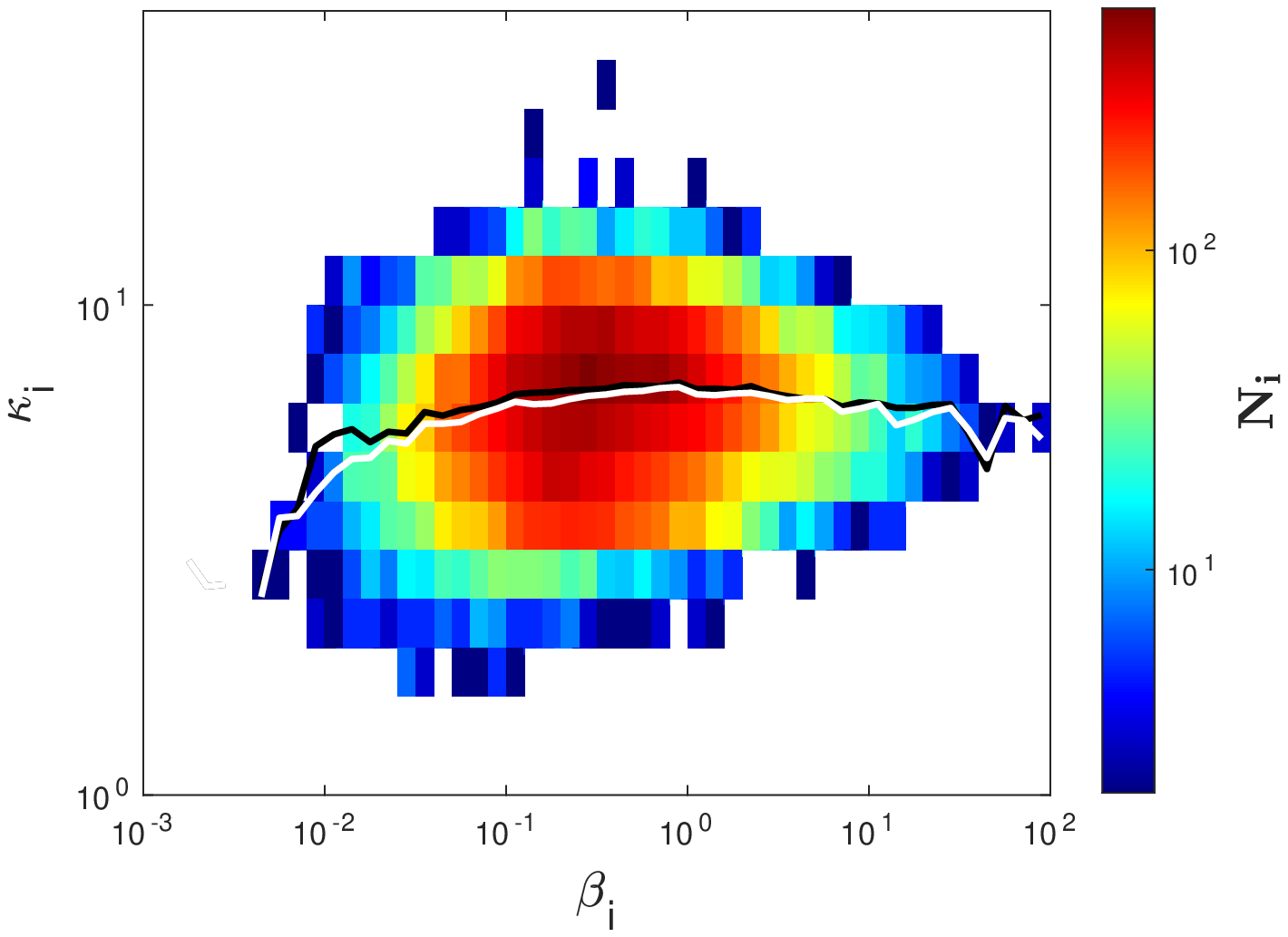}{0.5\textwidth}{(a)}
		  \fig{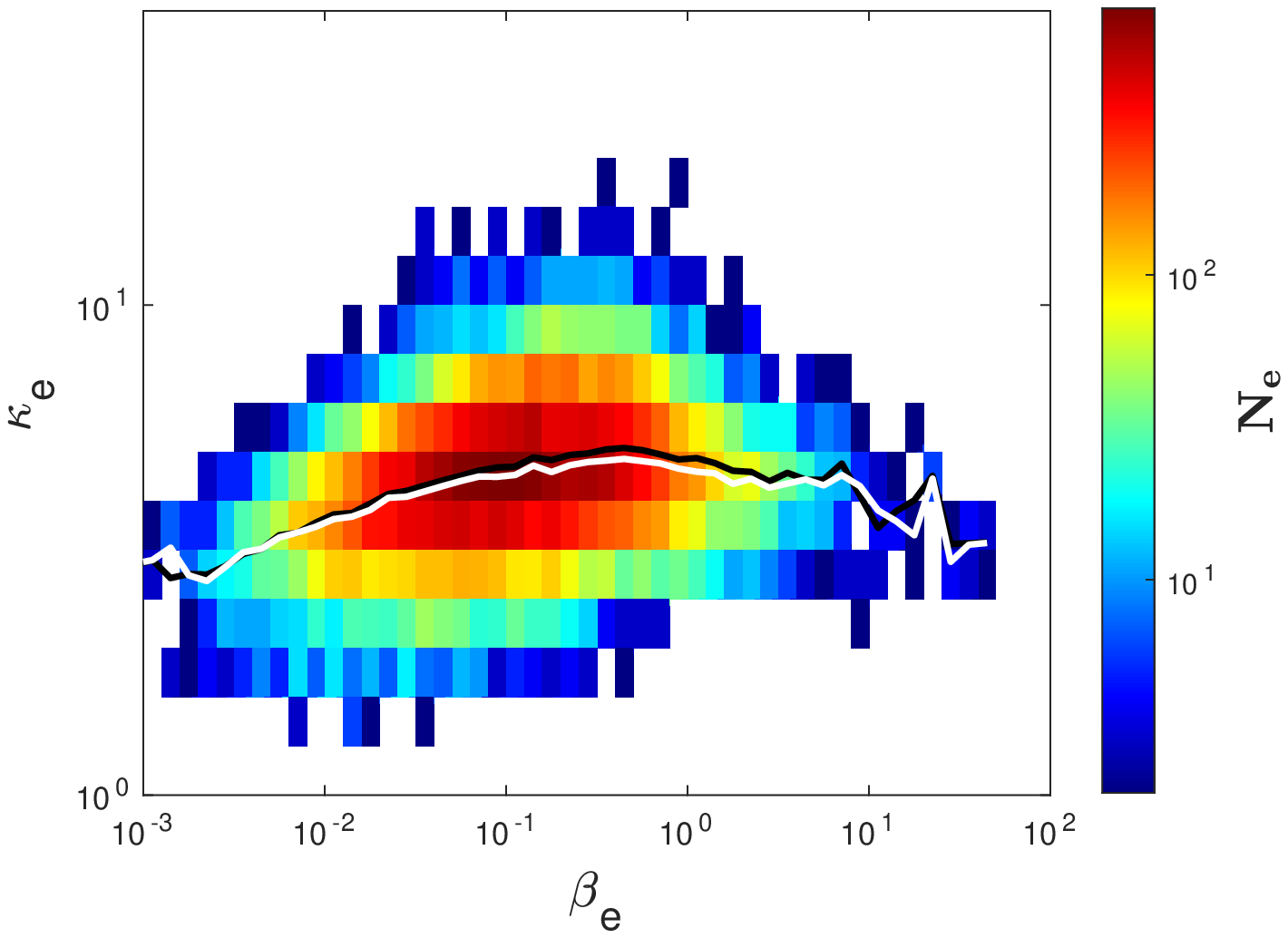}{0.5\textwidth}{(b)}
		  }
\caption{2-D histogram of the number of observations grouped into bins of sizes $0.1$ in the $\beta - \kappa$ plane, for (a) ions, and (b) electrons. The colorbar indicates the  number of observations $\langle N \rangle$ in logarithmic (base 10) color scale. The black and white solid lines represent Mean and Median  kappa values in each beta bin, respectively.}
\label{fig:numofOBS}
\end{figure*}

To assess the relation between beta, kappa, and core energy, we define a grid in the logarithmic ($\beta, \kappa $) space using a cell size of $\Delta \log_{10} \beta =\Delta \log_{10} \kappa_{i} = 0.1$. The grid is defined for the range -3 $< \log_{10} \beta <$ 2 and -0.6 $< \log_{10} \kappa <$1.6, and used to create the color-coded plots frequently presented in this study. Figure \ref{fig:numofOBS} shows the number of measurements in each bin ($N$), where empty bins contain less than ten measurements.

The observed ion and electron kappa indices vary from 1.5 (lowest possible state) to a little above $40$, and the observed beta values are from $10^{-3}$ to nearly $10^2$ (Tables \ref{table:ionstarvalues} and \ref{table:electronstarvalues}). 
The distribution of the observations in this space, however, presents a clear maximum, for both ions and electrons. 
While for the ions, most cases are in the range $5 \leq \kappa_{i} \leq 8$ and $0.1 \leq \beta \leq 0.7$, 
the electrons exhibit slightly smaller kappa and beta values in general: 
$4 \leq \kappa_{e} \leq 5$ and $0.04 \leq \beta_{e} \leq 0.3$  (see Tables \ref{table:ionstarvalues} and \ref{table:electronstarvalues}). 

Interestingly, for both species there are combinations of kappa and beta that are not observed. 
For example, plasmas with $\beta_{i}<0.01$ and $\kappa_{i}>10$, or plasmas with $\beta_{e}>30$ and $\kappa_{e}>6$ appear not to occur often in the magnetosphere. As illustrated in Figure \ref{fig:numofOBS}, the mean and median (black and white solid lines, respectively) kappa values increase with  beta, for both species, up to $\beta\sim1$. 
For $\beta>1$, the mean (and median) kappa is almost constant, with a slight decrease towards larger beta values. This feature is similar to the result obtained by \citet{kirpichev2020dependencies} (see their Figure 3) in the case of ions, but with smaller kappa values. 

\begin{figure*}
\gridline{\fig{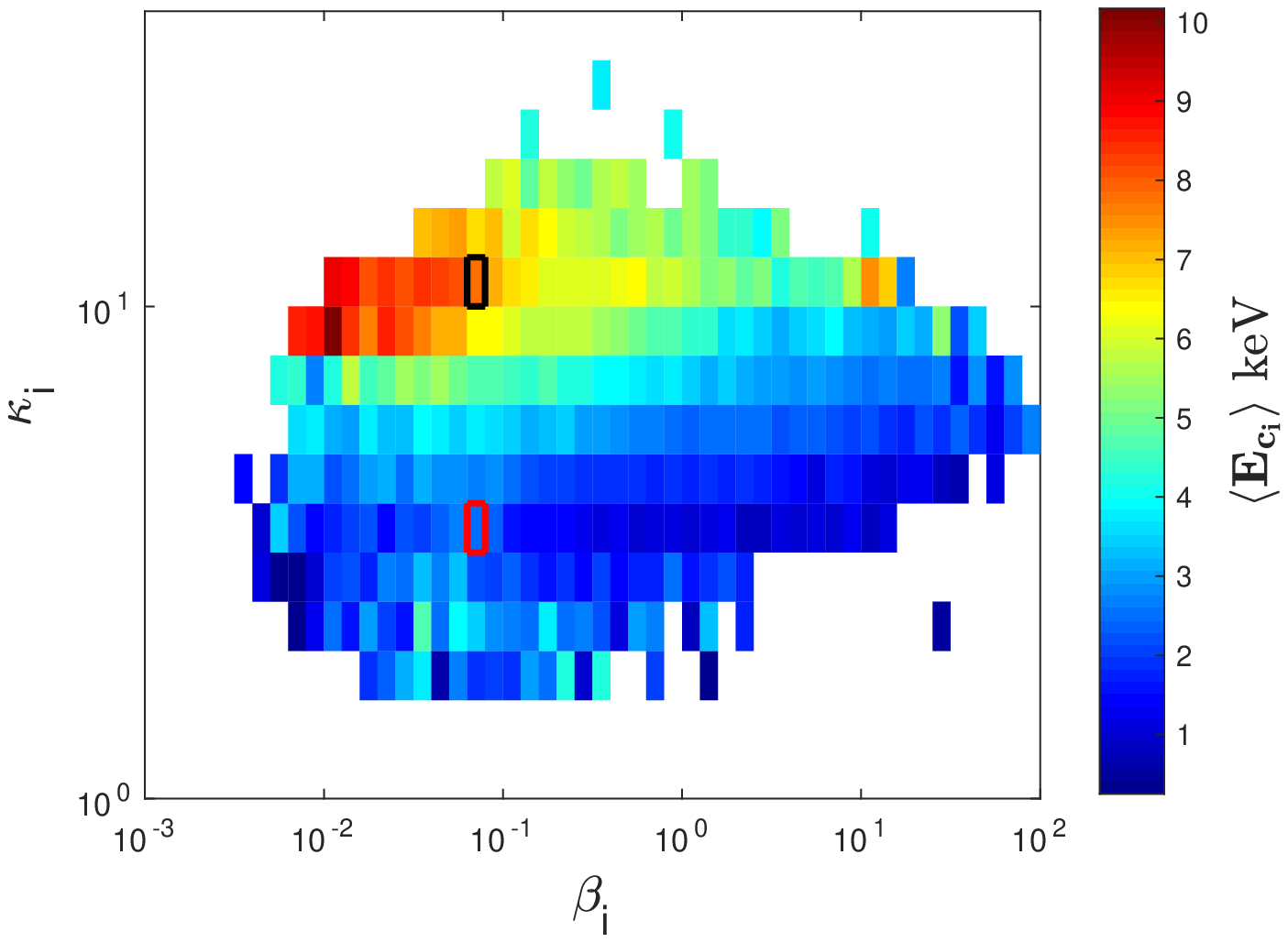}{0.48\textwidth}{(a)}
          \fig{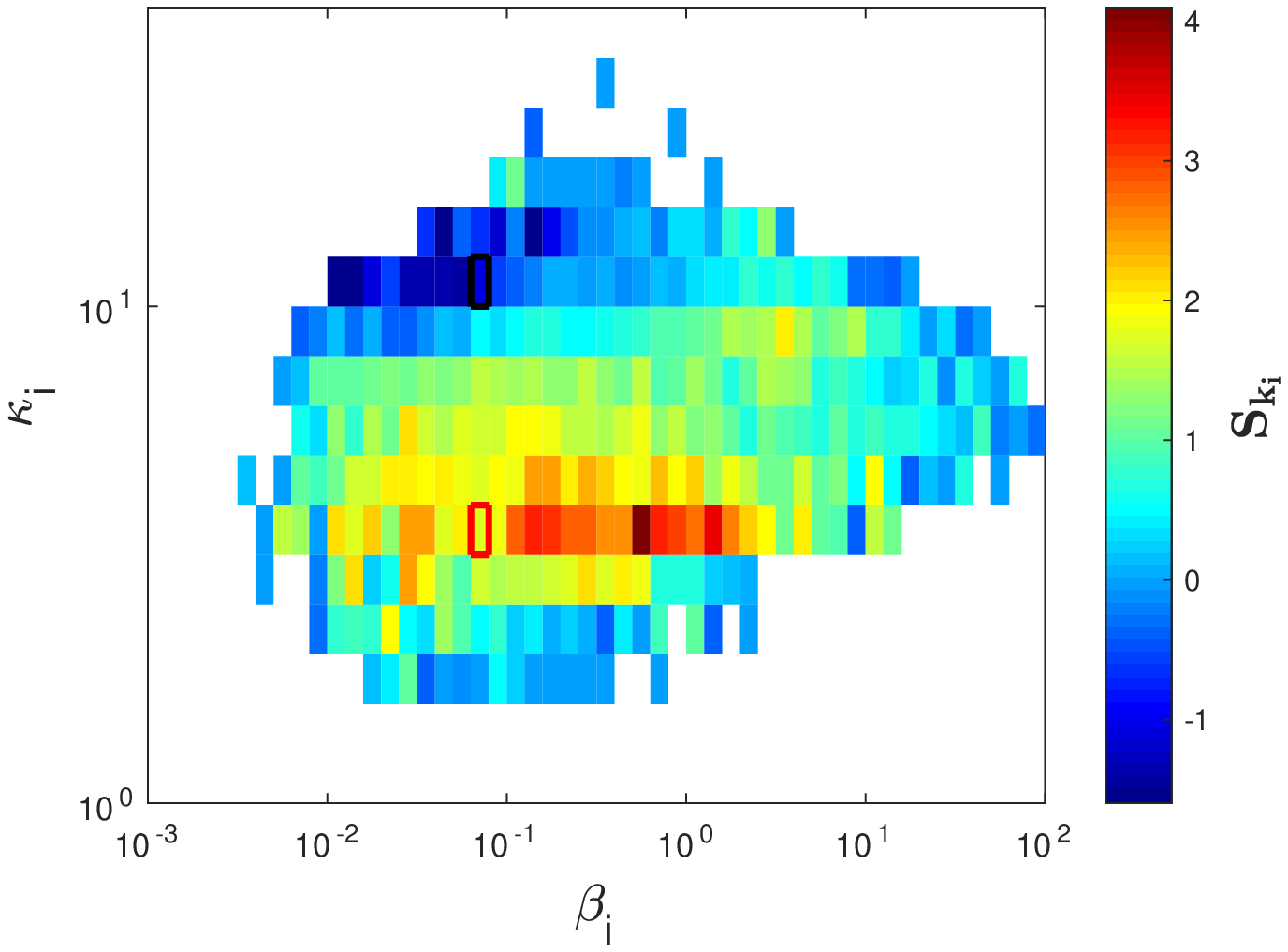}{0.48\textwidth}{(b)}
		  }
\gridline{\fig{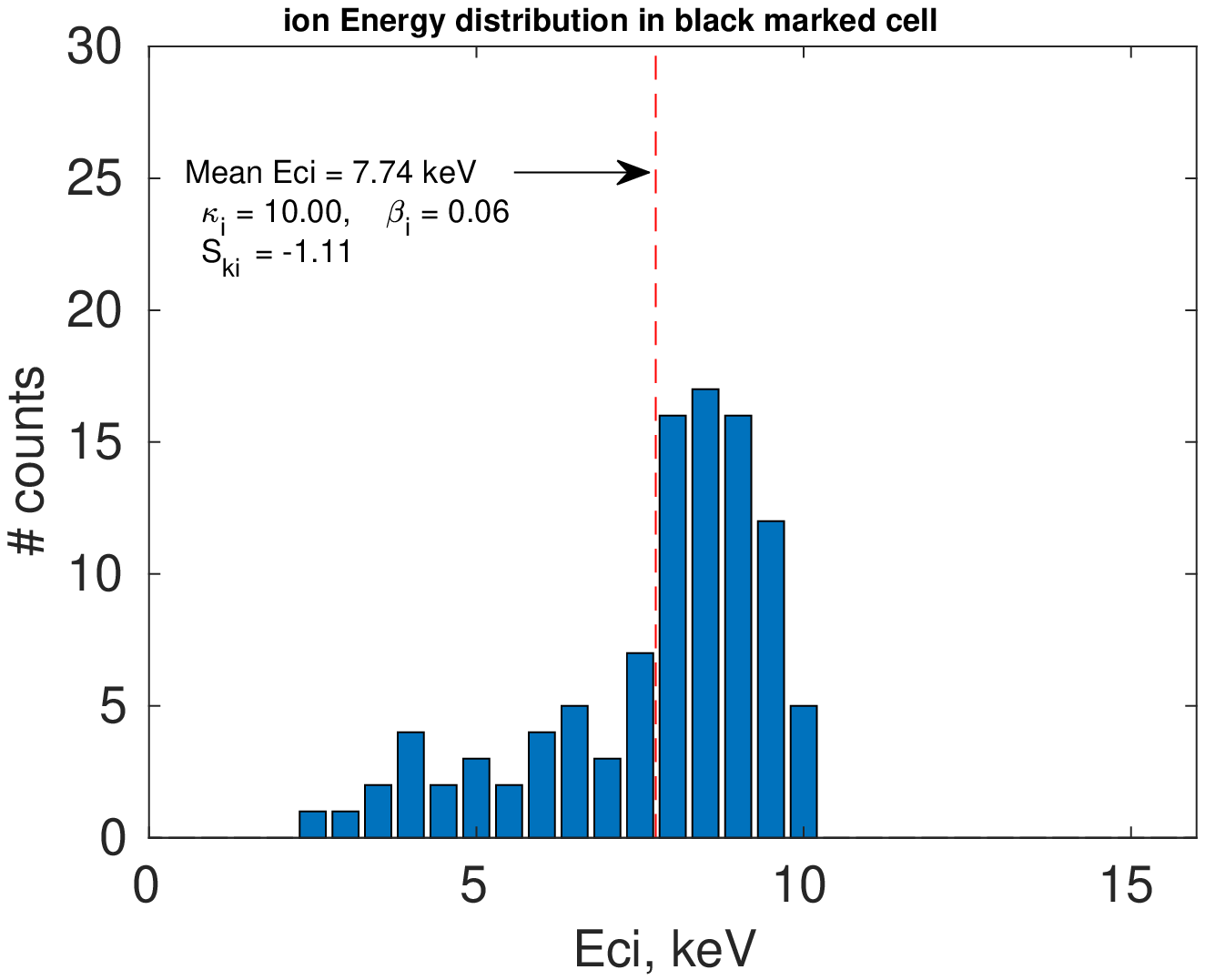}{0.48\textwidth}{(c)}
		  \fig{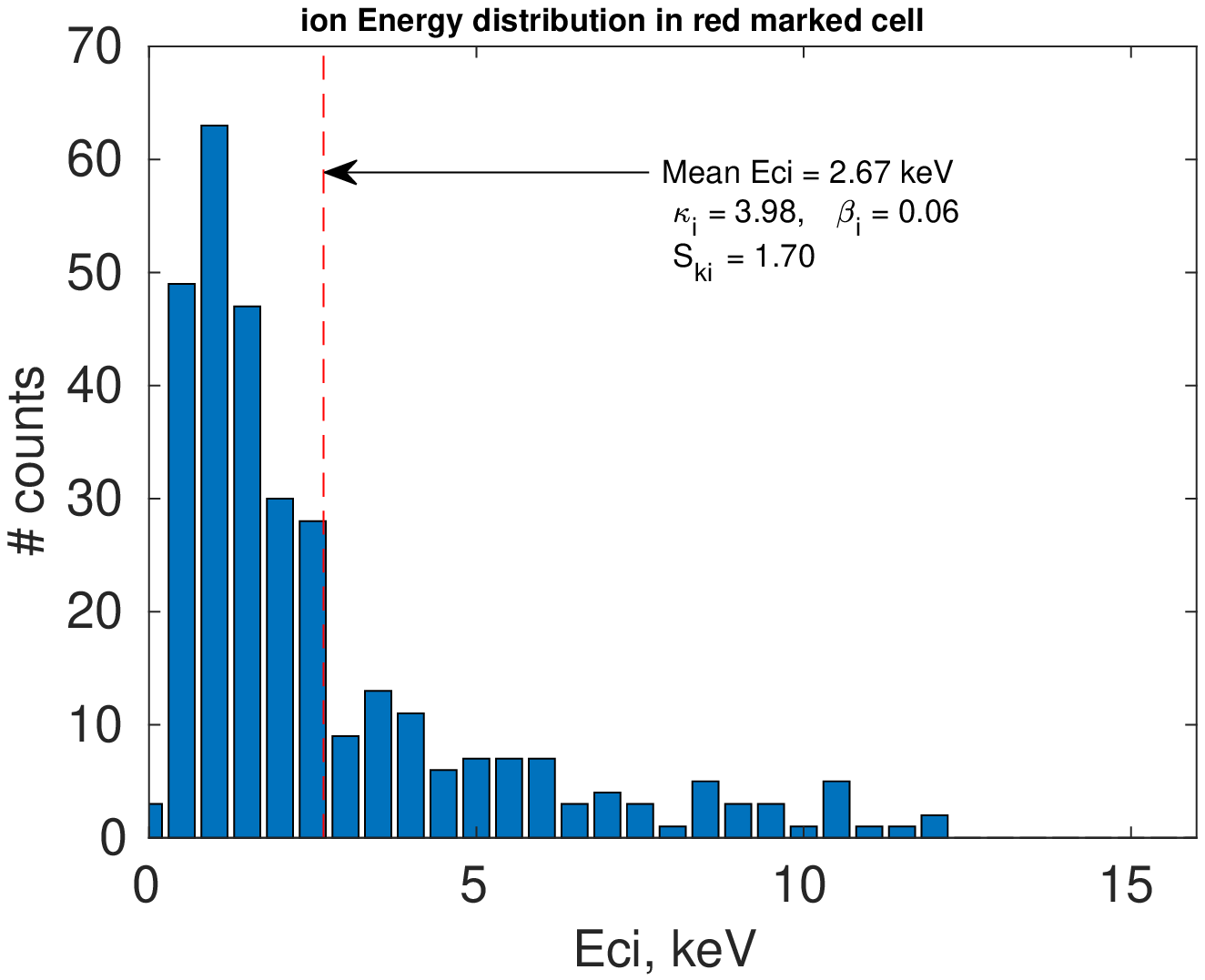}{0.48\textwidth}{(d)}
          }
\caption{Core energy distributions and skewness examples.
(a) 2-D plot of average ion core energy $\langle E_{ci} \rangle$ (color bar) in the $\beta_{i} - \kappa_{i}$ plane. 
(b) The skewness $S_{ki}$ of the distributions of $E_{ci}$ in each cell of the $\beta_{i} - \kappa_{i}$ plane. 
(c) Histogram of one $E_{ci}$ distribution to illustrate a negative skewness. This particular distribution corresponds to the bin marked with a black cell in panels (a) and (b).
(d) Histogram of one $E_{ci}$ distribution to illustrate a positive skewness. This particular distribution corresponds to the bin marked with a red cell in panels (a) and (b).
}
\label{fig:Eciandskew}
\end{figure*}

\begin{figure*}
\gridline{\fig{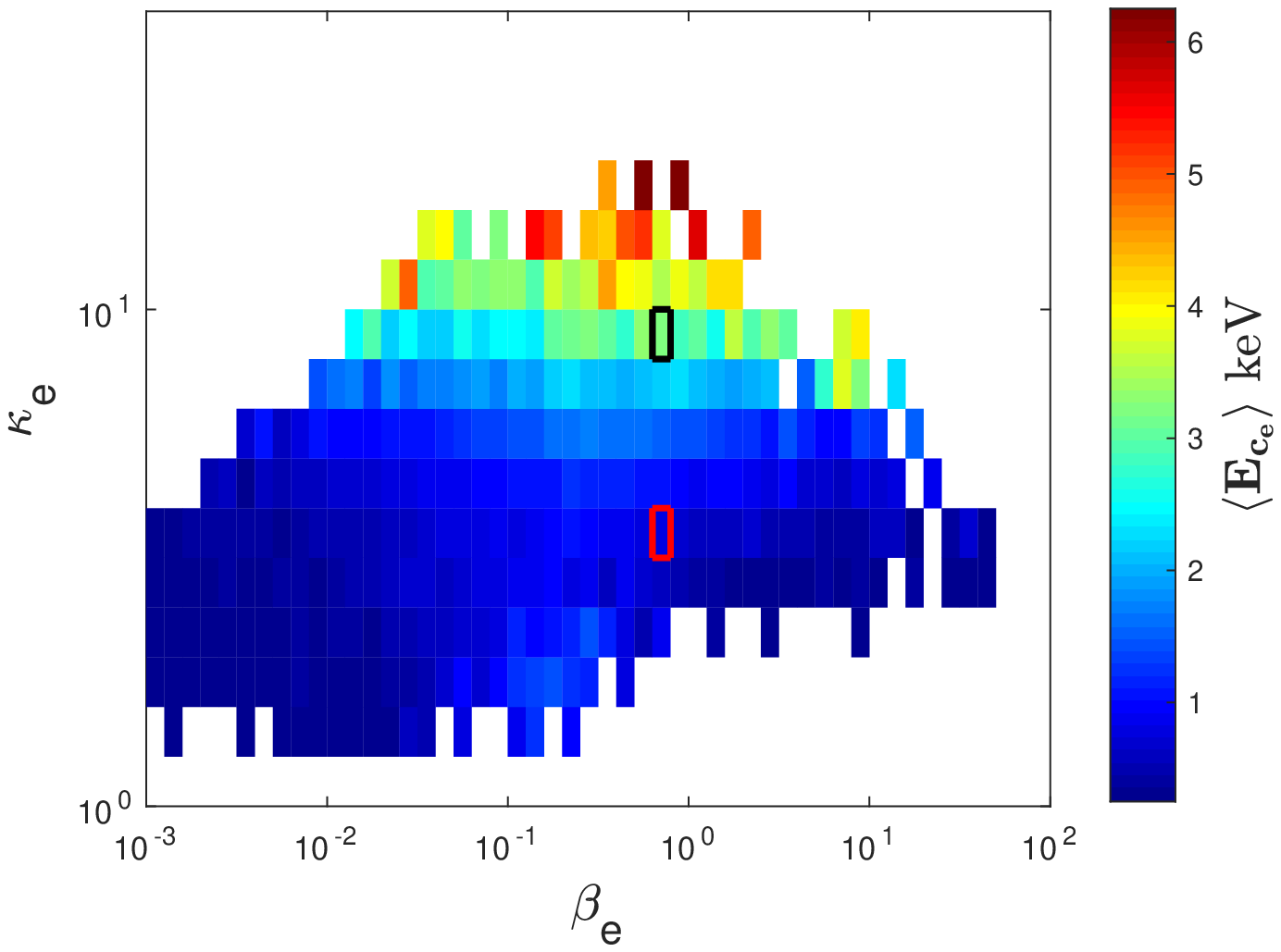}{0.48\textwidth}{(a)}
          \fig{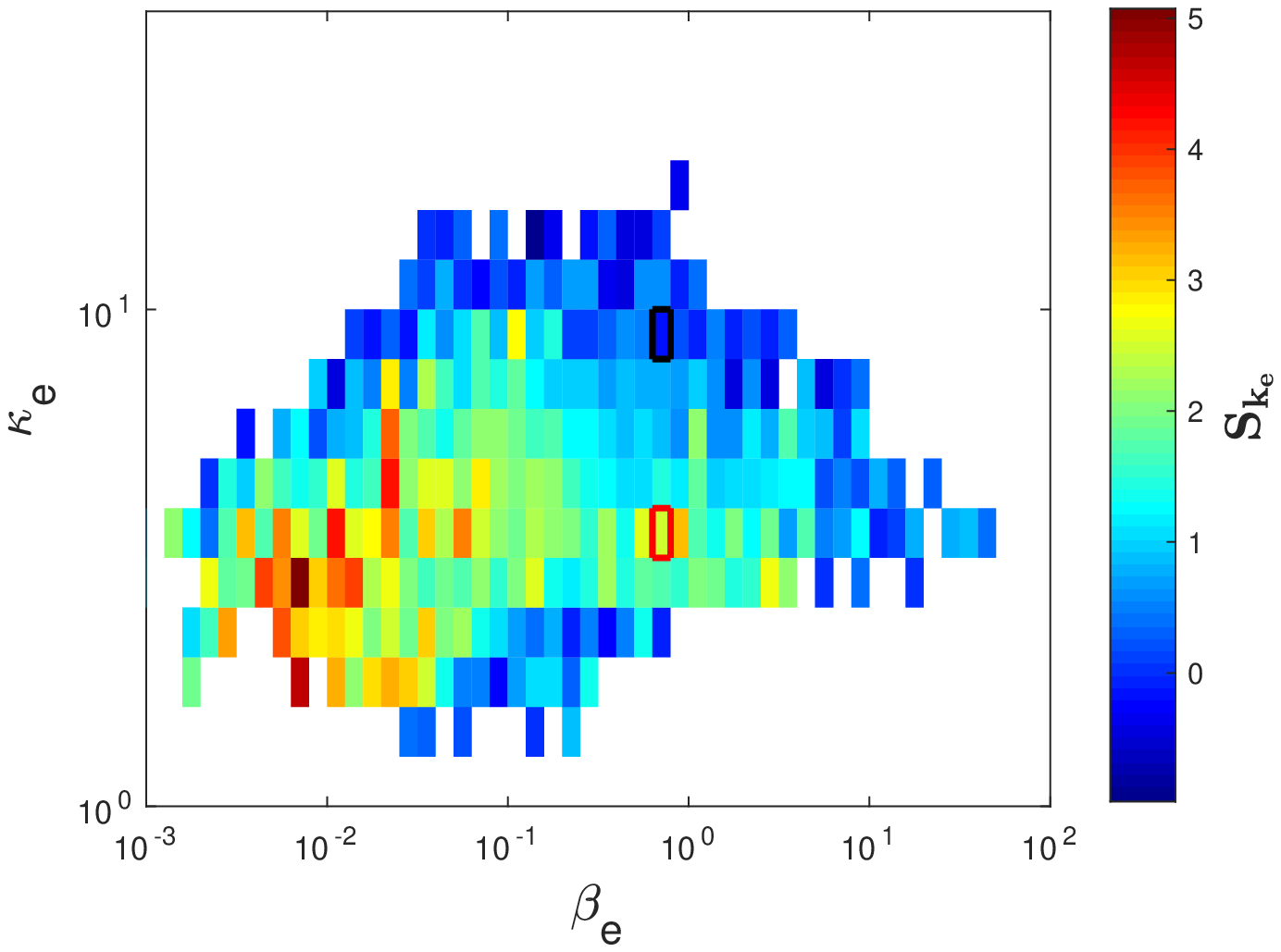}{0.48\textwidth}{(b)}
		  }
\gridline{\fig{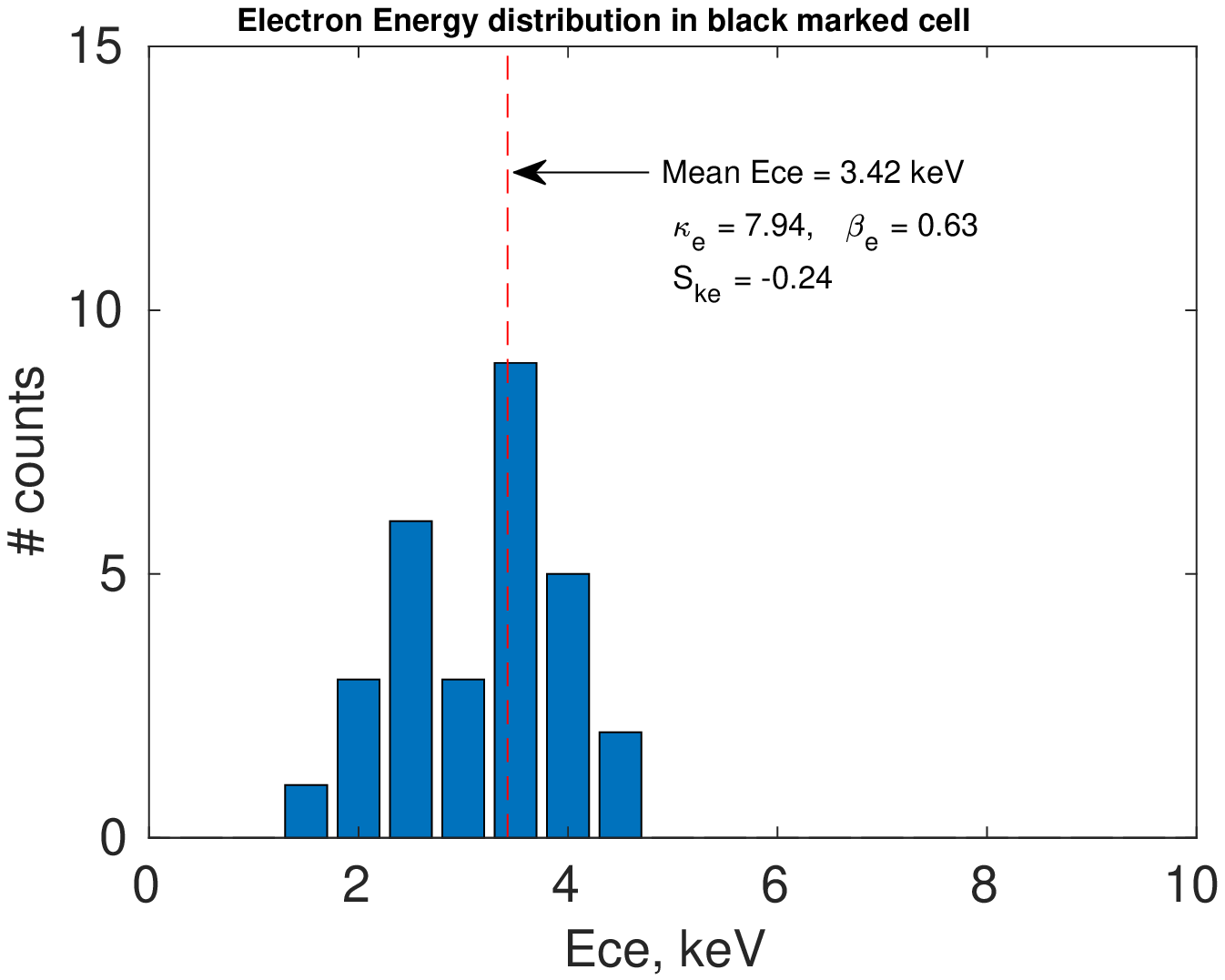}{0.48\textwidth}{(c)}
		  \fig{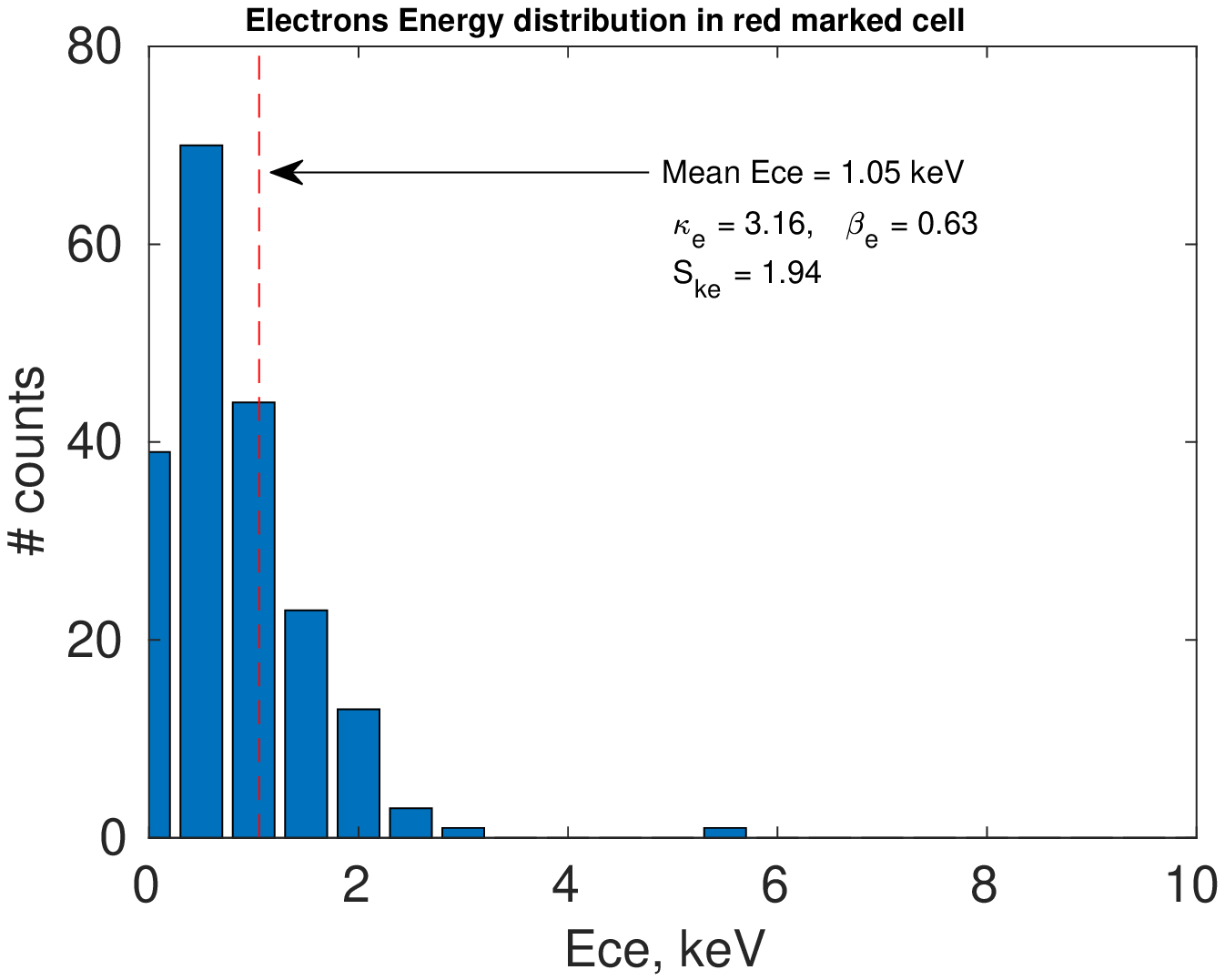}{0.48\textwidth}{(d)}
          }
\caption{ Core energy distributions and skewness examples.
(a) 2-D plot of average electron core energy $\langle E_{ce} \rangle$ (color bar) in the $\beta_{e} - \kappa_{e}$ plane. 
(b) The skewness $S_{ke}$ of the distributions of $E_{ce}$ in each cell of the $\beta_{e} - \kappa_{e}$ plane. 
(c) Histogram of one $E_{ce}$ distribution to illustrate a negative skewness. This particular distribution corresponds to the bin marked with a black cell in panels (a) and (b).
(d) Histogram of one $E_{ce}$ distribution to illustrate a positive skewness. This particular distribution corresponds to the bin marked with a red cell in panels (a) and (b).
}
\label{fig:Eceandskew}
\end{figure*}

The core energies obtained from the considered fits  cover the range $0.25$\,keV to $15.25$\,keV in the case of ions, and $0.25$\,keV to $9.25$\,keV for the electrons. 
In order to study the relation between $\kappa$ and  $E_c$ for a fixed $\beta$, it is necessary to obtain the most representative $E_{c}$ for each ($\beta,\kappa$) bin. A popular candidate would be the mean value, which works well in the case of normal distributions. In order to determine the degree to which they depart from a normal distribution, we analyze the distribution of core energies in each bin. As observed in Figures \ref{fig:Eciandskew}(b) and \ref{fig:Eceandskew}(b), the obtained distributions are not always Gaussian. To characterize this, we determined the mean, median, and skewness for the $E_{c}$ distribution in each ($\beta,\kappa$) bin. Figures \ref{fig:Eciandskew} and \ref{fig:Eceandskew} show the mean $E_{c}$ values and the skewness for ions and electrons. 
As seen in both species, the hot plasma is negatively skewed; meanwhile, the cold plasma is positively skewed. 
Therefore, to ensure the robustness of our study, the analyses were performed twice: once for the mean $E_{c}$ values and once for the median $E_{c}$ values. 
We can confirm that the results are qualitatively very similar for both cases.
Median and also mode $E_{c}$ values as a function of $\kappa$ and $\beta$, for both species, can be seen in the Appendix (Figures \ref{fig:medianEnergy} and \ref{fig:modeEnergy}).

Previous studies of Kappa distributions for ions have reported that kappa increases with core energy in a linear fashion \citep{Christon_et_al_1989, Collier_1999}. Meanwhile, \citet{kirpichev2020dependencies} has recently established that a power-law function of the form $\kappa = AE^{\gamma}$ can be used to describe the relationship between kappa and core energy in the case of ions. 
Here we use the same function as \citet{kirpichev2020dependencies} for both species.

\begin{figure*}
\gridline{\fig{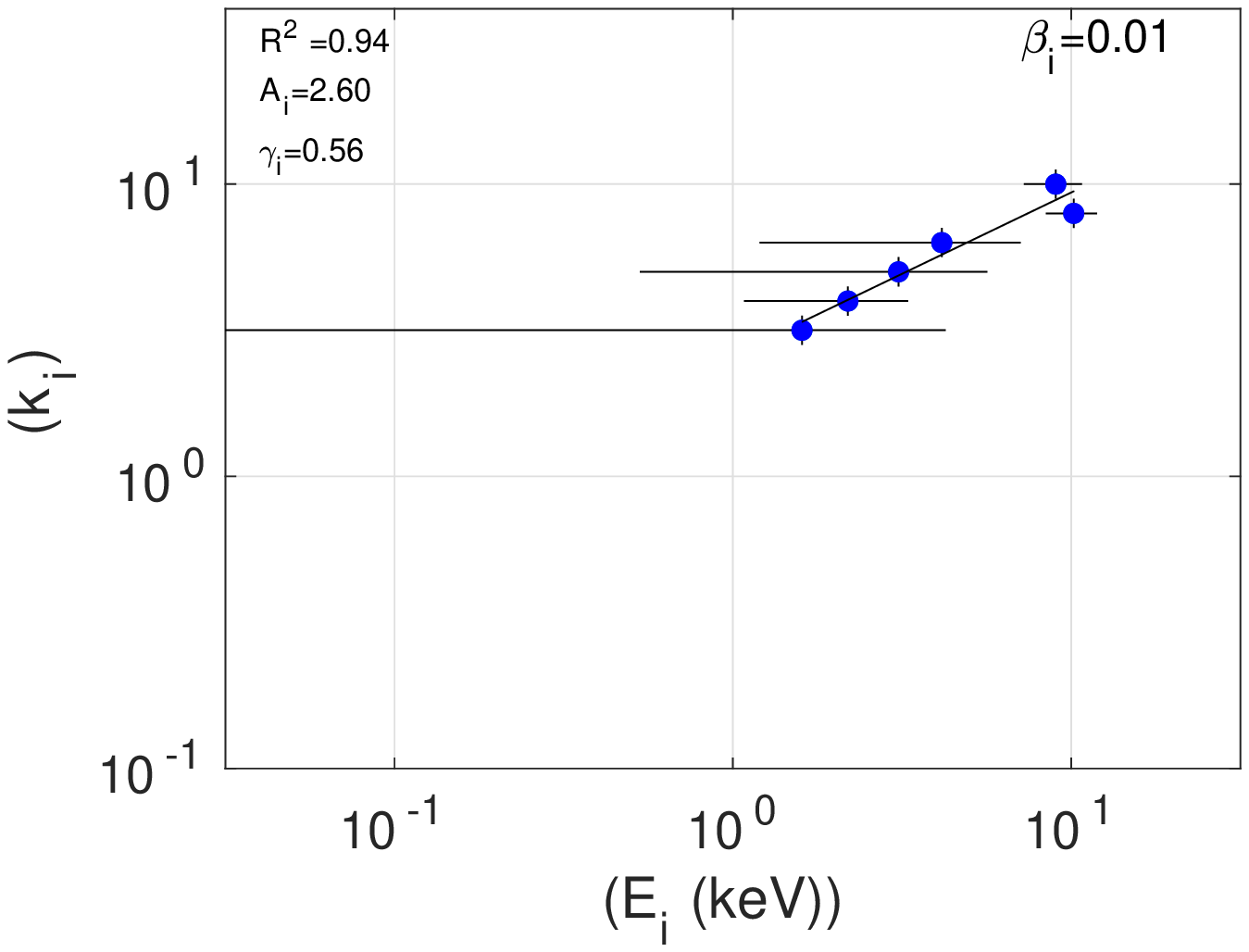}{0.35\textwidth}{(a)}
          \fig{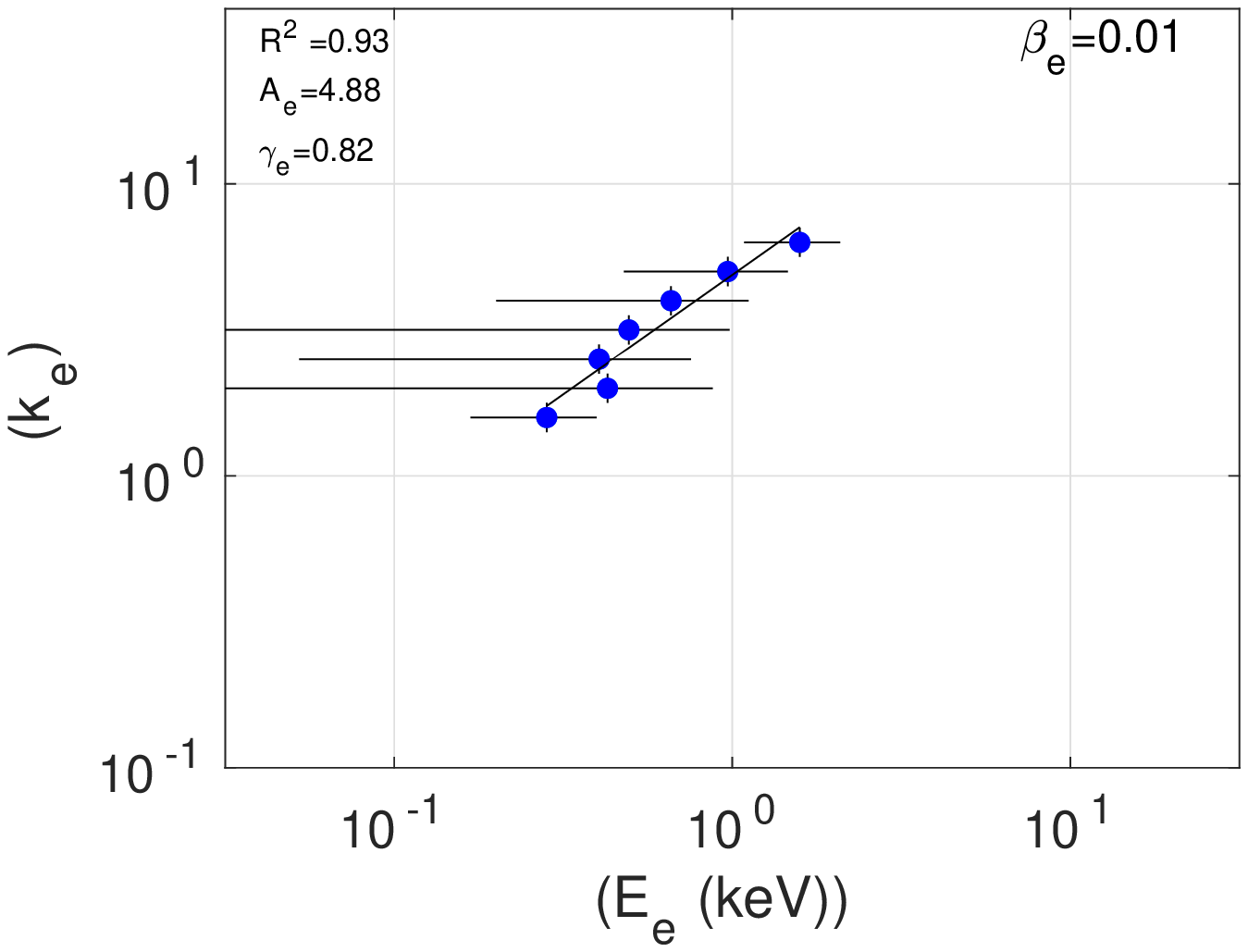}{0.35\textwidth}{(b)}
          }
\gridline{\fig{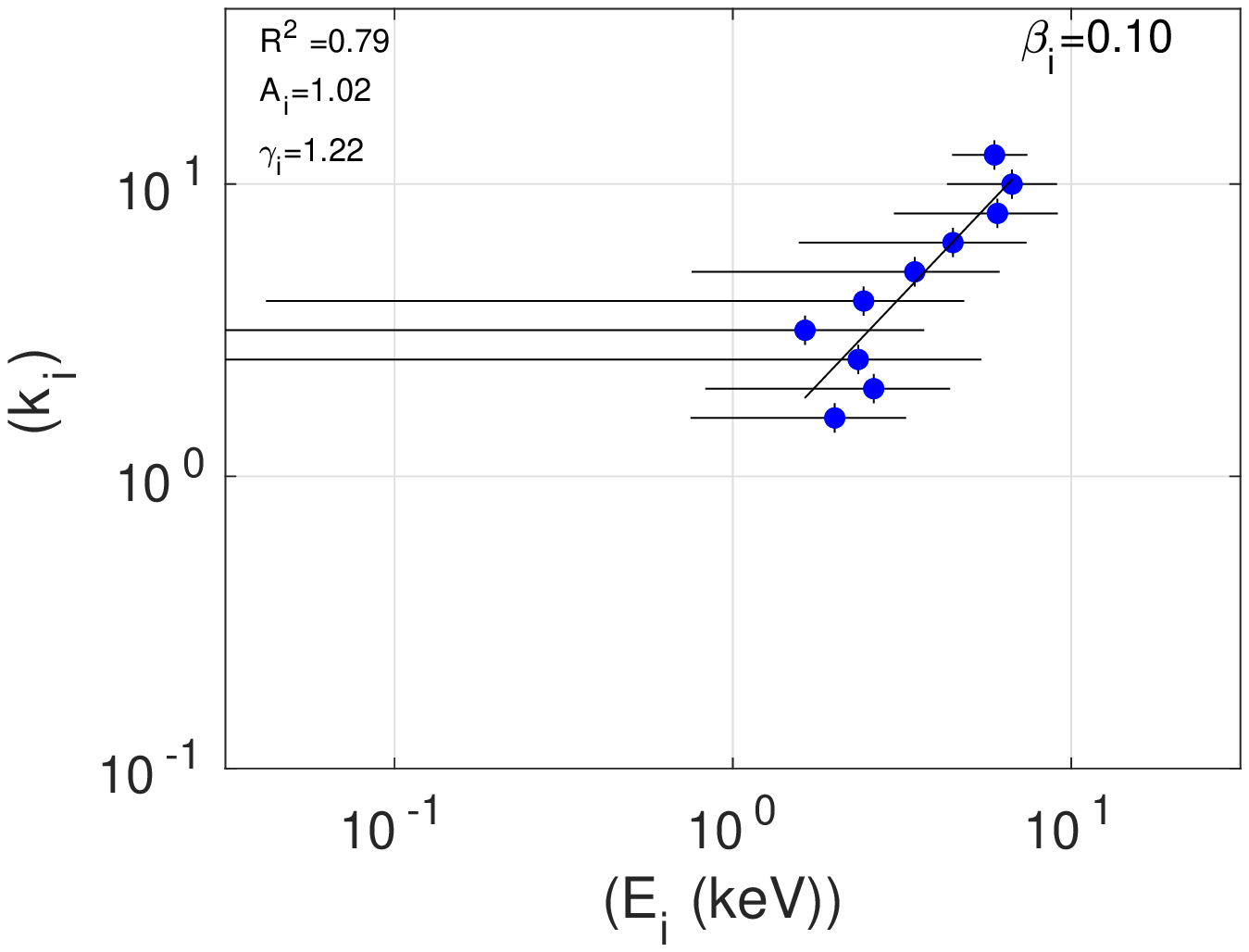}{0.35\textwidth}{(c)}
          \fig{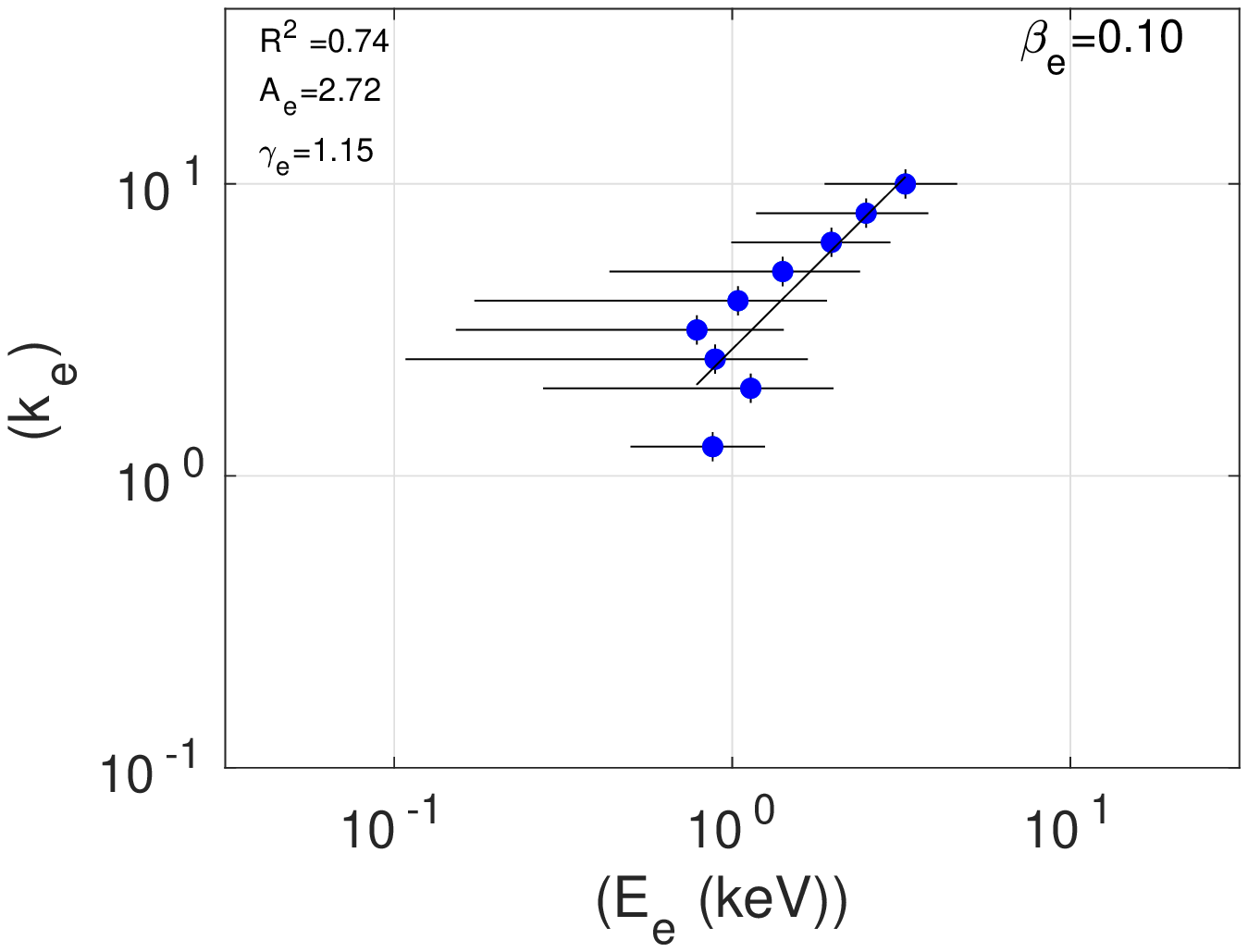}{0.35\textwidth}{(d)}
          }
\gridline{\fig{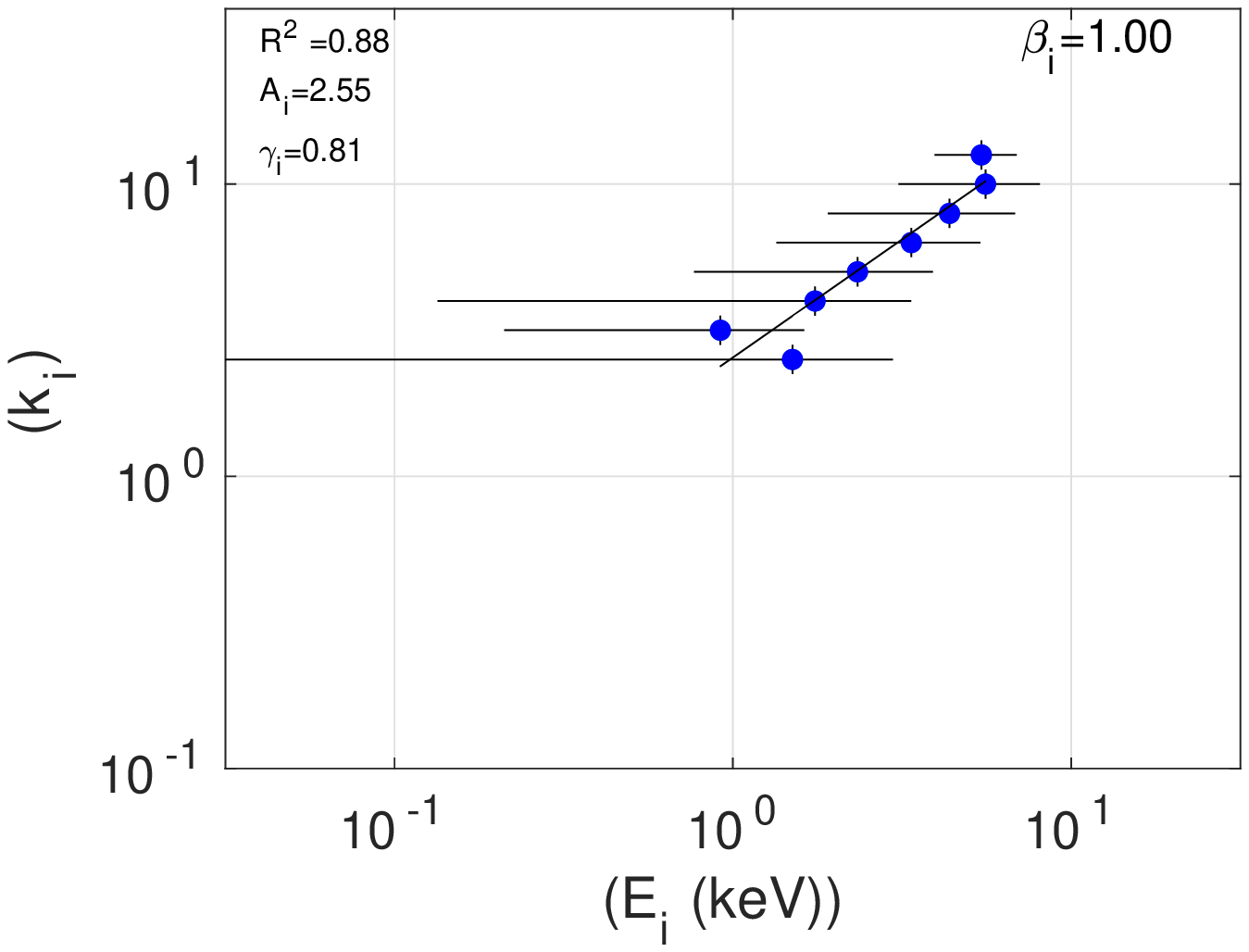}{0.35\textwidth}{(e)}
		  \fig{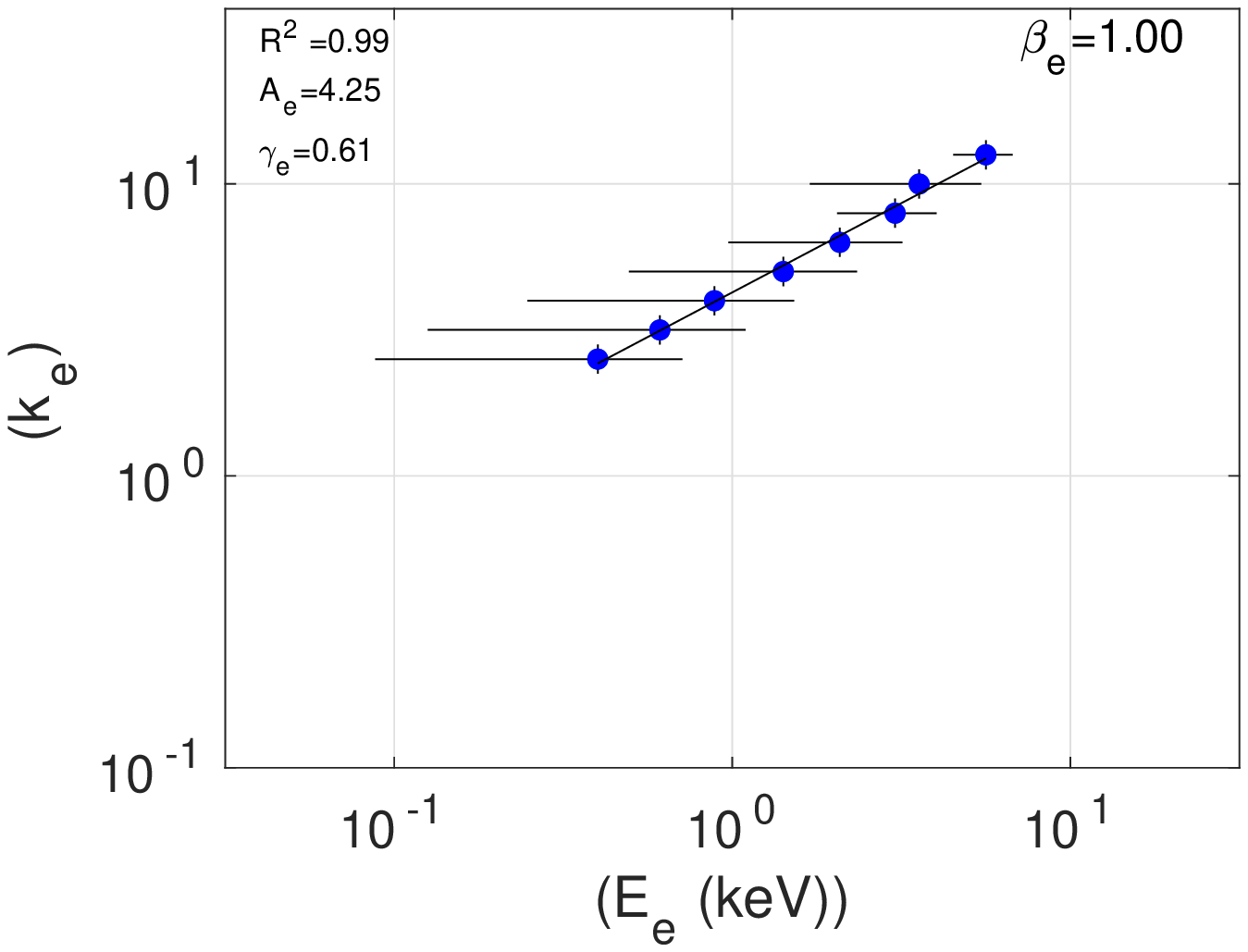}{0.35\textwidth}{(f)}
		  }
\gridline{\fig{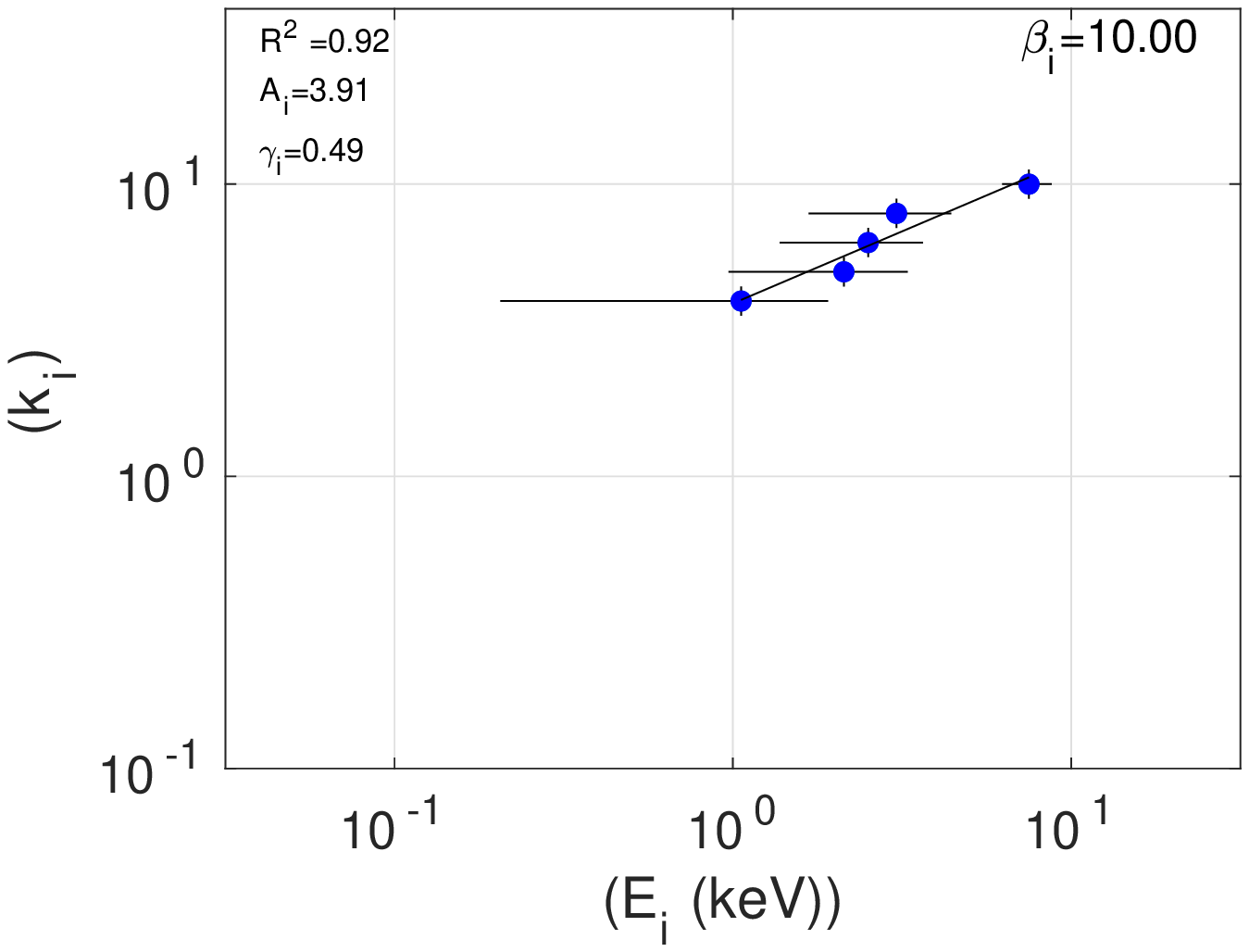}{0.35\textwidth}{(g)}
		  \fig{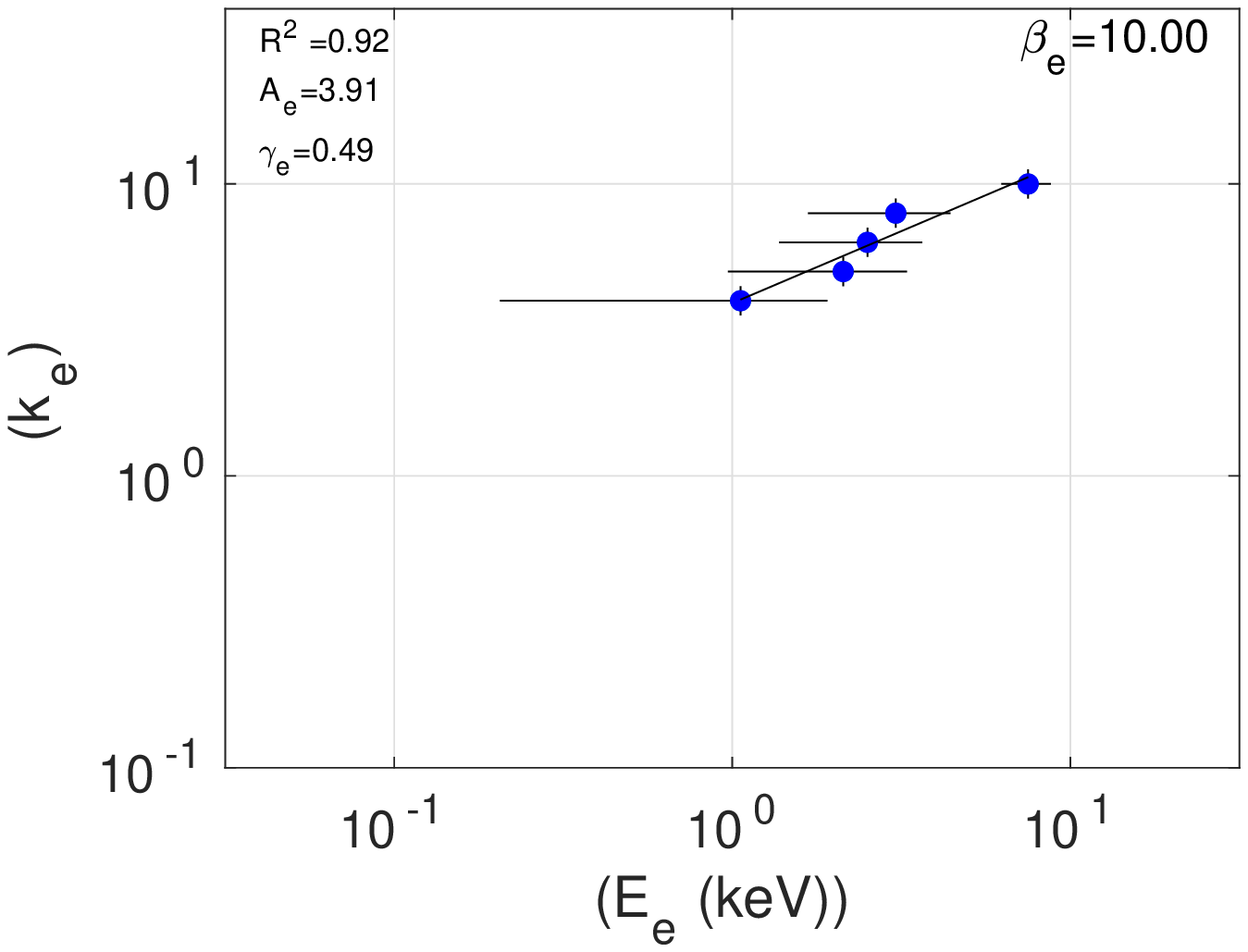}{0.35\textwidth}{(h)}  		  
          }
\caption{Plots of the dependence of ion (left) and electron (right) core energies with $\kappa$, for different constant $\beta$ values. 
Correlation coefficients and power-law fitting coefficients $A$ (Amplitude), and $\gamma$ (power-law index) are shown on the plots.}
\label{fig:iANDedependences}
\end{figure*}

\begin{figure*}
\gridline{\fig{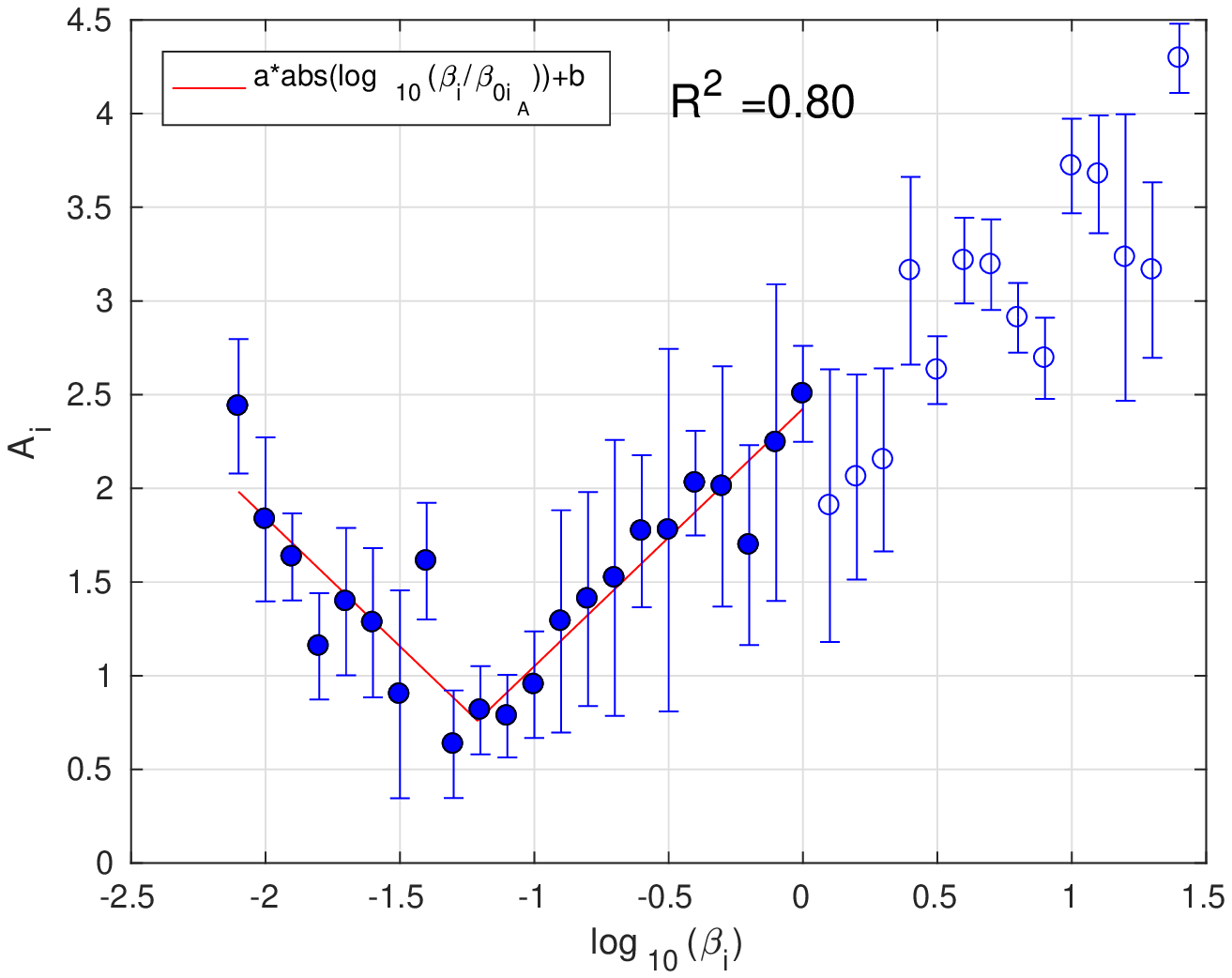}{0.5\textwidth}{(a)}
          \fig{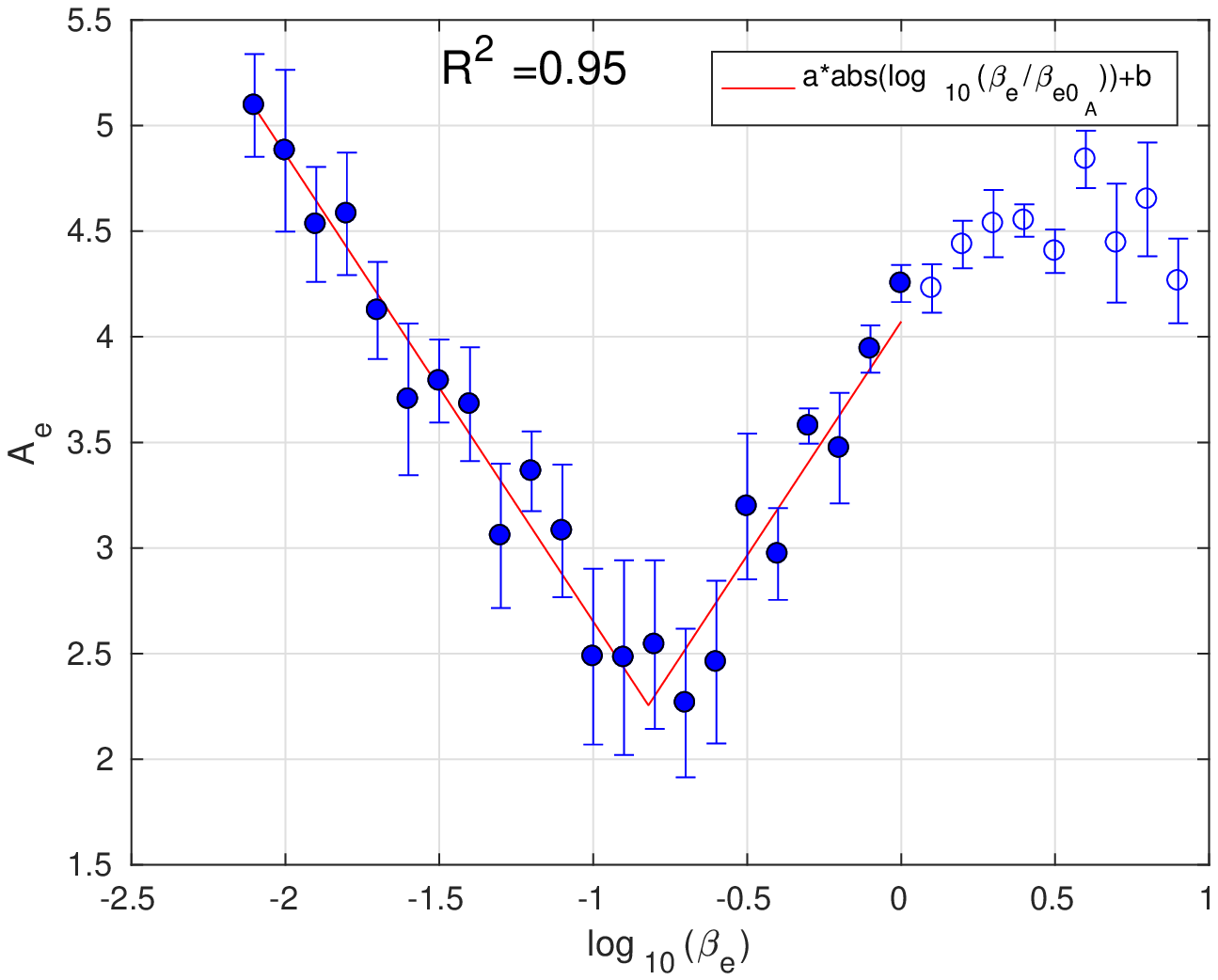}{0.5\textwidth}{(b)}
          }
\gridline{\fig{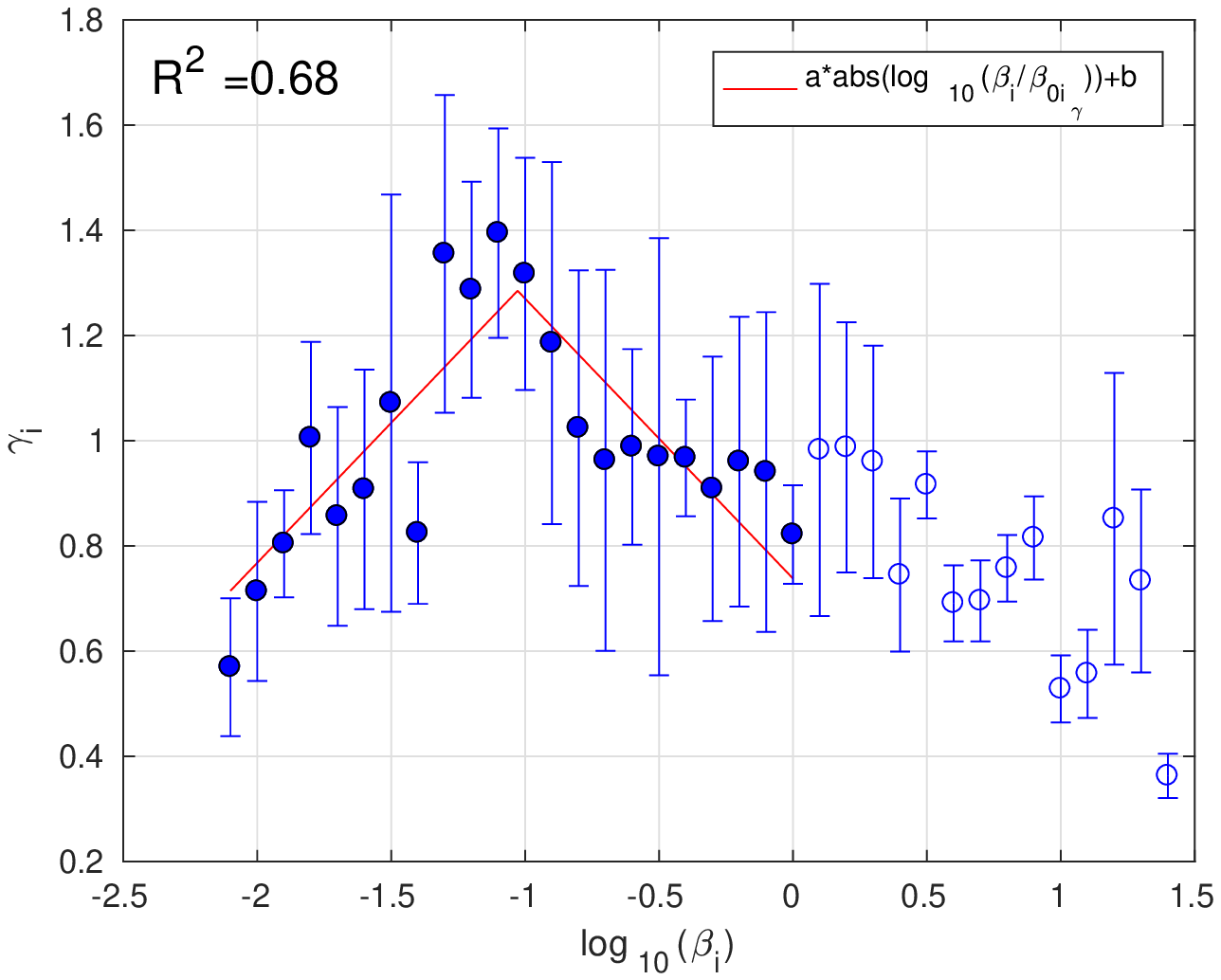}{0.5\textwidth}{(c)}
          \fig{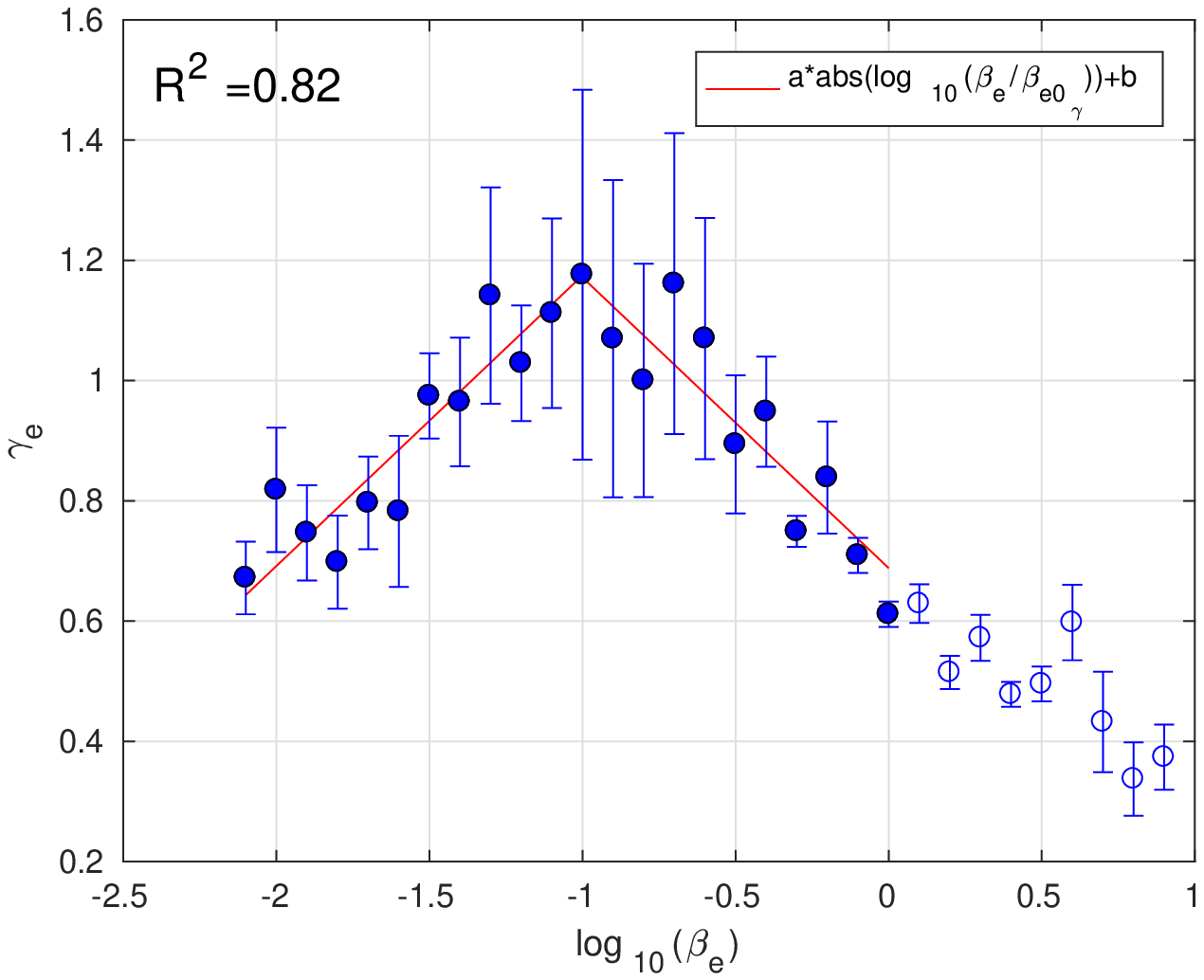}{0.5\textwidth}{(d)}
          }
\caption{ Dependency of the fitted power-law coefficient $A$ (top) and $\gamma$ (bottom) with $\beta$. 
The left panels are for ions and the right panels for electrons.
}
\label{fig:interceptslopecurves}
\end{figure*}

Examples of kappa versus core energy, for some selected beta values, are shown in Figure \ref{fig:iANDedependences}, where the left column (Panels (a),(c),(e), and (g)) show the results for ions, and the right column (Panels (b), (d), (f), and (h)) for electrons. The horizontal error bars correspond to the standard deviation of the core energy $E_{c}$ in each ($\kappa, \beta$) bin. 
The fits to log-log data have linear correlation coefficients $R^2>0.7$. 
Thus we conclude that under fixed $\beta $ conditions, the $\kappa$-index increases with $E_{c}$ for both ions and electrons, and that they relate via a power-law. 
Table \ref{table:fittingvalues} details the results of these fits for a few selected beta values.
In order to establish whether there is a dependence of $A$ and $\gamma$ on $\beta$ we plot all obtained fitting coefficients as shown in Figure \ref{fig:interceptslopecurves}. It was found that for both species, the relation between $A$ and beta has a clear minimum near $\beta=0.1$, and that is symmetric with respect to it up to at least $\beta=1$. A similar relation is observed for $\gamma$, which exhibits a maximum at approximately the same value of $\beta$. 
To characterise this behaviour we use an empirical relation between the power-law coefficients $A$ or $\gamma$ and $\beta$ of the form $a\vert \log_{10}(\beta/\beta_{0}) \vert + b$, where $\beta_{0}$ is the location of the extremum. 
This function was fitted to $A$ and $\gamma$ data around the minimum or maximum (as examples, the data used for the fits in the cases shown in Figure \ref{fig:interceptslopecurves} are plotted with filled circles). 
The results of these fits can be found in Table \ref{table:coeffvalue}. 
In both cases, we find that $\beta_0 \sim 0.1$ and the slopes involved are different for $ A $ and $ \gamma $, for ions and electrons.

\begin{figure*}[h!]
\gridline{\fig{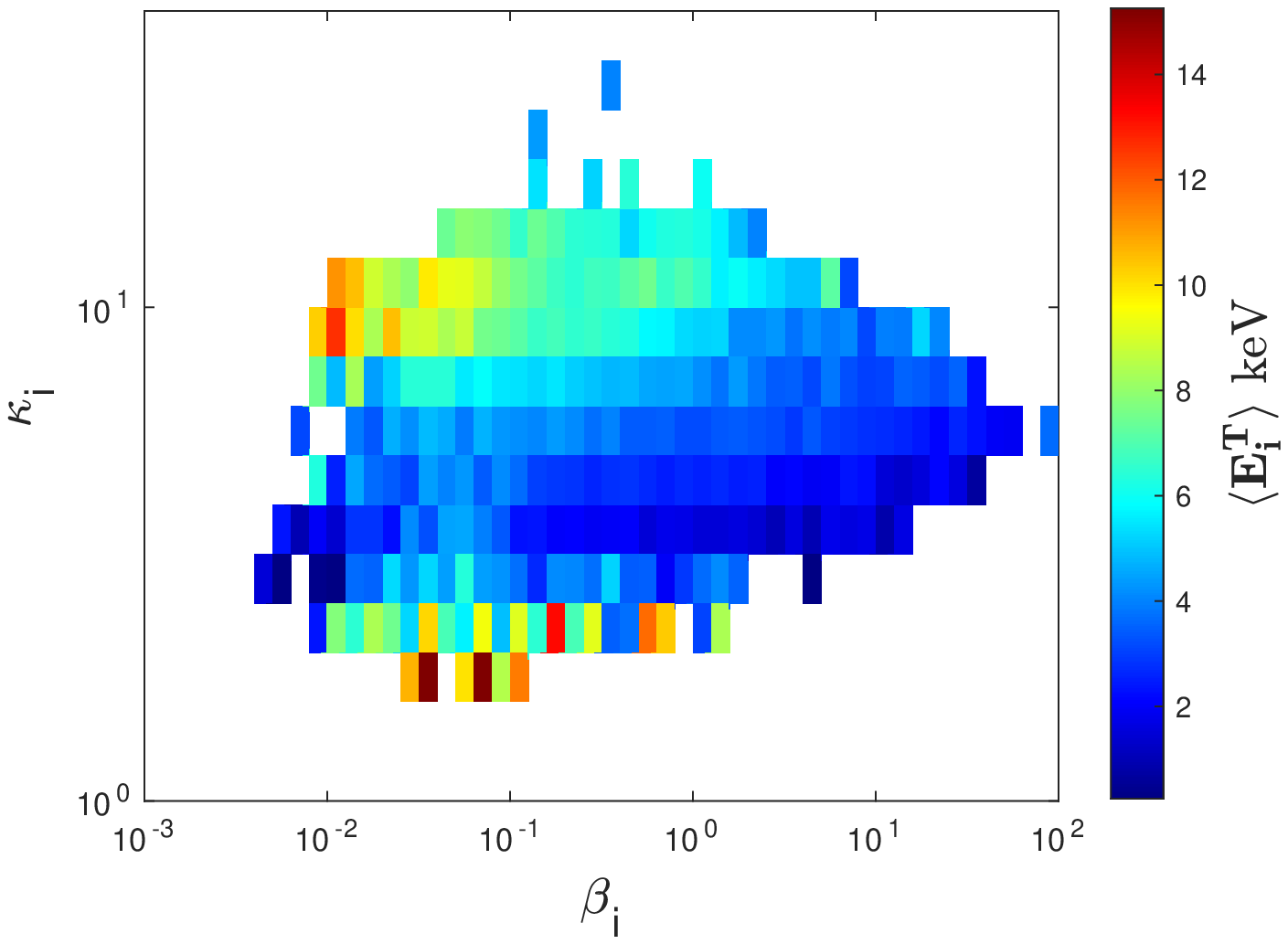}{0.53\textwidth}{(a)}
		  \fig{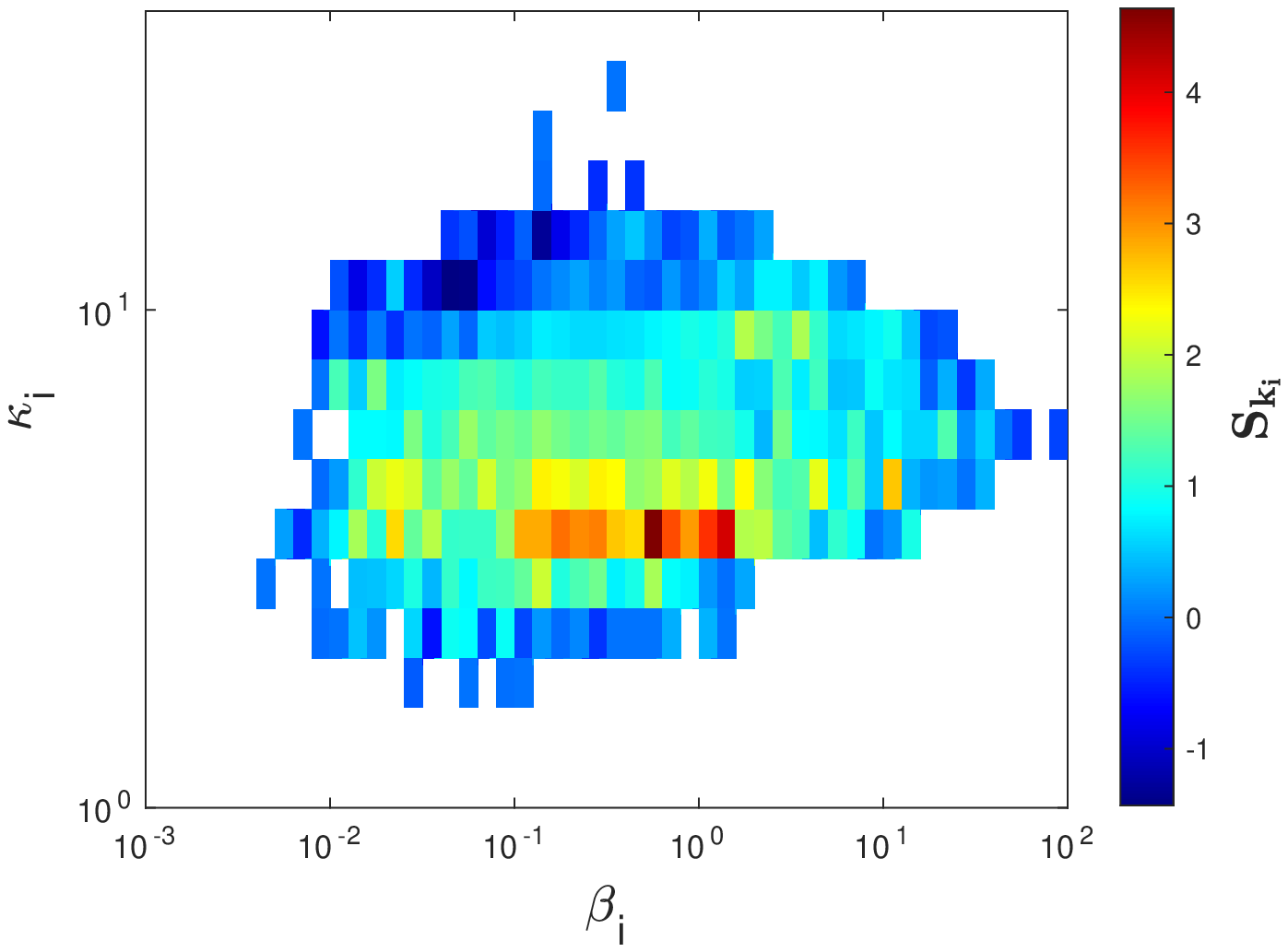}{0.53\textwidth}{(b)}
          }
\gridline{\fig{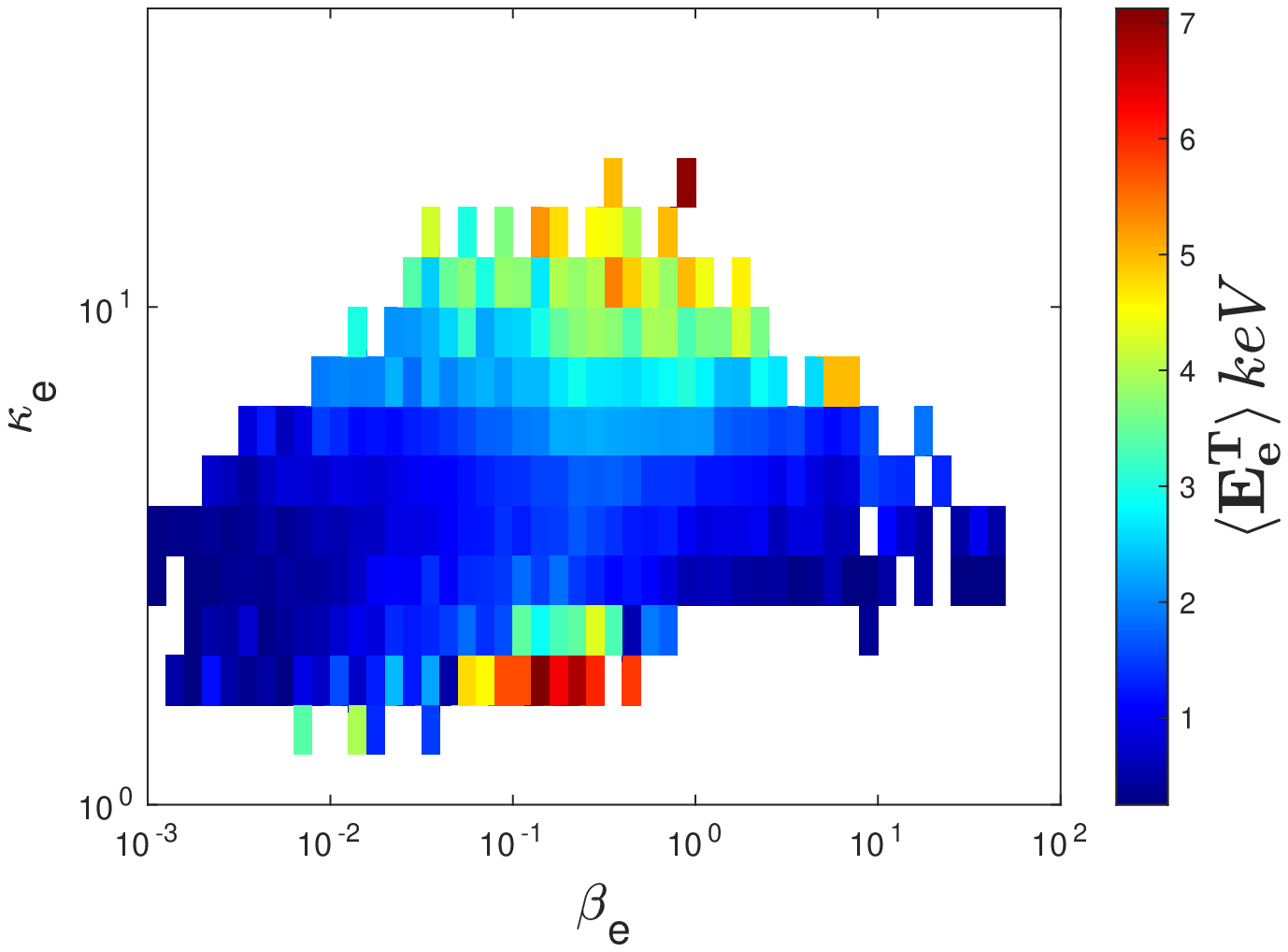}{0.53\textwidth}{(c)}
		  \fig{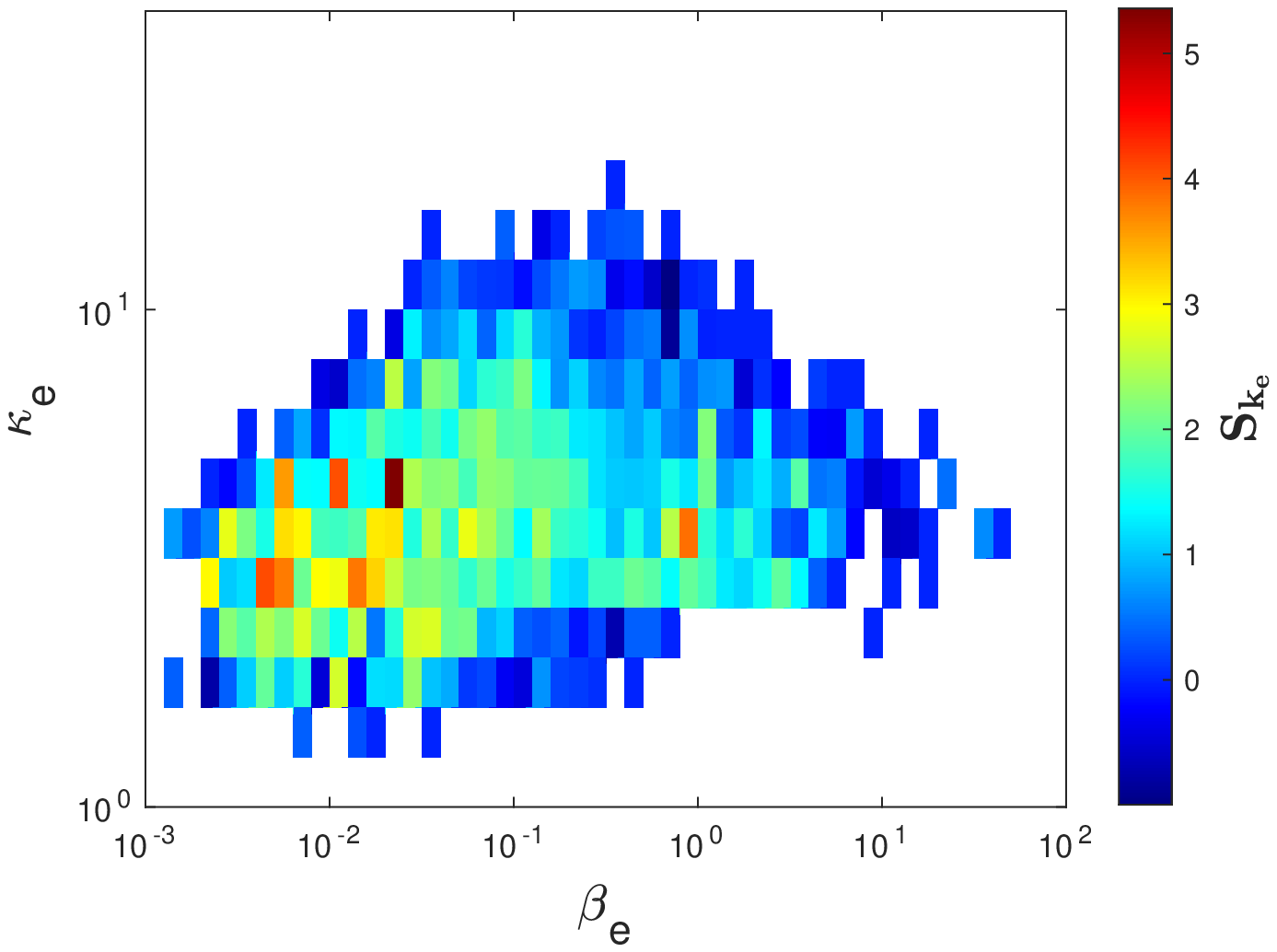}{0.53\textwidth}{(d)}
          }
\caption{
(a) 2-D plot of ion total energy $E^T_{i}$ (color bar) in the $\beta_{i} - \kappa_{i}$ plane. 
(b) The skewness $S_{ki}$ of the distributions of $E^T_{i}$ in each cell of the $\beta_{i} - \kappa_{i}$ plane. 
(c) 2-D plot of electron total energy $E^T_{e}$ (color bar) in the $\beta_{e} - \kappa_{e}$ plane.
(d) The skewness $S_{ki}$ of the distributions of $E^T_{e}$ in each cell of the $\beta_{e} - \kappa_{e}$ plane.}
\label{fig:total_energy}
\end{figure*}

\begin{figure*}
\gridline{\fig{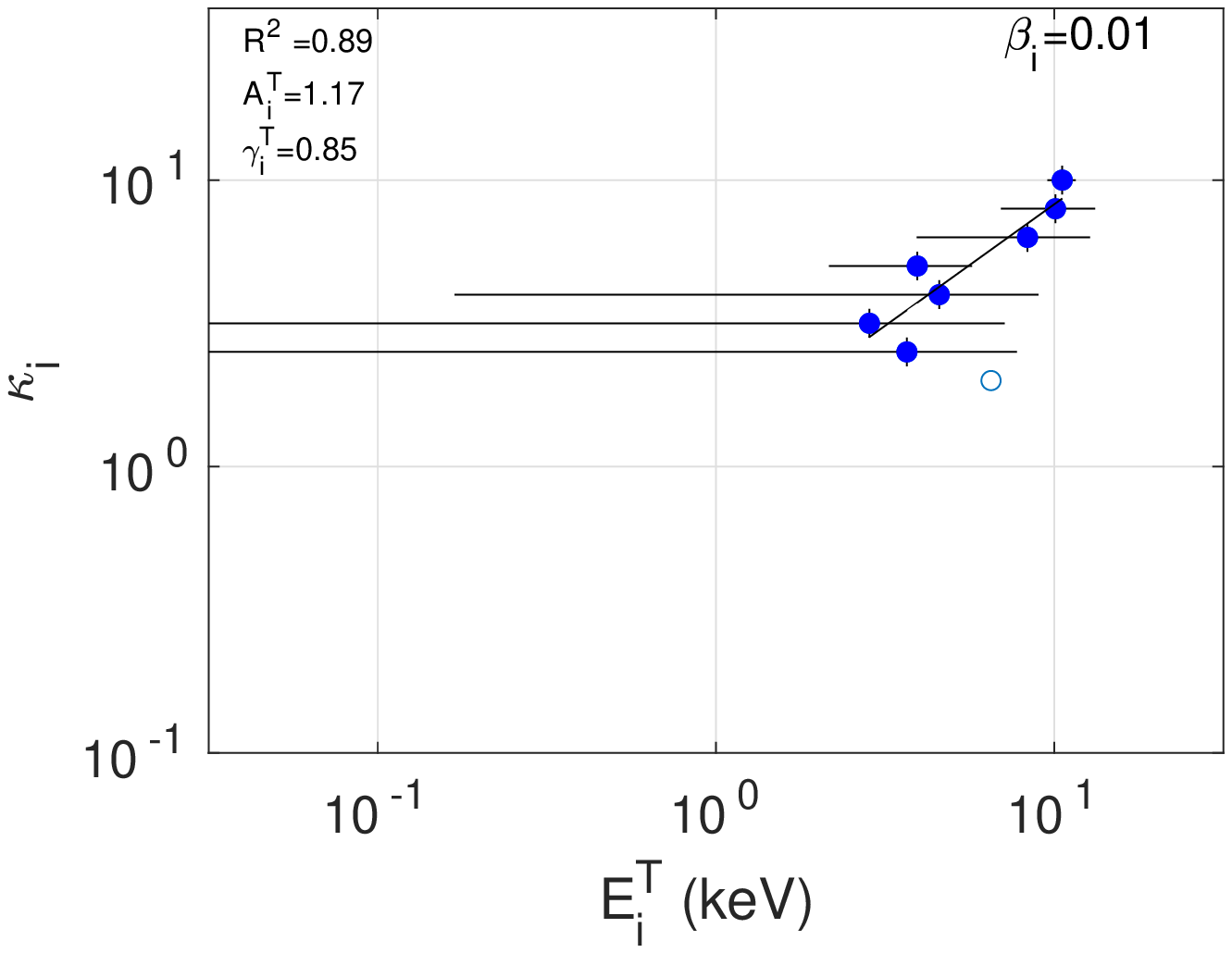}{0.35\textwidth}{(a)}
          \fig{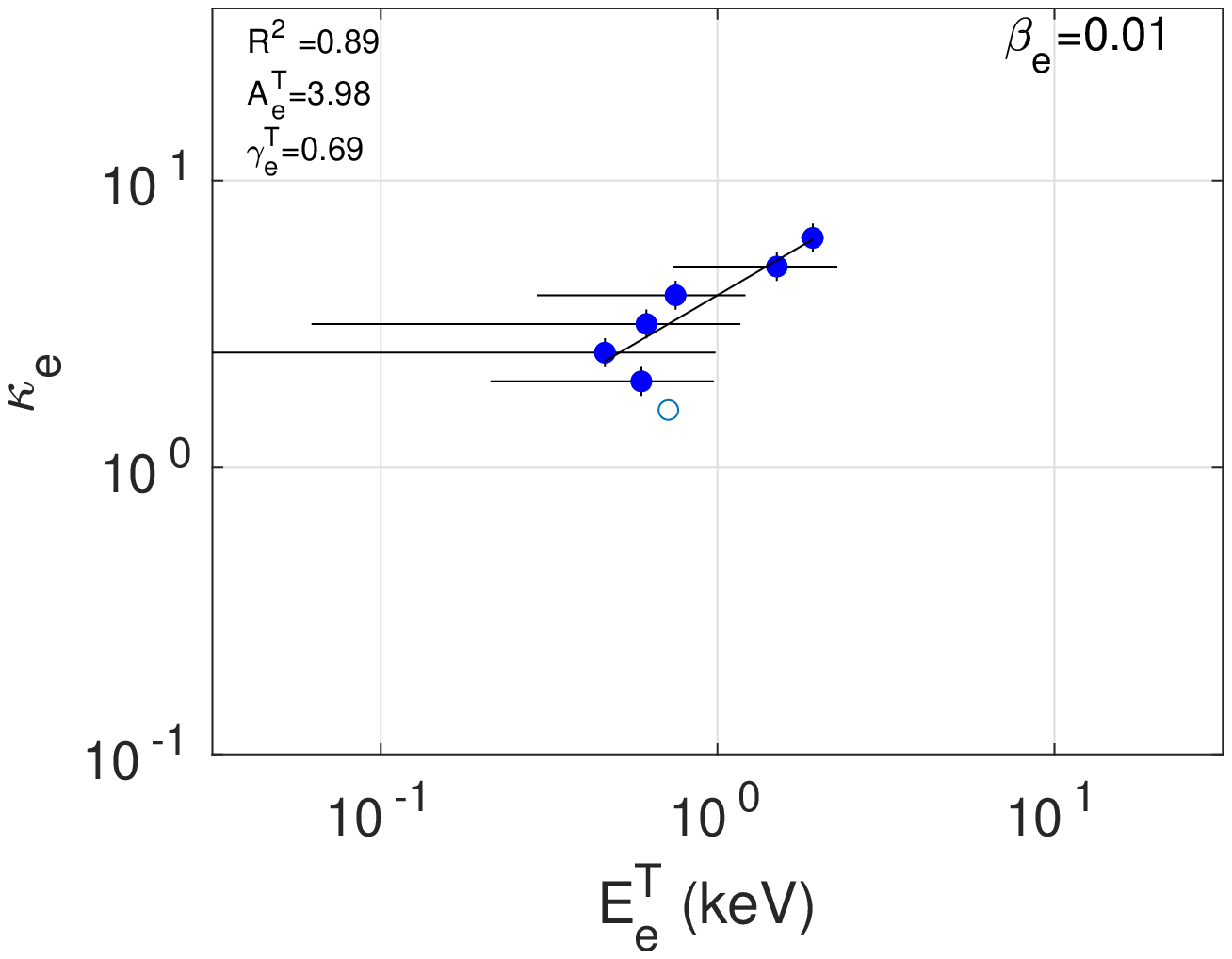}{0.35\textwidth}{(b)}
          }
\gridline{\fig{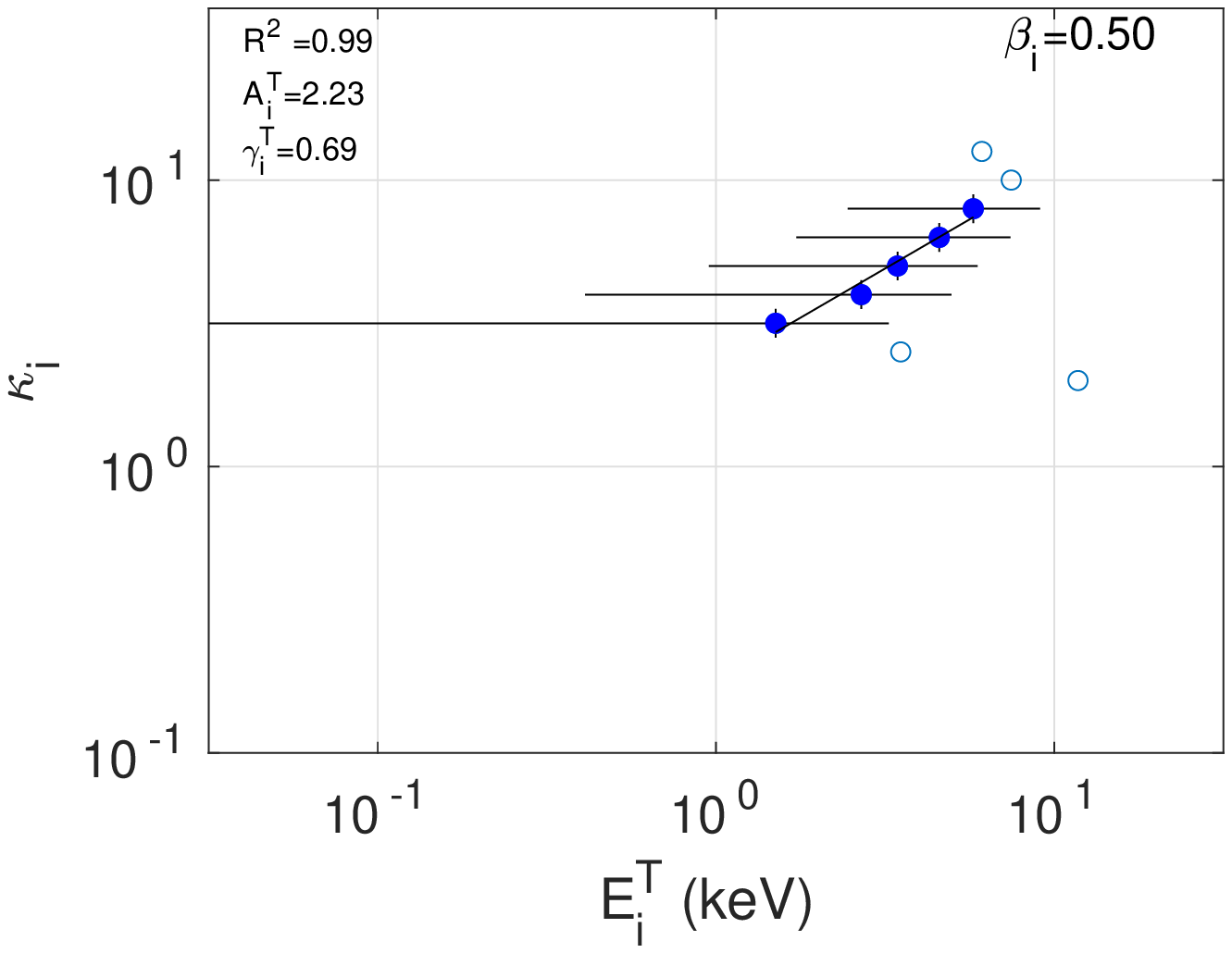}{0.35\textwidth}{(c)}
          \fig{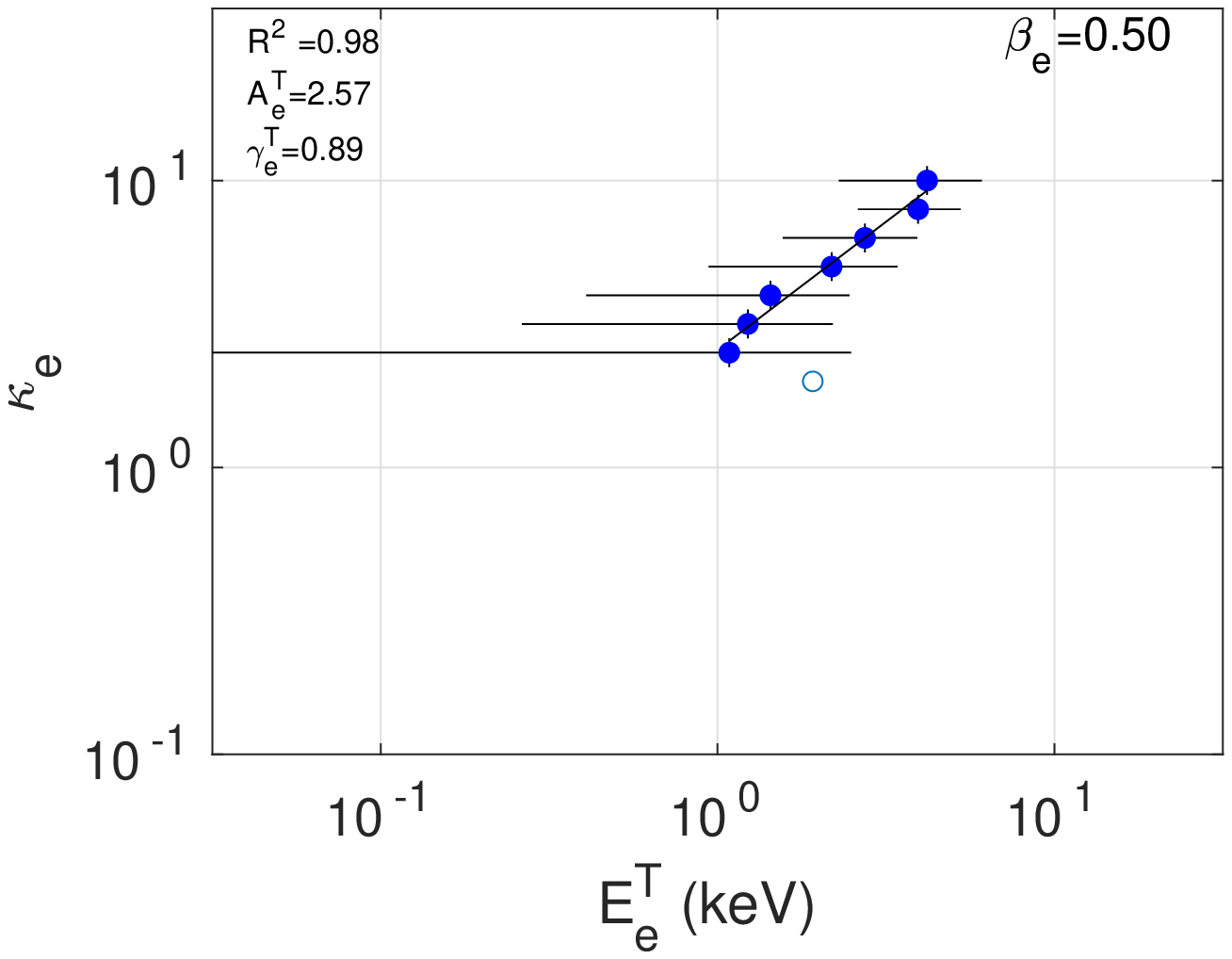}{0.35\textwidth}{(d)}
          }
\gridline{\fig{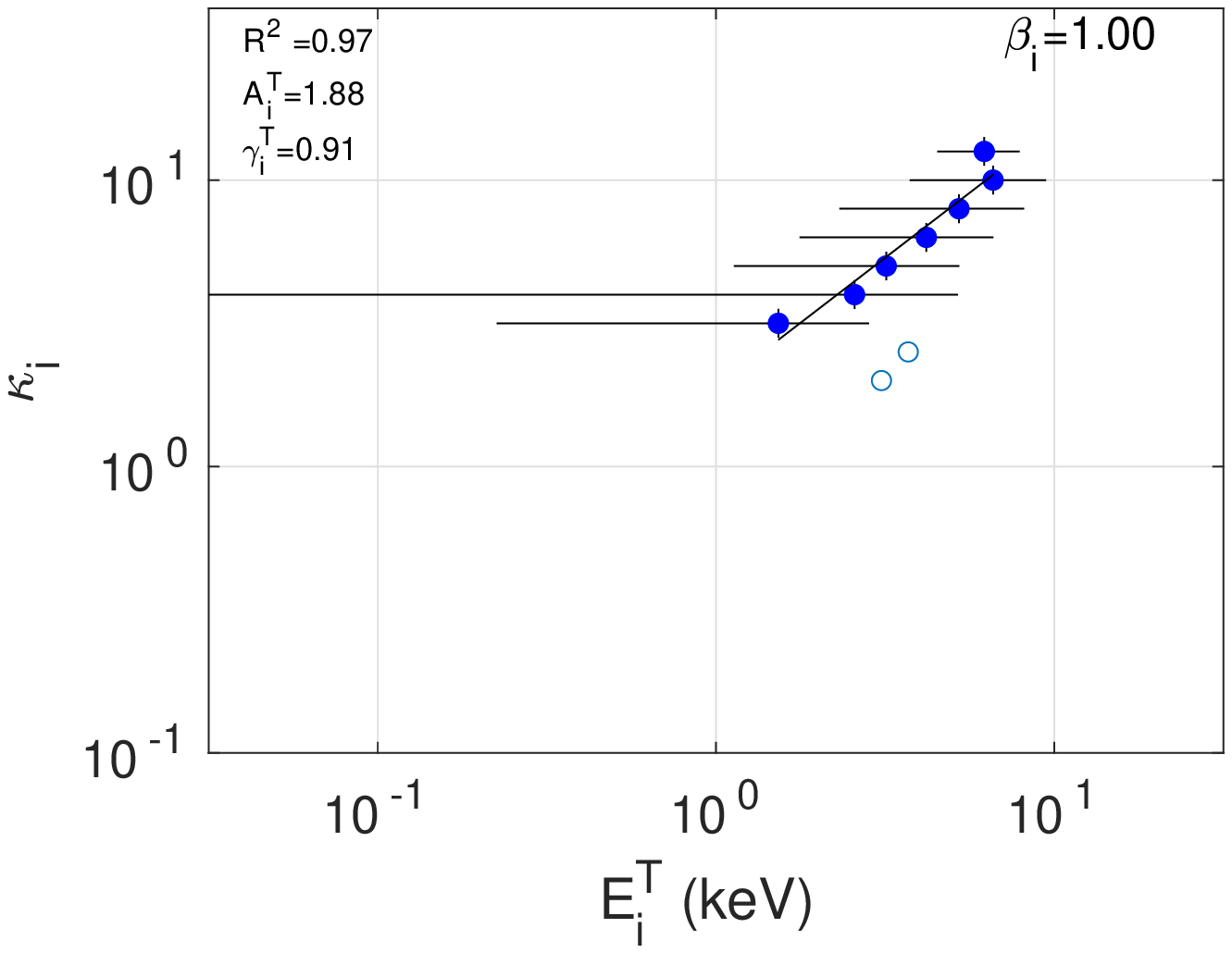}{0.35\textwidth}{(e)}
		  \fig{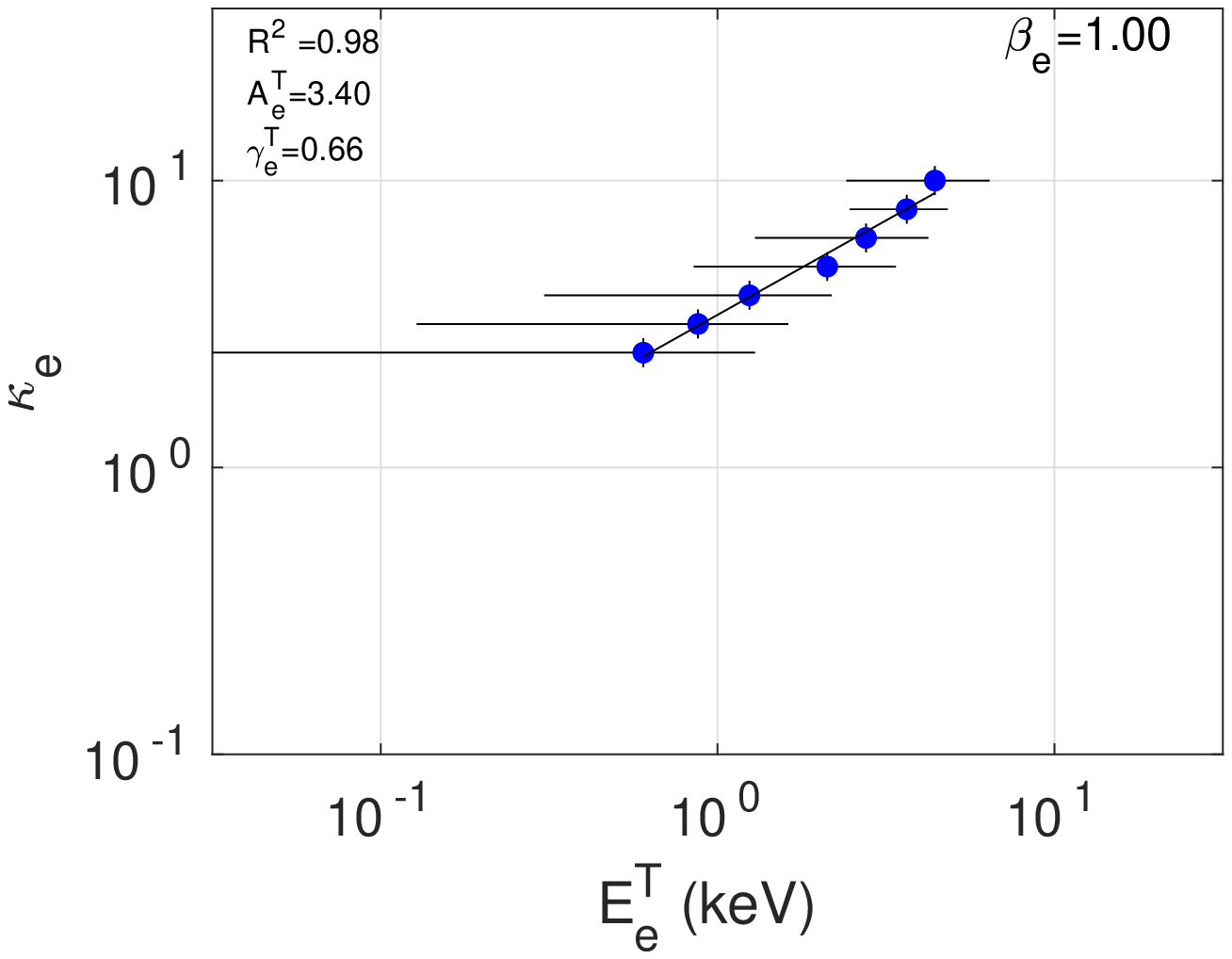}{0.35\textwidth}{(f)}
		  }
\gridline{\fig{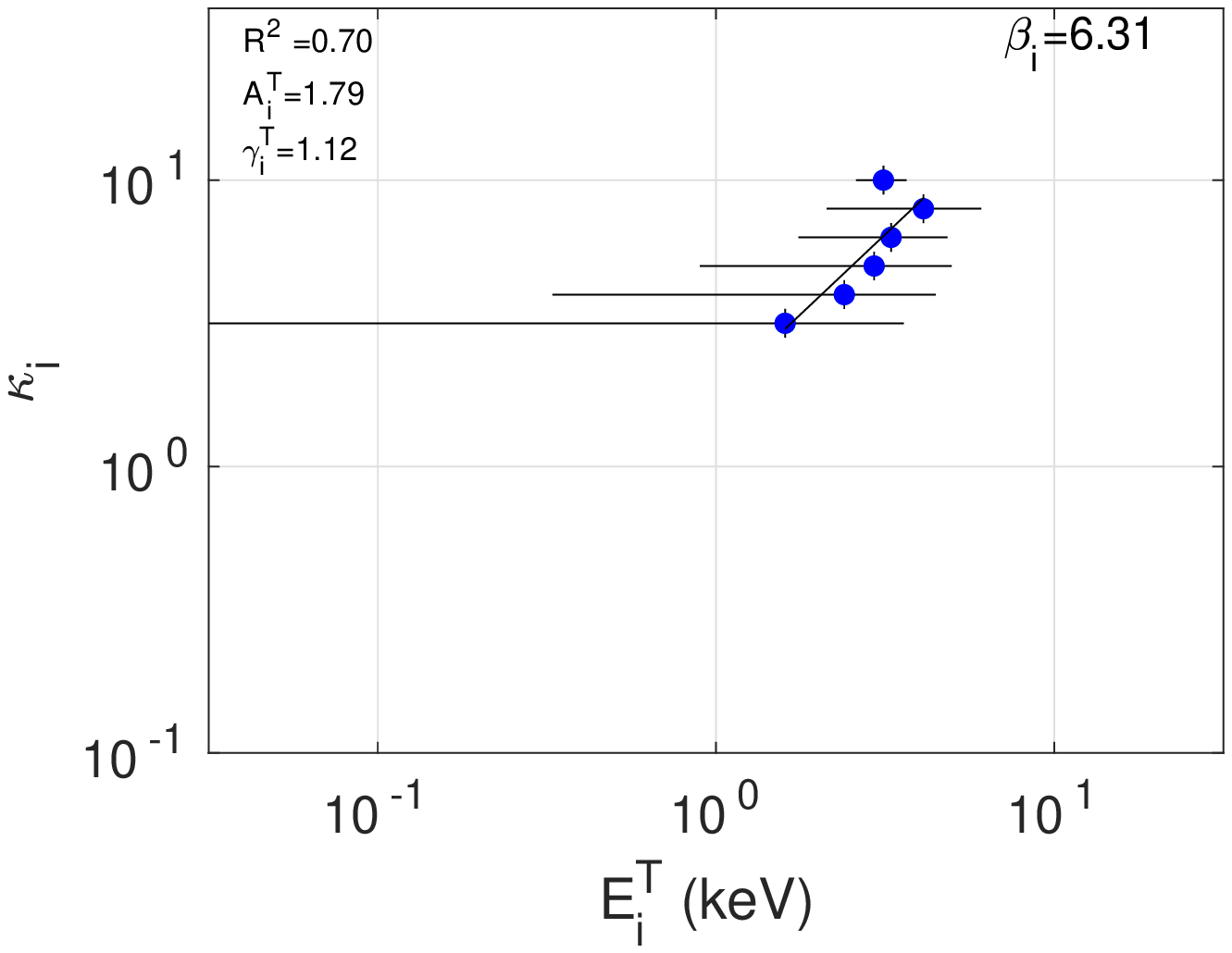}{0.35\textwidth}{(g)}
		  \fig{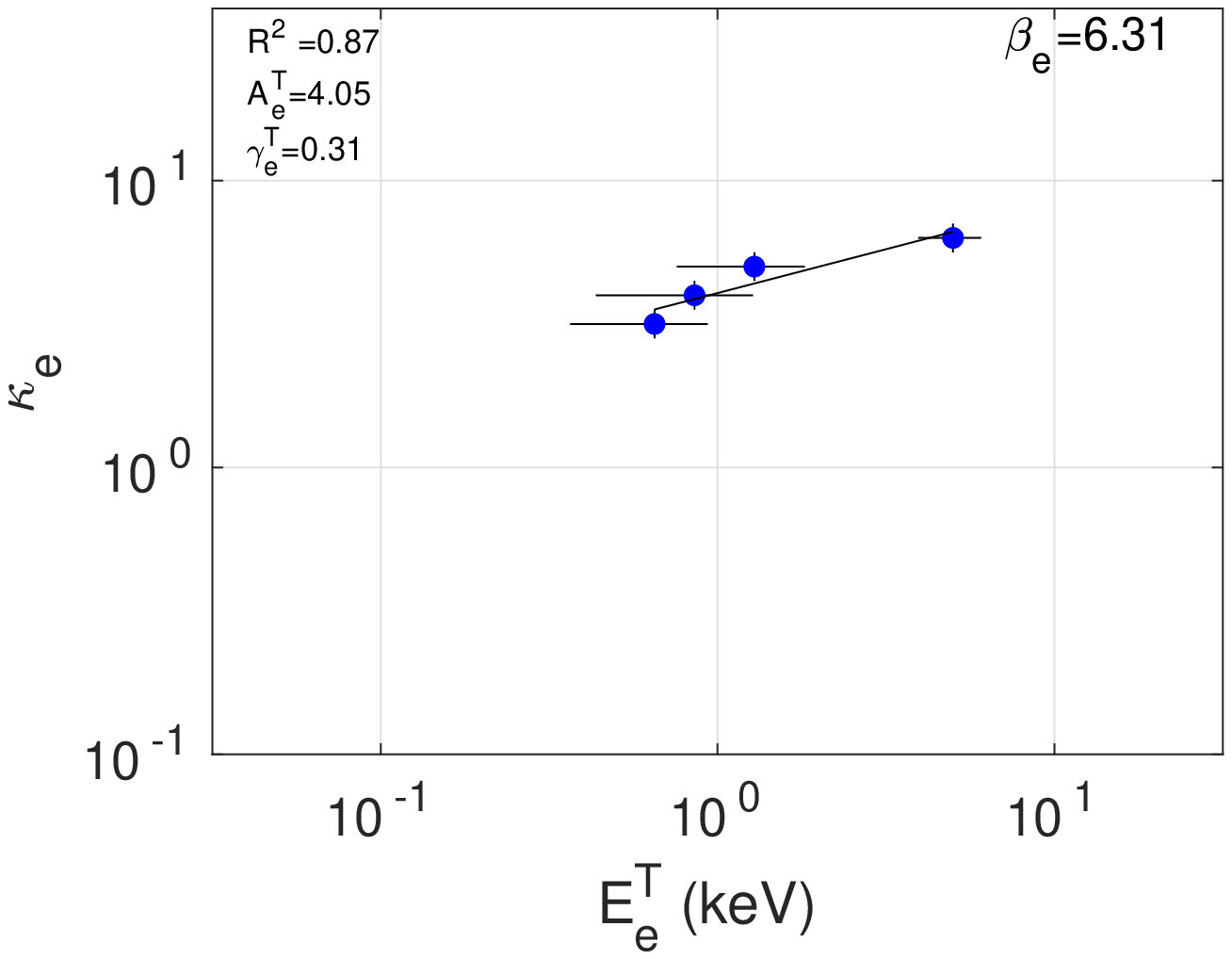}{0.35\textwidth}{(h)}  		  
          }
\caption{Plots of the dependence of ion (left) and electron (right) Total energies with $\kappa$, for different constant $\beta$ values. 
Correlation coefficients and power-law fitting coefficients $A^T$ (Amplitude), and $\gamma^T$ (power-law index) are shown on the plots. }
\label{fig:iANDedependences_total}
\end{figure*}

\begin{figure*}
\gridline{\fig{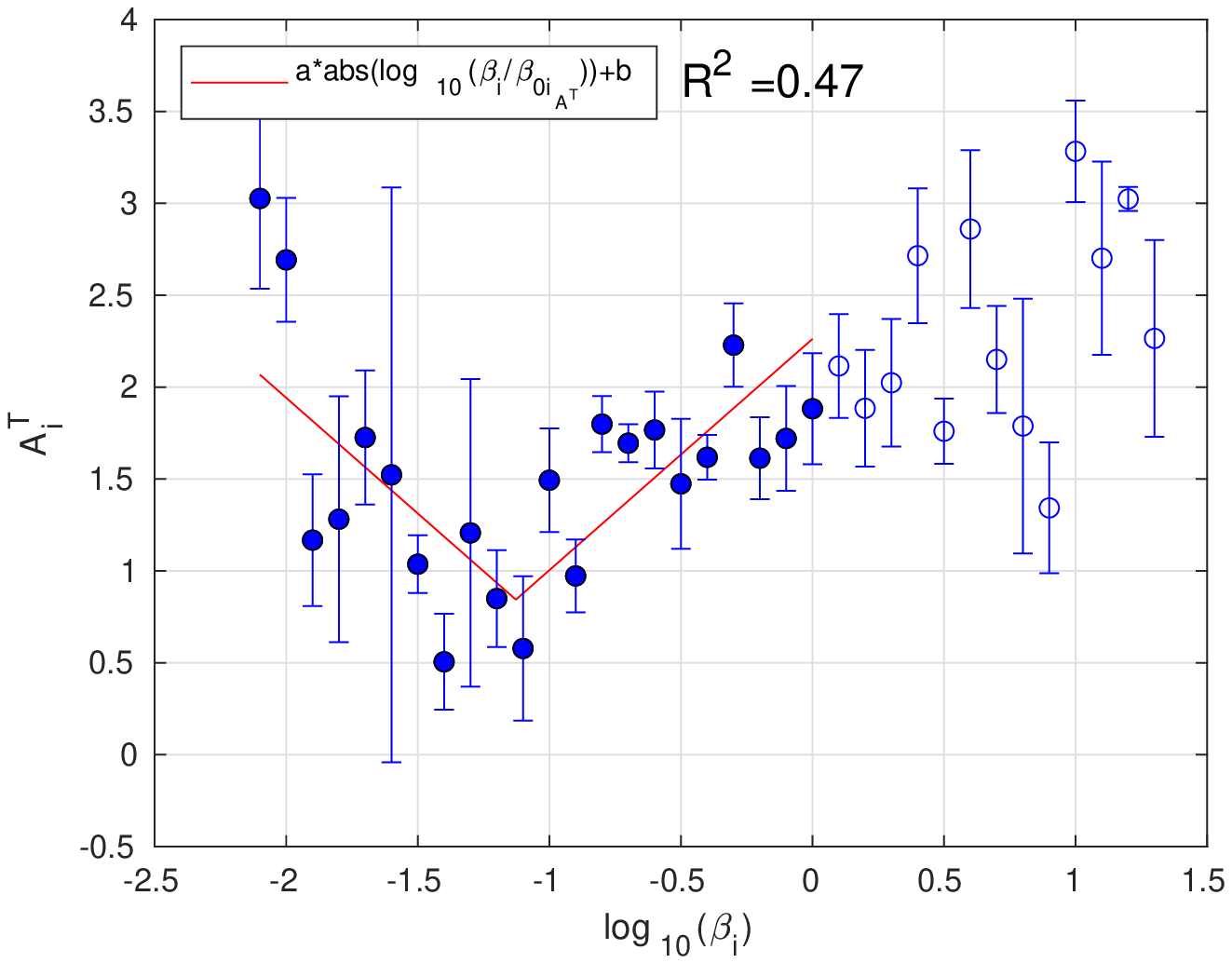}{0.5\textwidth}{(a)}
          \fig{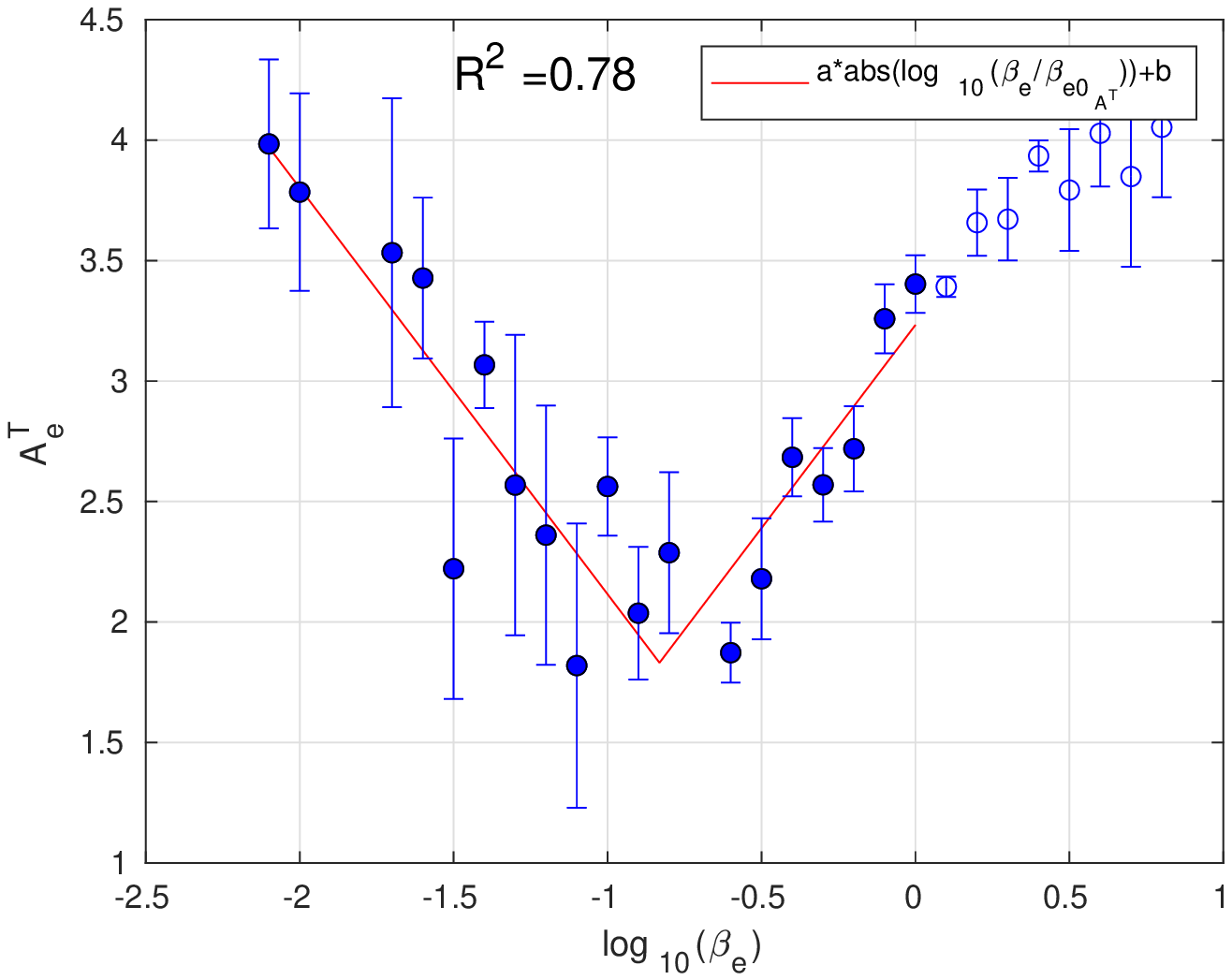}{0.5\textwidth}{(b)}
          }
\gridline{\fig{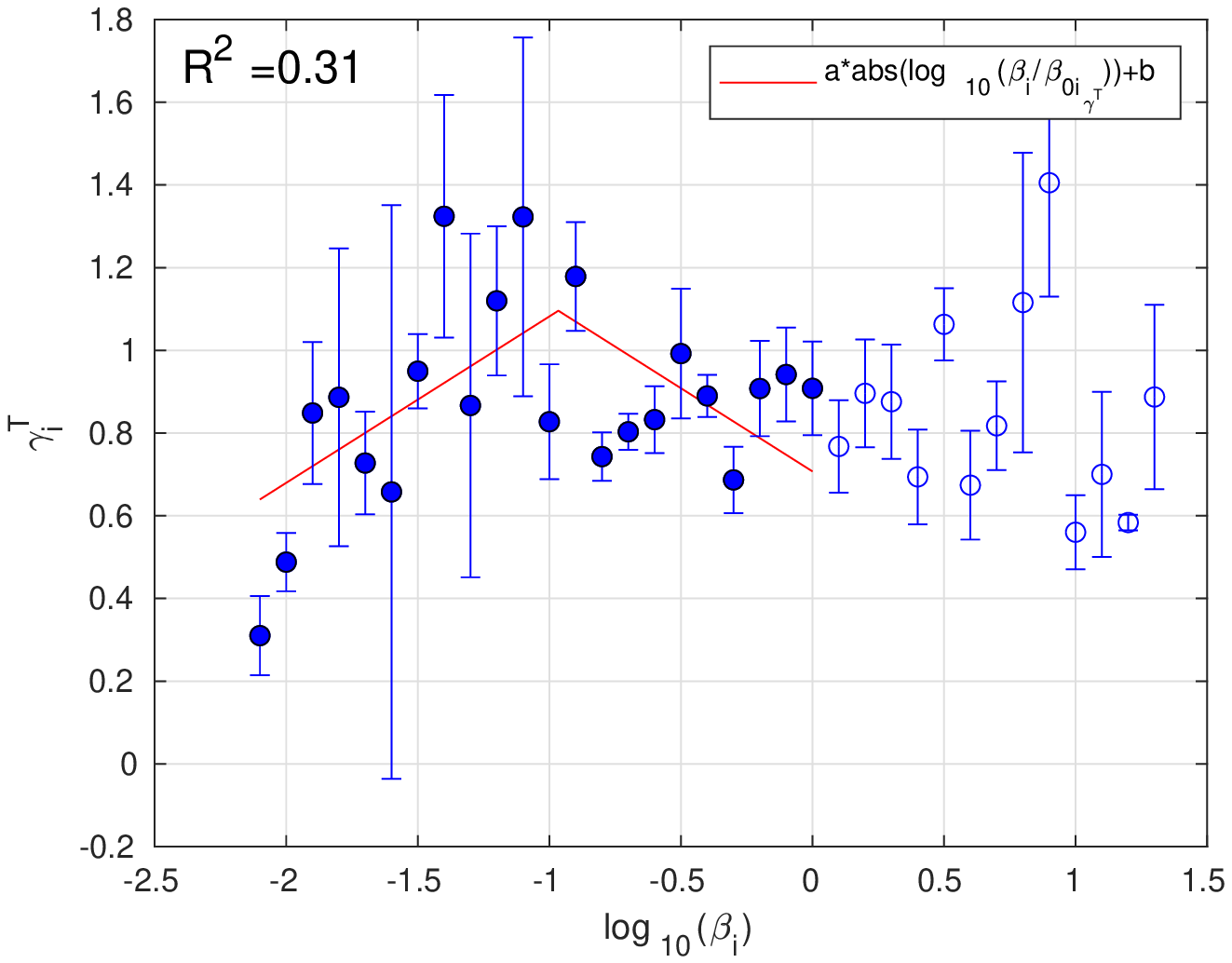}{0.5\textwidth}{(c)}
          \fig{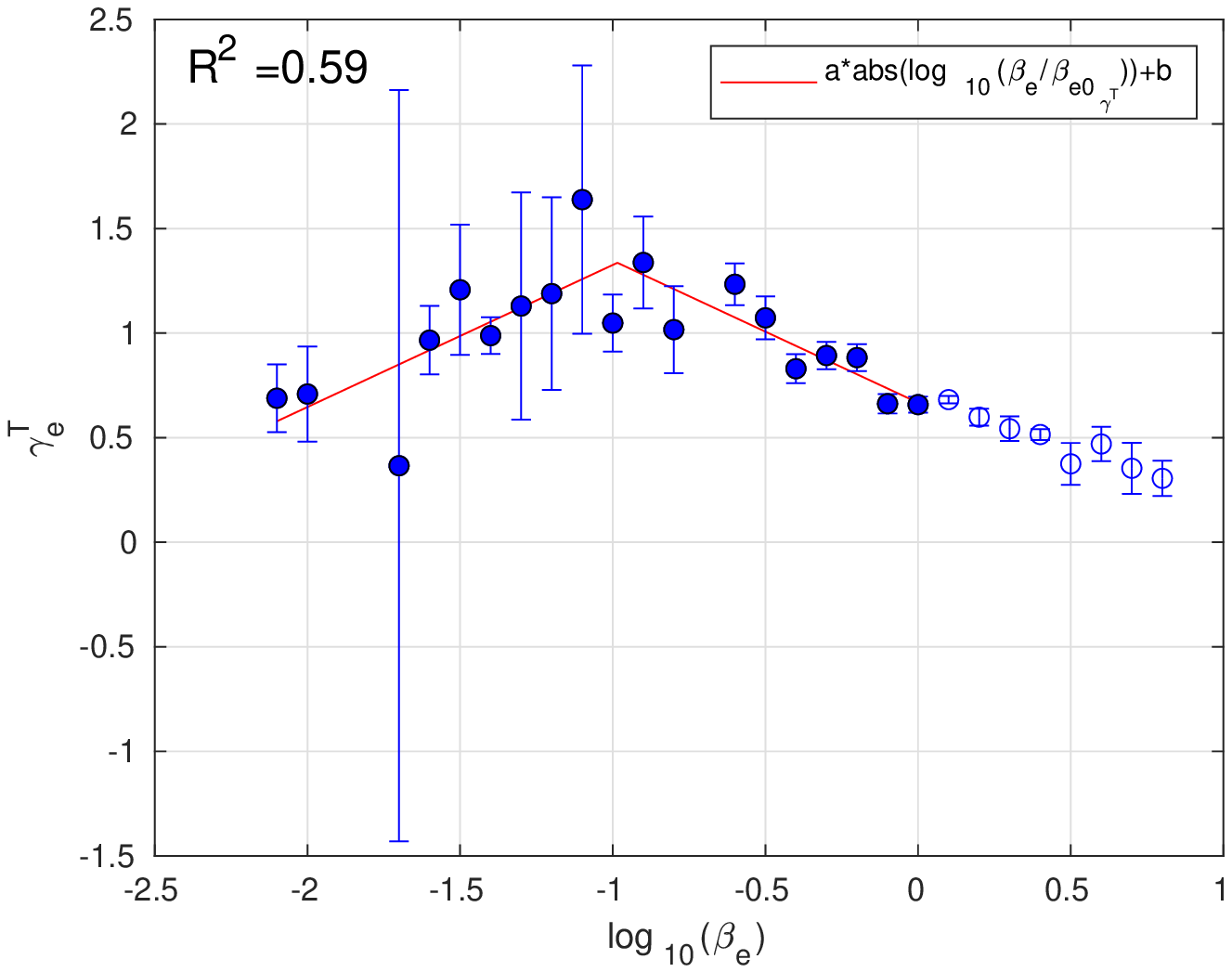}{0.5\textwidth}{(d)}
          }
\caption{Dependency of the fitted power-law coefficient $A^{T}$ (top) and $\gamma^{T}$ (bottom) with $\beta$. 
The left panels are for ions and the right panels for electrons.}
\label{fig:interceptslopecurves_total}
\end{figure*}

As we have mentioned in section \ref{sec:intro}, the total characteristic particle kinetic energy given by Equation~(\ref{eq:2}), enables a straightforward comparison between Kappa and Maxwellian distributions, as it also considers the effect of the suprathermal within the Kappa distribution model. 
In order to see how the inclusion of this effect may alter our conclusions, we have repeated our analysis using $E_{total}$ instead of $E_c$. 
Comparison of Fig. \ref{fig:total_energy} with Figs. \ref{fig:Eciandskew} and \ref{fig:Eceandskew} shows that for every combination between $\beta$ and $\kappa$, $\langle E_{total}\rangle$ is larger than $\langle E_c\rangle$ (see also Tables \ref{table:ionstarvalues} and \ref{table:electronstarvalues}). 
Nevertheless the statistical results are similar and, as expected, noticeable differences appear only for very small $\kappa$ values close to 3/2, where $E_{total} \gg E_{c}$. In addition, we have corroborated that for both species $\kappa$ and $E_{total}$ also follow a power-law relation $\kappa = A^T E^{\gamma^T}_{total}$, with different $A^T$ and $\gamma^T$ parameters for different plasma beta (see Figure \ref{fig:iANDedependences_total}). Further, computing the value of our $A^T$ and $\gamma^T$ as function of beta, our results show that the relation between $E_{total}$ and $\kappa$ is qualitatively the same as the in the case of $E_{c}$ (see Figure \ref{fig:interceptslopecurves_total}).

\begin{deluxetable}{lrcccrccc}
  \tablecaption{Power-law fitting coefficients $A$ (Amplitude) and $\gamma$ ( power-law index) obtained from the dependence of $\kappa$ on $E_c$, and their respective correlation coefficient ($R^{2}$). Values are given for both ion and electron and for some selected values of $\beta$.\label{table:fittingvalues}
  }
  \tablehead{
    \colhead{$\beta$} & & \colhead{$A_{i}$} & \colhead{$\gamma_{i}$} & \colhead{$R^2$} &   & \colhead{$ A_{e}$} & \colhead{$\gamma_{e}$} & \colhead{$R^2$}
  }
  \startdata
0.01 && 2.60 & 0.56 & 0.94 & & 4.88 & 0.82 & 0.93 \\
0.05 && 0.63 & 1.36 & 0.72 & & 3.06 & 1.14 & 0.82 \\
0.1 && 1.02 & 1.22 & 0.79 & & 2.72 & 1.15 & 0.74 \\
0.5 && 2.02 & 0.90 & 0.60 &&  3.65 & 0.73 & 0.99 \\
1 && 2.55 & 0.81 & 0.88 &  & 4.25 & 0.61 & 0.99 \\
5 && 3.10 & 0.72 & 0.94 & & 4.44 & 0.43 & 0.90 \\
10 && 3.91 & 0.49 & 0.92 & & 3.91 & 0.49 & 0.92\\ 
\enddata
\end{deluxetable}

\begin{deluxetable}{lcccc}
  \tablecaption{Measured parameters for the empirical relation $a\vert \log_{10}(\beta/\beta_{0}) \vert + b$ that describes the dependency between $A$ and $\beta$ or between $\gamma$ and $\beta$.\label{table:coeffvalue}
  }
  \tablehead{
    \colhead{Fitting coefficient} & \colhead{$a$} & \colhead{$\beta_{0}$} & \colhead{$b$} & \colhead{$R^2$}
  }
  \startdata
$A_{i}$ & 1.37 & 0.06 & 0.76 & 0.80 \\
$A^T_{i}$ & 1.26 & 0.07 & 0.84 & 0.46 \\
$\gamma_{i}$ & -0.53 & 0.09 & 1.28 & 0.68 \\
$\gamma^T_{i}$ & -0.40 & 0.11 & 1.10 & 0.31 \\
$A_{e}$ & 2.21 & 0.15 & 2.25 & 0.95 \\
$A^T_{e}$ & 1.69 & 0.15 & 1.83 & 0.78 \\
$\gamma_{e}$ & -0.48 & 0.09 & 1.17 &0.82 \\
$\gamma^T_{e}$ & -0.68 & 0.10 & 1.33 & 0.59 \\
\enddata
  \tablecomments{The last column gives the correlation coefficient. $A^T$ and $\gamma^T$ are amplitude and power index obtained from kappa and total energy relation.
  }
\end{deluxetable}

\section{Discussion}
\label{discussion}
The most evident result to emerge from our observation is the clear positive correlation between $\beta_{e}$ and $\beta_{i}$ (see Figure \ref{fig:kvEscatter}(a)), which would be impossible in the absence of an ion and electron temperature correlation, hence a pressure correlation. 
In this case, we expect the ion temperature to be higher than the electron temperature, with a ratio of $ T_{i}/ T_{e}$ between $4$ and $6$ \citep{baumjohann1993near, borovsky1997earth, espinoza18GRL, wang2012spatial}.
Also, there is a lack of correlation between $\kappa_{e}$ and $\kappa_{i}$, which implies that the processes that regulate the macroscopic properties of each species are not related. 
Perhaps the time scales for the energy relaxation of electrons and ions are different. 
Figure \ref{fig:kvEscatter}(c) clearly emphasizes this as $E_{c_{e}}$ versus $E_{c_{i}}$ are only slightly correlated. 


Despite the strong scattering observed for the $\kappa$ values, on average, $\kappa$ increases with $\beta$ up to $\beta=1$, and then it slowly decreases as $\beta$ increases beyond $1$. 
This trend is observed for both species, although the effect is more pronounced for the electrons. This type of analysis was carried out in the solar wind, and a similar trend was noted. Our results are consistent with the previous studies of ion distributions by \citet{livadiotis2018generation} (see their Figure 10(b)). The fact that the same relation between kappa and beta is observed in both systems, and at similar beta values, suggests this behavior is controlled by some universal property present in collisionless plasma systems.

In the case of the magnetosphere, comparing with similar previous studies, we note that in our case, ion $\kappa$ values are lower than those reported by \citet{kirpichev2020dependencies} (see their Figure 3), which may be related to the fact that they studied only some particular regions of the magnetosphere.
However, for both ions and electrons, the $\kappa$ values we found across different regions of the magnetosphere, are similar to those obtained by \citet{espinoza18GRL}, even though that study considered only the plasma sheet and restricted the dataset to $\beta > 1$. This good agreement also suggests that the relation between core energy, plasma beta and the kappa power-index may be explained by basic plasma physics processes rather than phenomenological or specific properties of the Earth's magnetosphere.  

Our analysis of the relationship between $\beta$, $\kappa$, and $E_{c}$ (section \ref{presentation}) shows that some combinations of $\kappa$ and $\beta$ are not present in the analysed plasmas.
For low $\beta$ regimes this may be due to plasma dynamics in which the magnetic field dominates and drives the system towards specific configurations. 
On the other hand, for large $\beta$ regimes, temperature fluctuations or kinetic instabilities are expected to dominate, which might make certain plasma configurations more stable than others. 
A similar phenomenon was previously reported by \citet{livadiotis2018generation} for solar wind observations. Thus, in low and high plasma $\beta$ regimes, the degrees of freedom decrease and make the plasma remain in just a few possible states.

 In Figure \ref{fig:Eciandskew}(a) we observe that on average, ion core energies ($E_{c_{i}}$) are more enhanced as the plasma becomes closer to a Maxwellian: $\kappa_{i} > 8$ for $\langle E_{c_{i}} \rangle > 7 \,$keV. 
The figure also shows that the colder plasma corresponding to $\kappa_{i} \leq 8 $ with energy value $\langle E_{c_{i}} \rangle \leq 3 \,$keV are dominant across all plasma $\beta_{i}$ regimes. In the case of electrons, as shown in Figure \ref{fig:Eceandskew}(a), electron core energies $E_{c_{e}} > 4 \,$keV correspond to larger $\kappa$ values ($\kappa_{e} \geq 10$), thus tend to be closer to thermodynamic equilibrium. Meanwhile, smaller kappa values ($\kappa_{e} \leq 8$) are typical for smaller energies $E_{c_{e}} \leq 2 \,$keV across all plasma $\beta_{e}$ regimes. 
This correlation found between high energies and high values of kappa may suggest that plasma heating and the thermalization of the distribution may be in agreement with \citet{Collier_1999}, who suggested that the increment in kappa depends on ion core temperature, indicating that particle distributions of hotter plasmas tend to Maxwellian functions.



Previous studies have also shown that the $\kappa$-index usually increases with core energy. For instance, \citet{Collier_1999} proposed that these two parameters have a linear relation; meanwhile, \citet{kirpichev2020dependencies} opted for a power-law dependency but using only ion measurements. Our study strongly agrees with the latter, as power-law functions fit well the relation between $\kappa$ and $E_{c}$ for all values of $\beta$ parameter, and for both species, as illustrated in Figure \ref{fig:iANDedependences}. Similar results are obtained for the relation between $\kappa$ and $E_{total}$ (see Figure \ref{fig:iANDedependences_total}). 
Moreover, we have found that the coefficients of power-law fitting $A$ (amplitude) and $\gamma$ (power-law index) depend strongly on $\beta$, for both species, such that different $\beta$ values yield different values for $A$ and $\gamma$ (see Figures \ref{fig:interceptslopecurves} and \ref{fig:interceptslopecurves_total}). 
These results are in contrast with \citet{kirpichev2020dependencies}, who obtained practically constant values for the power index $\sim 0.5$. 
The discrepancy can be attributed to the fact that they applied different, stronger restrictions to select plasmas for their analyses.


Further, the increment in kappa value with energy is attributed to positive values of the power-index ($\gamma$), consistent with the result showing that larger values of the core energy correspond to larger $\kappa$. 
However, we found the relation is non-linear. 
In low beta plasma ($\beta < 0.1$), the relation between kappa and core energy exhibits a small $\gamma$ value for ions and electrons, and the same happens in the high beta regime  $(\beta > 0.1)$ as seen in Figure \ref{fig:interceptslopecurves}. Meanwhile, for a specific range of beta value ($\beta \sim 0.1-0.3$) the relation between kappa and energy is stronger showing a maximum value of $\gamma$ for both species. As previously mentioned, this is consistent with~\citet{livadiotis2018generation}. Thus, the presence of extremums near $\beta \sim 0.1$, as shown in Figure \ref{fig:interceptslopecurves}, indicates the existence of two regimes. Regardless the details of the solar wind or the magnetosphere, in poorly collisional space plasma systems it is expected plasma magnetization to dominate in the low beta regime, and therefore the relation between $\kappa$ and core energy should be stronger. On the other hand, in the case of a large beta plasma correlations are mainly due to temperature fluctuations such that the magnetic field becomes less relevant for the determination of $\kappa$. 

Between these two separated regimes there is an intermediate range in which other plasma effects should be relevant. For instance, the difference between cold and hot plasma, or the effectiveness of instabilities that was responsible for the power-law coefficients trend in  Figure \ref{fig:interceptslopecurves}. This is consistent with theoretical linear analysis done by \citet{mace2010parallel} in which deviations from the cold plasma approximation of whistler waves propagating through a bi-Kappa distributed plasma were found to be relevant for $\beta>0.1$, such that $\beta \sim 0.1$ may correspond to a transition value between cold plasmas ($\beta <0.1$), where the magnetic field controls the dynamics, and hot plasmas ($\beta >0.1$) in which temperature effects become relevant. 

Recently a similar behavior was described by \citet{lopez2020alternative}, in which different regimes of plasma waves and instabilities were found to be strongly dependent on plasma beta. 
For instance, at $\beta < 0.1$ there is a dominance of propagating parallel or obliquely to the magnetic field; while  at $\beta > 0.1$ the characteristics of different relative drifts of suprathermal electrons are dominant, which can be in electrostatic mode. 
Moreover, this is also  consistent with \citet{moya2020towards}, where they allowed kappa to evolve in time due to the quasi-linear relaxation of electron cyclotron instability. 
Their results suggest that kappa is not affected by the instability in the low plasma beta regime ($\beta < 0.1$), that for $\beta\sim0.3$ the relaxation of the instability resulted in a larger variation of $\kappa$, while in the high plasma beta regime ($\beta > 0.5$) kappa remains more or less the same.
 
\section{Summary and Conclusion}
\label{conclusion}
The observations of ion and electron energy fluxes, made by the THEMIS mission instruments (section \ref{sec:dataanalysis}), have provided the opportunity to study the behavior of the Kappa distribution parameters for 47,058 cases, in which the spectra of both species were successfully modelled with Kappa functions. 
The plasma measurements were made in the geomagnetic tail and in the plasma that surrounds the Earth beyond 7 R$_{E}$. 
More specifically, the studied region corresponds to $-35 \leq X \leq 7$ R$_{E}$, $-30 \leq Y \leq 30$ R$_{E}$, and $-10 \leq Z \leq 10$ R$_{E}$ and only plasmas that satisfy the restrictions summarised in Table \ref{table:parametersrestrictions} were considered.
We studied the relationship between $\kappa$ and $E_c$ for a wide range of the plasma $\beta$ parameter, from $10^{-3}$ to $10^{2}$. 

This is the first time this type of research is carried out using THEMIS data covering different regions of the magnetosphere, and one important finding of our research is the  presence of Kappa distributions in many regions of the magnetosphere. On average, the $\kappa$ indices of ions and electrons were found to increase with $\beta$, up to $\beta\sim1$ (Figure \ref{fig:numofOBS}). 
This tendency seems more pronounced for the electrons. 
However, for $\beta>1$ the $\kappa$ indices tend to decrease slowly. 
In addition, certain combinations of $\kappa$ and $\beta$  are absent in the studied plasmas (for instance $\beta_{i}<0.01$ and $\kappa_{i}>10$ for ions, or $\beta_{e}>30$ and $\kappa_{e}>6$ for electrons; see Fig. \ref{fig:numofOBS}). 
Such division means that the magnetic field plays a crucial role in the relative number of energetic particles and the presence of high energy tails in the distributions. Our results show also that systems with large ion kappa indices ($\kappa \geq 10$) appear to have higher core energies, $E_{c_i}\geq5$\,keV (Figure \ref{fig:Eciandskew}(a)). 
The same trend is observed for electrons, as systems with $\kappa_e>8$ have energies $E_{c_e}\geq3$\,keV (Figure \ref{fig:Eceandskew}(a)). Thus, the hottest plasmas tend to be in states closer to thermodynamic equilibrium (as their distributions approach a Maxwellian). 

A more detailed study for both species shows a robust correlation of the form $\kappa=AE_c^\gamma$, showing that kappa increases with energy, and the relation between $\kappa$ and core energy is stronger for a specific range of beta value ($\beta \sim 0.1-0.3$), for both ions and electrons. 

This observation may reflect the level influence of the magnetization of the system on the propagation of plasma waves and instabilities \citep[as suggested by][on the basis of Vlasov linear theory]{mace2010parallel,lopez2020alternative}; or the effectiveness of the relaxation process of kinetic instabilities to induce changes on plasma parameters, including $\kappa$ \citep{moya2020towards}.
%

%
Moreover, we found that both $A$ and $\gamma$ depend on $\beta$, the values of $A$ are found to exhibit a minimum near $\beta\sim0.1$; while the power-law indices $\gamma$ exhibit a maximum at around the same value of $\beta$. The observed trend for both species suggests a universal plasma transition at around $\beta\sim0.1$.

Waves, instabilities, and temperature fluctuations may have stronger or lesser effect on the observed $\kappa$ indices
depending on the value of $\beta$.
%
%
When $\beta$ is small (i.e. in a strongly magnetized plasma), $\kappa$ is more dependant on the magnetic field than on instabilities or fluctuations; and therefore plasma waves and instabilities have little effects on kappa.
There exist a transition near $\beta \sim 0.1-0.5$ (i.e. for less magnetized plasmas), in which the system becomes more susceptible to plasma waves and instabilities, which after relaxation can result in significant changes of plasma parameters, including $\kappa$, which is consistent with the results presented in Figure \ref{fig:interceptslopecurves}. Finally, for $\beta>0.5$, kinetic instability thresholds prevent the plasma parameters to deviate from a quasi-equilibrium state, and the value of $\kappa$ is expected to be determined mostly by temperature fluctuations. 
Consistently, Figure \ref{fig:interceptslopecurves} shows that for $\beta>1.0$, both $A$ and $\gamma$ deviate from the observed trend and tend to take nearly constant values. 
This effect is particularly noticeable for $A_{e}$, which may be evidence of electron demagnetization, which often takes place in highly fluctuating electric and magnetic fields \citep{antonova1999generation}, and that are commonly observed at high plasma beta ($\beta\geq1$) \citep{valdivia2016magnetic}.

The results we have obtained are remarkably similar to the case of solar wind, suggesting a universal behaviour of Kappa distributions in poorly collisional plasmas. The relation we have found between kappa, energy, and beta with the transition at a particular value of beta is a significant result that can be used by theoretical models in the future. Besides, these relation is much more complex than what was reported in the previous studies. A comprehensive insight into the behavior of these parameters should be investigated further using theoretical models. We expect our results to motivate the space and astrophysical plasma physics community to consider more in-depth studies using more realistic theoretical models and simulations to unravel the dynamics responsible for such behavior.



\acknowledgments
This work was supported by Agencia Nacional de Investigación y Desarrollo de Chile (ANID) grants 21181777 and 1191351 (P.S.M.). M.S. acknowledges support from  Universidad de Santiago de Chile through the grant DICYT 042031S. C.M.E. and M.S. acknowledge support from AFOSR (FA9550-19-1-0384).
We acknowledge NASA contract
NAS5-02099 and V. Angelopoulos for the use of data from the THEMIS mission, specifically C.W. Carlson and J.P. McFadden for the use of ESA data, D. Larson for the use of SST data, and K.H. Glassmeier, U. Auster, and W. Baumjohann for
the use of FGM data. The data of the THEMIS satellite mission used in this paper are available on THEMIS mission website: http://themis.ssl.berkeley.edu/index.shtml.
\appendix
\section{Median and Mode $E_c$ values as a function of $\kappa$ and $\beta$}
\begin{figure*}
\gridline{\fig{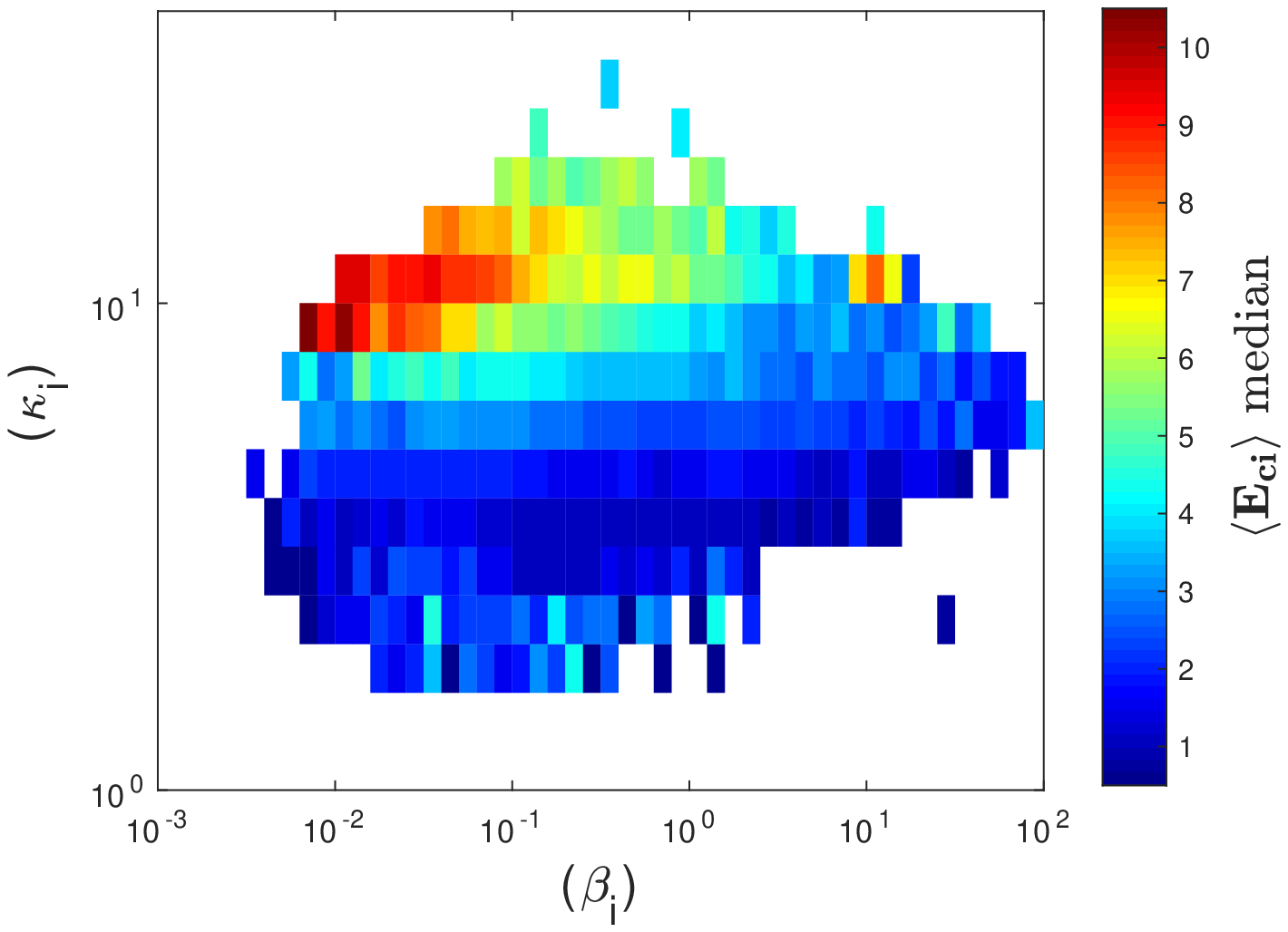}{0.5\textwidth}{(a)}
		  \fig{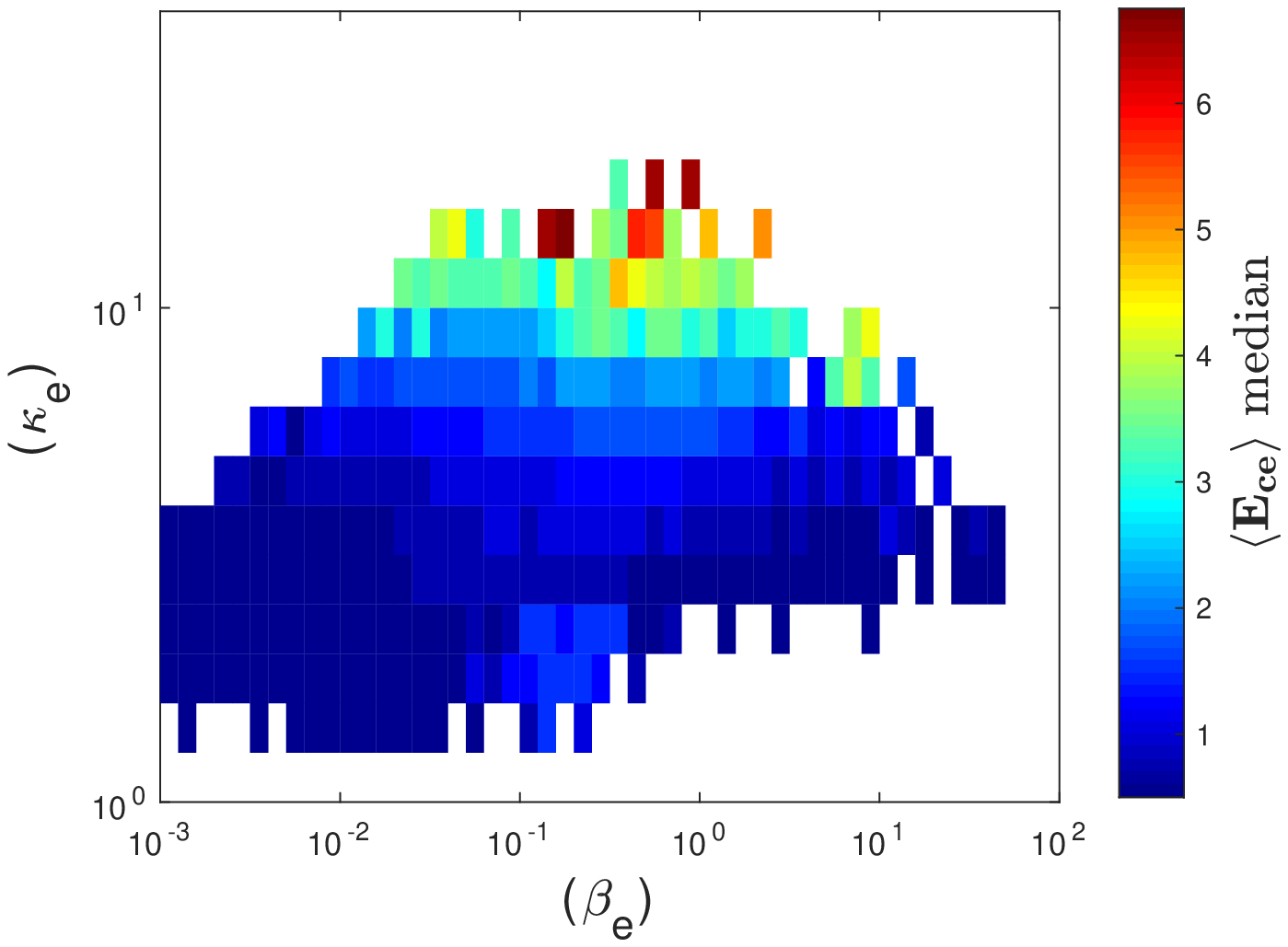}{0.5\textwidth}{(b)}
          }
\caption{
2-D distribution of the medians core energy values for each cell in the $\beta_{i} - k_{i}$ plane for ions (left) and electrons (right). 
}
\label{fig:medianEnergy}
\end{figure*}

\begin{figure*}
\gridline{\fig{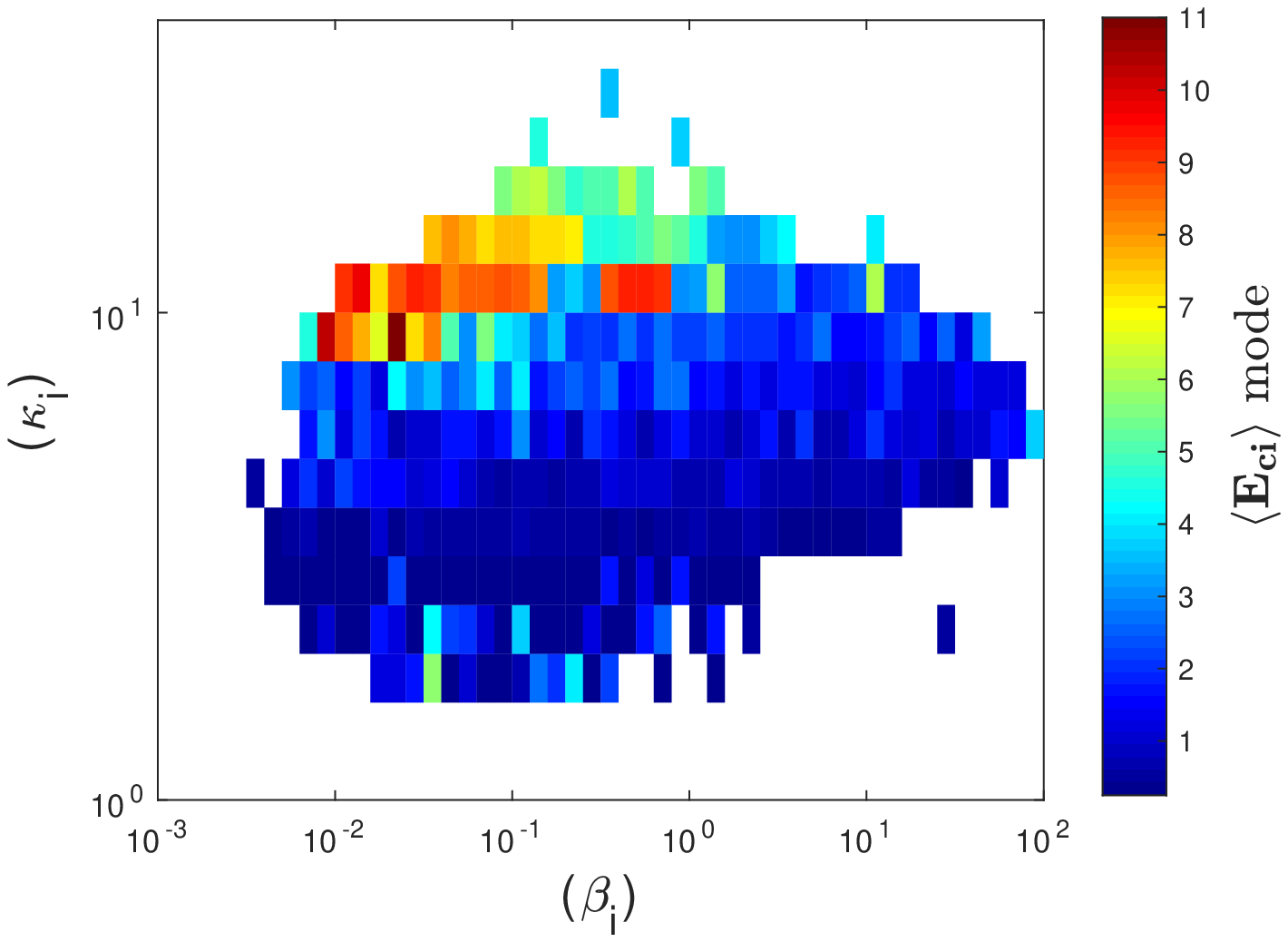}{0.5\textwidth}{(a)}
		  \fig{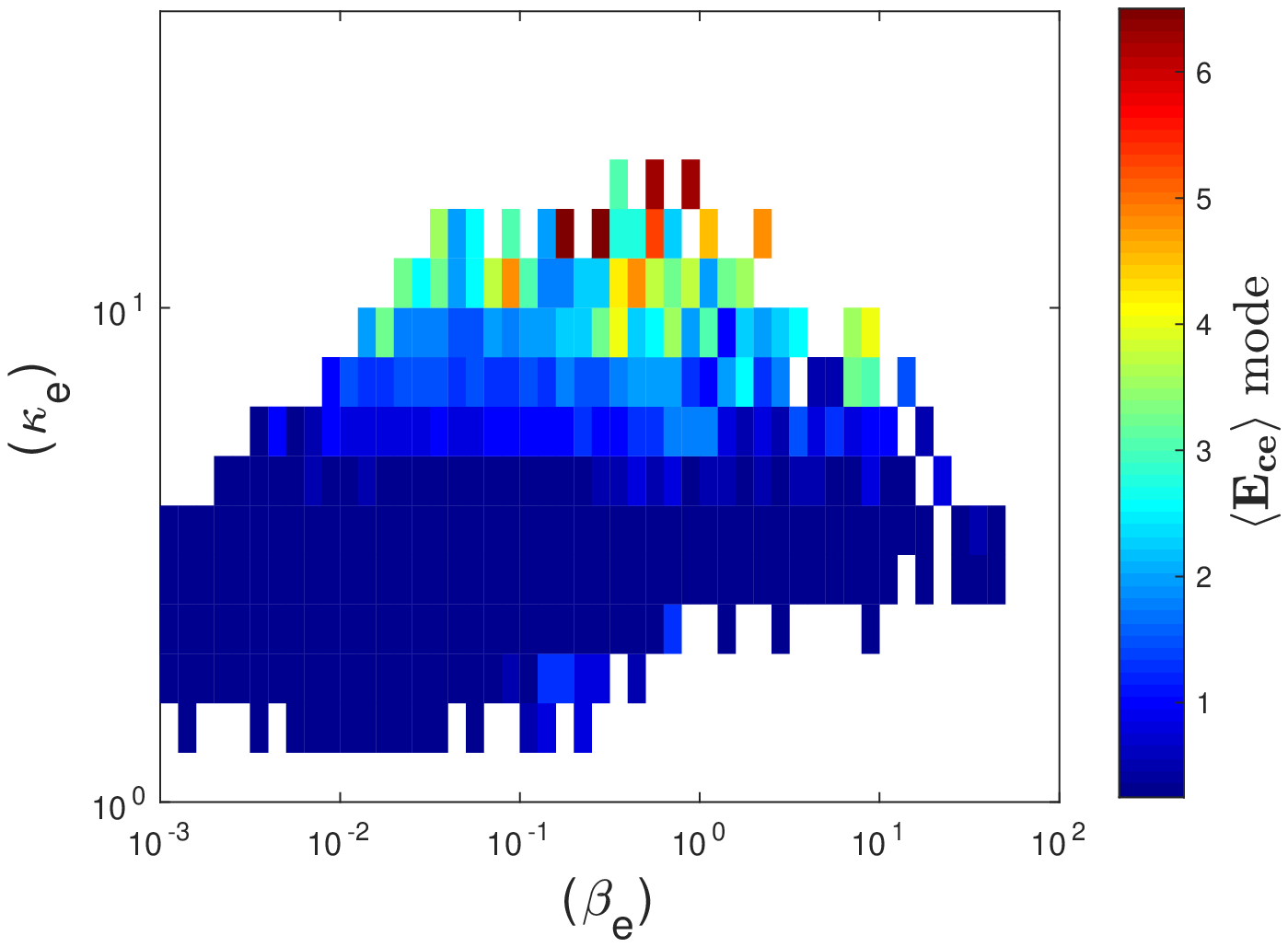}{0.5\textwidth}{(b)}
          }
\caption{
2-D distribution of the modes core energy values for each cell in the $\beta_{i} - k_{i}$ plane for ions (left) and electrons (right). 
}
\label{fig:modeEnergy}
\end{figure*}

Figs. \ref{fig:medianEnergy} and \ref{fig:modeEnergy} show the distribution of Median and Mode $E_c$ values, respectively, in the $\kappa$ and $\beta$ space. 
These figures can be compared to Figs. \ref{fig:Eciandskew} and \ref{fig:Eceandskew}.

\bibliography{new_kappa_beta_ref}{}
\bibliographystyle{aasjournal}



\end{document}